\documentclass[12pt, a4paper]{article}

\usepackage[height=25cm,width=16cm]{geometry}
\usepackage{feynmp}
\usepackage{amsmath}
\usepackage{ascmac}
\usepackage{dcolumn}
\usepackage{bm,here}
\usepackage{subfig}
\usepackage{comment}
\usepackage{ifpdf}
\usepackage{slashed}
\usepackage{colortbl}
\usepackage{color}
\usepackage{comment}
\usepackage[mathscr]{eucal}
\usepackage[sort&compress, numbers, merge]{natbib}
\usepackage{booktabs}

\usepackage{cancel}
\usepackage{rotating}

\ifpdf
\usepackage{graphicx}     
\usepackage[bookmarksopen,colorlinks=true,linkcolor=bblue,citecolor=ppink,urlcolor=ppink]{hyperref}
\else     
\usepackage[dvipdfmx]{graphicx}     
\usepackage[dvipdfmx,bookmarksopen,colorlinks=true,linkcolor=bblue,citecolor=ppink,urlcolor=ppink]{hyperref}
\fi

\usepackage{tikz}
\usetikzlibrary{shapes.geometric, arrows}

\usepackage{multicol}
\definecolor{red}{rgb}{1,0,0}
\definecolor{ppink}{rgb}{0.921545,0.440586,0.687243}
\definecolor{bblue}{rgb}{0.400000,0.400000,1.000000}
\usepackage[charter]{mathdesign}
\usepackage{soul}
\usepackage{wrapfig}

\usepackage[capitalise]{cleveref}
\usepackage{physics}
\usepackage{aastex}
\usepackage{orcidlink}

\newcommand\blfootnote[1]{%
	\begingroup
	\renewcommand\thefootnote{}\footnote{#1}%
	\addtocounter{footnote}{-1}%
	\endgroup
}
\newcommand{\tableandfigurefontsize}{\small}
\newcommand{\smalltableandfigurefontsize}{\footnotesize}
\DeclareCaptionFormat{cont}{#1 (cont.)#2#3\par}
\newcommand{\posteriorfigwidth}{4.8cm}

\newcommand{\mydefcitealias}[3]{\defcitealias{#1}{#2}\newcommand{#3}{\citetalias{#1}}}
\mydefcitealias{2013ApJ...770...57B}{PB13}{\pbi}
\mydefcitealias{2013MNRAS.428.3121M}{BM13}{\bmi}
\mydefcitealias{2019MNRAS.488.3143B}{PB19}{\pbii}
\mydefcitealias{2018MNRAS.477.1822M}{BM18}{\bmii}
\mydefcitealias{Ando:2020yyk}{SA20}{\ando}

\newcommand{\Vpeak}{V_\text{peak}}
\newcommand{\sigmalossq}{\sigma_\text{los}^2}

\begin{document}
	
\begin{titlepage}
		
\begin{flushright}
    \hfill IPMU22-0038 \\
\end{flushright}
		
\begin{center}

	\vskip 1.5cm
	{\Large \bf Cosmological prior for the $J$-factor estimation of dwarf spheroidal galaxies}

	\vskip 2.0cm
	{\large Shun'ichi Horigome\,\orcidlink{0000-0001-5251-2284}$^{1,a}$\blfootnote{$^a$shunichi.horigome@ipmu.jp},
			Kohei Hayashi\,\orcidlink{0000-0002-8758-8139}$^{2,3,4,b}$\blfootnote{$^b$kohei.hayashi@g.ichinoseki.ac.jp}, 
			and
			Shin'ichiro Ando\,\orcidlink{0000-0001-6231-7693}$^{5,1,c}$\blfootnote{$^c$s.ando@uva.nl}
			}
			
	\vskip 2.0cm
	{\sl $^1$  Kavli IPMU (WPI), UTIAS, University of Tokyo, Kashiwa, Chiba, 277-8583, Japan} \\ 
	{\sl $^2$ National Institute of Technology, Ichinoseki College, Takanashi, Hagisho, Ichinoseki, Iwate, 021-8511, Japan} \\ 
	{\sl $^3$ Astronomical Institute, Tohoku University, 6-3 Aoba, Aramaki, Aoba-ku, Sendai, Miyagi 980-8578, Japan} \\
    {\sl $^4$ Institute for Cosmic Ray Research, The University of Tokyo, Chiba, 277-8582, Japan} \\
	{\sl $^5$ GRAPPA Institute, Institute of Physics, University of Amsterdam, 1098 XH Amsterdam, The Netherlands} \\
	[.3em]
	
    \vskip 1.5cm
    \begin{abstract}
        \noindent
        Dark matter halos of dwarf spheroidal galaxies (dSphs) play important roles in dark matter detection. Generally we estimate the halo profile using a kinematical equation of dSphs but the halo profile has a large uncertainty because we have only a limited number of kinematical dataset. In this paper, we utilize cosmological models of dark matter subhalos to obtain better constraints on halo profile of dSphs. The constraints are realized as two cosmological priors: satellite prior, based on a semi-analytic model of the accretion history of subhalos and their tidal stripping effect, and stellar-to-halo mass relation prior, which estimates halo mass of a galaxy from its stellar mass using empirical correlations.
        In addition, we adopt a radial dependent likelihood function by considering velocity dispersion profile, which allows us to mitigate the parameter degeneracy in the previous analysis using a radial independent likelihood function with averaged dispersion. 
        Using these priors, we estimate the squared dark matter density integrated over the region-of-interest (so-called $J$-factor) of 8 classical and 27 ultra-faint dSphs. Our method significantly decreases the uncertainty of $J$-factors (upto about $20\%$) compared to the previous radial independent analysis. We confirm the model dependence of $J$-factor estimates by evaluating Bayes factors of different model setups and find that the estimates are still stable even when assuming different cosmological models.

    \end{abstract}
			
\end{center}
		
\end{titlepage}
	
\tableofcontents
\newpage
\setcounter{page}{1}
	
\section{Introduction}
\label{sec: intro}

The presence of dark matter in our universe is one of the most important open questions in the current physics. Even though cosmological observations agree with predictions of $\Lambda$-cold dark matter (CDM) model with surprising accuracy~\cite{Planck:2018vyg}, we still do not know what dark matter is. In order to answer this question, many candidates and detection methods have been proposed~\cite{Feng:2010gw}.
Among various detection methods, those using dwarf spheroidal galaxies (dSphs) are interesting. DSphs are a kind of satellite galaxy of the Milky Way with large mass-to-light ratio, which implies that they are dark matter rich objects. Such a large amount of dark matter enables us to explore the nature of dark matter. In the indirect detection method, we can constrain the annihilation cross section by observing signal flux of annihilation products. This method is very useful to detect well-motivated dark matter candidates such as the weakly interacting massive particles (WIMPs)~\cite{Bhattacherjee:2014dya,Ando:2021jvn}, because they have large annihilation cross section thanks to a non-relativistic quantum effect, so-called the Sommerfeld effect~\cite{Hisano:2003ec,Hisano:2004ds}. 
The sensitivity of the detection depends on the estimated amount of signal flux. This amount is proportional to an astrophysical quantity, so-called $J$-factor:
\begin{equation}
    J(\Delta\Omega) \equiv \int\dd{\Omega}\int\dd{l}\rho^2(r)
\end{equation}
where $\Omega$ is the solid angle, $\Delta\Omega$ is the region-of-interest, $l$ denotes the line-of-sight distance, $r$ is the radius from the center of a target dSph, and $\rho(r)$ denotes the dark matter density profile. In order to obtain reliable and useful results, we need to know the accurate and precise value of $J$-factor, that is, the shape of dark matter density profile $\rho(r)$.


Although dSphs are useful tools to study the dark matter, their dark matter density profiles have large astrophysical uncertainty compared to other uncertainties from particle physics models. 
Generally, the dark matter density profile is estimated by fitting spectroscopic data set of dSph member stars using the Jeans equation~\cite{1987gady.book.....B}.
However, the stellar data set cannot completely determine the dark matter profile because we generally use empirical models of the dark matter profile through the fitting and their parameters are degenerated. 
Fortunately, from the viewpoint of cosmology, structure formation models predict the distribution of dSph profiles in the universe, which is useful to select theoretically favored density profile model parameters and mitigate the problematic degeneracy.
For instance, Ref.~\cite{Geringer-Sameth:2014yza} use the conventional theory of spherical collapse to roughly constrain the parameter space of dark matter density profile.

Recently, semi-analytic models of the tidal mass striping effect on the cold dark matter halo are developed~\cite{Hiroshima:2018kfv,Ando:2019xlm}, which allows us to construct a multivariate distribution of dSph mass and its tidal truncation radius. This probability distribution was applied as a prior (called \emph{satellite prior}) for the $J$-factor estimation of dSphs by fitting their averaged velocity dispersion~\cite{Ando:2020yyk} and it was shown that the satellite prior has a potential to break the degeneracy among parameters of the NFW profile~\cite{1997ApJ...490..493N}.

While the satellite prior gives a statistical trend for the whole dSphs, cosmology offers us another way to constrain the dark matter mass of individual dSph. Since the dark matter plays an essential role to form the structure of the universe including dSphs, the stellar components of dSphs has a strong relation to their dark matter halo, known as the stellar-to-halo mass relation (SHMR)~\cite{Wechsler:2018pic}. The dark matter halo mass in each dSph can be therefore constrained by its stellar mass by using this relation.

In this paper, we perform a more detailed analysis of the satellite prior in Ref.~\cite{Ando:2020yyk} (hereafter \ando) by considering the radial dependence of the velocity dispersion to optimize given kinematical data sets. The radial dependency weakens the degeneracy among dark matter halo parameters and gives more precise estimation than the radial independent analysis.
Moreover, we consider some SHMR models to obtain more reasonable estimation results from the viewpoint of cosmology. These models help us to obtain more accurate constraints of the halo parameters than the previous satellite prior only analysis. From the other point of view, our analysis also provides a method for evaluating the credibility of each SHMR model using dark matter halos of dSphs.
This paper is organized as follows: In \cref{sec:method}, we discuss our analysis method. In \cref{subsec:likelihood}, we describe our model setups and assumptions on the dSph system and introduce our likelihood function. In \cref{subsec:priors}, we explain the construction of the satellite prior and the stellar-to-halo mass relation prior. We set up several choice of these prior reflecting the uncertainty of cosmological models. In \cref{subsec:data}, we show the table of dSphs including their half-light radii and discances. The kinematical data set and preanalysis method for each dSph is also described.
In \cref{sec:results}, we show results of the analysis and estimated $J$-factor values. Here we compare results of different priors to verify the stability of estimated $J$-factor by changing cosmological models.
In \cref{sec:summary}, we discuss and summarize our results.

\section{Method}
\label{sec:method}
\subsection{Models and likelihood}
\label{subsec:likelihood}
We assume that dSphs are spherical and steady systems according to conventional analyses\footnote{The sphericity of dSphs is also assumed when deriving the satellite prior~\cite{Ando:2020yyk}. However, Ref.~\cite{Hayashi:2016kcy} suggests that axisymmetric models show better fitting performance than spherical models in terms of Bayes factor comparison. Hence, further study on the axisymmetric formalization of the satellite prior would be useful for future $J$-factor analyses, which is however beyond the scope of this paper.}~\cite{Geringer-Sameth:2014yza,Bonnivard:2015xpq}. Their velocity dispersions are determined by the spherical Jeans equation
\begin{equation}
    \frac{1}{\nu(r)}\pdv{(\nu(r)\sigma_r^2(r))}{r} + 2\sigma_r^2(r)\beta(r) = - \frac{G M(r)}{r^2} ,
\end{equation}
where $G$ is the gravitational constant, $r$ denotes the distance from the dSph center, $\nu(r)$ is the stellar number density and $M(r)$ denotes the dark matter mass enclosed within the radius $r$. The anisotropy of the stellar motion $\beta(r)$ is defined by the ratio of the velocity dispersions $\sigma_r^2(r)$, $\sigma_\theta^2(r)$ and $\sigma_\phi^2(r)$ as 
$
    \beta(r) \equiv 1 - \tfrac{\sigma_\theta^2 + \sigma_\phi^2}{2\sigma_r^2}.
$
By definition, $\beta(r)$ satisfies $-\infty < \beta(r) \leq 1$. In this paper, we assume that $\beta(r) = \beta_\text{ani}\,\text{(const.)}$ for simplicity.

We describe the stellar profile $\nu(r)$ as the Plummer profile, a widely-used fitting function of the stellar number density of dSphs~\cite{1911MNRAS..71..460P}.\footnote{Note that the stellar profile fitting by using generalized function suggests that some dSphs have multiple components.~\cite{Hayashi:2018uop,Pace:2020jln}}. Its stellar number density $\nu(r)$ and surface density $\Sigma(R)$ are given by
\begin{align}
    \nu(r) &= \frac{3}{4\pi R_e^3} \qty(1 + \qty(\frac{r}{R_e})^2)^{-5/2},
    \\ \Sigma(R) &= \frac{1}{\pi R_e^2}\qty(1+\frac{R^2}{R_e^2})^{-2},
\end{align}
where $R$ denotes the radius projected on the celestial sphere and $R_e$ is the half-light radius of the surface density profile. 

For the dark matter density profile, there are many discussions and no consensus exists yet. This is known as the core-cusp problem~\cite{2010AdAst2010E...5D}; N-body simulation shows the cuspy dark matter density profile ($\rho(r)\propto r^{-1}$ around the center), while observations suggest the cored profile ($\rho(r)\propto \mathrm{const.}$).
In this paper, we assume the cold dark matter model, then the dark matter density profile can be well described by the truncated NFW profile~\cite{1997ApJ...490..493N}, whose mass density $\rho(r)$ and enclosed mass $M(r)\equiv \int\dd{r}4\pi r^2 \rho(r)$ are respectively written as
\begin{equation}
    \rho(r) = \begin{cases}
    \rho_s \qty(\frac{r}{r_s})^{-1}\qty(1+\frac{r}{r_s})^{-2}  & (0 \leq r \leq r_t) \\
    0 & (r_t < r)
    \end{cases},
\end{equation}
\begin{equation}
    M(r) = \begin{cases}
    4\pi \rho_s r_s^3 f_\text{NFW}(r/r_s) & (0 \leq r\leq r_t)
    \\ 4\pi \rho_s r_s^3 f_\text{NFW}(r_t/r_s) & (r_t < r),
    \end{cases}
\end{equation}
where $\rho_s$ and $r_s$ is scale density and radius of the profile, respectively, $r_t$ denotes the truncation radius, and an auxiliary function $f_\text{NFW}(x)$ is defined as follows:
\begin{equation}\label{eq:f_nfw}
    f_\text{NFW}(x) = \ln\qty(1+x)-\frac{1}{1+x^{-1}}.
\end{equation}

Using these quantities, we define our likelihood function as follows:
\begin{equation}\label{eq:likelihood}
    \mathcal{L}(\Theta) = \prod_i \mathcal{N}[v_i;v_\text{dSph},\sigmalossq(R_i)+\delta\sigma_i^2],
\end{equation}
where $v_\text{dSph}$ is the systemic velocity of each dSph, $\Theta \equiv \qty{R_e,r_s,\rho_s,r_t,\beta_\text{ani},v_\text{dSph}} $ represents the parameter set in our model, $\mathcal{N}[x;\mu,\sigma^2]$ denotes the normal (Gaussian) distribution with the mean $\mu$ and the variance $\sigma^2$, $v_i$ is the observed velocity of $i$-th star and $\delta\sigma_i$ is its observational error. The function $\sigma_\text{los}^2(R)$ is the projected velocity dispersion along the line-of-sight at projected radius $R$, given by the following formula:
\begin{equation}\label{eq:sigmalossq}
    \sigmalossq(R) = \frac{2}{\Sigma(R)}\int_R^\infty\dd{r}\qty(1-\beta(r)\frac{R^2}{r^2})\frac{\nu(r)\sigma_r^2(r)}{\sqrt{1-R^2/r^2}}.
\end{equation}
When $\beta(r) = \beta_\text{ani}\ (\text{const.})$, \cref{eq:sigmalossq} is simplified to \cite{2005MNRAS.363..705M}
\begin{gather}
    \sigmalossq(R) = \frac{1}{\Sigma(R)}\int_R^\infty \dd{s}\,\nu(s) \frac{GM(s)}{s} {K}(s/R),\\
    {K}(u) = \sqrt{1-u^{-2}}\qty[u^2\qty(\frac{3}{2}-\beta_\text{ani})\,_2F_1 \qty(1,\frac{3}{2}-\beta_\text{ani};\frac{3}{2};1-u^2)-\frac{1}{2}].
\end{gather}
where ${}_2F_1(a,b;c;z)$ is the Gaussian hypergeometric function.

We note that our likelihood function has $R$-dependence in contrast with that \ando\ used the averaged ($R$-independent) velocity dispersion $\overline{\sigmalossq} = \frac{4\pi G}{3}\int_0^\infty\dd{r}r\nu(r)M(r)$. The advantage of the our $R$-dependent analysis is that it weakens the degeneracy between parameters by probing the shape of $\sigmalossq(R)$ even when $\overline{\sigmalossq}$ is not changed.

\subsection{Priors}
\label{subsec:priors}

\subsubsection{Photometry prior}
The half-light radius $R_e$ is constrained by the result of photometric observations, which is realized as a photometric prior. We adopt log-normal distribution for the half-light radius to construct the prior as follows:
\begin{equation}
    \pi_\text{photo.}(\log_{10}R_e/[\textrm{pc}]) = \mathcal{N}(\log_{10}R_e/[\textrm{pc}]|\log_{10}r_{e,\text{circ}}/[\textrm{pc}],\delta \log_{10}r_{e,\text{circ}}/[\textrm{pc}]),
\end{equation}
where we calculate the mean $\log_{10}r_{e,\text{circ}}$ and standard deviation $\delta \log_{10}r_{e,\text{circ}}$ based on the error propagation law by using $\hat{\theta}$ listed in Table.~1 on the supplement material of \ando.~\footnote{In \ando\ they assume the normal distribution for the half-light radius, but we found no significant difference between the normal and log-normal distributions because the standard deviation is so small that the log-normal distribution can be approximated by the normal distribution. In this paper, we adopt log-normal distribution, reflecting the fact that the radius must be positive.}

\subsubsection{Satellite prior}
\label{subsubsec:satellite_prior}
Structure formation models of subhalos in Milky Way predict structural parameters of subhalo profile $\rho(r)$: $r_s$, $\rho_s$ and $r_t$. In this paper we use the satellite prior proposed in \ando, briefly reviewed in the following:  
The formation of subhalos are well described by the extended Press-Schechter formalism~\cite{2011ApJ...741...13Y}, which gives the differential number of accreted subhalos $\frac{\dd[2]{N_\text{sh}}}{\dd{z_a}\dd{m_a}}$. Here $N_\text{sh}$ denotes the number of subhalo, $z_a$ and $m_a$ are the redshift and mass of a subhalo when the subhalo accreted onto its host. Here $m_a$ can be reinterpreted as halo parameters $\rho_{s,a}$, $r_{s,a}$ and $r_{200}$ by considering two conditions: i.) The subhalo is virialized $m_a=4\pi\rho_\text{crit}(z_a)200 r_{200}^3/3$, where the virial radius $r_{200}$ is calculated from $r_{s,a}$ by using the concentration parameter $c_a = r_{s,a}/r_{200}$, whose probability density distribution $P(c_a)$ is given by the log-normal distribution with mean $c_{200}(m_a,z_a)$~\cite{2015MNRAS.452.1217C} and standard deviation $\sigma_{\ln c} = 0.13$~\cite{2013ApJ...767..146I}. ii.) The dark matter distribution of the subhalo is given by the NFW profile $m_{200}=4\pi\rho_{s,a}r_s^3 f_\text{NFW}(r_{200}/r_{s,a})$, where $f_\text{NFW}$ is the same as defined in \cref{eq:f_nfw}.

After the accretion, the tidal force of Milky Way starts striping subhalo mass. In a semi-analytic strategy, the mass-loss rate through this process is modeled as
\begin{equation}
    \dv{m}{t} = -A\frac{m(z)}{\tau_\text{dyn}(z)}\qty[\frac{m(z)}{M(z)}]^\zeta
\end{equation}
where $\tau_\textrm{dyn}(z)$ denote the dynamical timescale~\cite{2016MNRAS.458.2848J}, $m(z)$ and $M(z)$ are subhalo and host halo mass at redshift $z$, respectively. The two parameters $A$ and $\zeta$ are calibrated by the results of N-body simulations. The solution of this equation with the initial condition $m(z_a) = m_a$ gives current ($z=0$) subhalo mass $m_0 = m(0)$. As the subhalo mass evolve, structural parameters $\rho_{s,a}$ and $r_{s,a}$ also evolve to $\rho_{s,0}$ and $r_{s,0}$ (or simply $\rho_s$ and $r_s$ ) according empirical fitting formula~\cite{2013ApJ...767..146I}. Finally, current truncation radius $r_{t,0}$ (or simply $r_t$) are determined by the NFW condition $m_{0}=4\pi\rho_{s,0}r_{s,0}^3 f_\text{NFW}(r_{t,0}/r_{s,0})$.

We combine the two distributions of parameters at accretion $\frac{\dd[2]{N_\text{sh}}}{\dd{z_a}\dd{m_a}}$ and $P(c_a)$ with the parameter evolution model to obtain the distribution of parameters at present.\footnote{Here we assume that there are no correlation between $(m_a,z_a)$ and $c_{t,a}$.} Instead of calculating the distribution of current parameters directly by using the Jacobian of the evolution formula, we obtain finite samples of the parameters. We subdivide $(\ln m_a, z_a, c_a)$ linearly and calculate the weight of $i$-th grid according to
\begin{equation}
    w_i = N \eval{\frac{\dd[2]{N_\text{sh}}}{\dd{z_a}\dd{m_a}}}_{z_a = z_{a,i}, m_a = m_{a,i}} (\Delta z_a)_i (\Delta m_a)_i \eval{P(c_a)}_{c_a=c_{a,i}} (\Delta c_a)_i
\end{equation}
where $N$ is a normalization factor to satisfy the condition $\sum_i w_i = N_\text{sh,tot} \equiv \iint\dd{z_a}\dd{m_a}\frac{\dd[2]{N_\text{sh}}}{\dd{z_a}\dd{m_a}}$. Each point $(\ln m_{a,i}, z_{a,i}, c_{a,i})$ is interpreted to $(\rho_{s,i},r_{s,i},c_{t,i})$ according to the stripping model, then we obtain finite samples of $(\rho_{s},r_{s},c_{t})$ with its weight.\footnote{Available in \url{https://github.com/shinichiroando/dwarf_params}. More integrated version named \textrm{SASHIMI} is in \url{https://github.com/shinichiroando/sashimi-c}.}

Some subhalos do not host any stars because baryons in too small a halo cannot lose their energy due to its ionizing background, known as reionization suppression~\cite{1992MNRAS.256P..43E,2019MNRAS.488.4585G}. In order to consider the effect, we multiply $w_i$ by the formation probability of a satellite for the given subhalo $P_\text{form}$, as follows:
\begin{equation}
    P_\text{form}(V_\text{peak}) = \frac{1}{2}\qty[1+\mathrm{erf}\qty(\frac{V_\text{peak}-V_{50}}{\sqrt{2}\sigma})],
\end{equation}
where $\Vpeak$ denotes the maximum circular velocity of the satellite at accretion time, given by $\Vpeak = (4\pi G\rho_{s,a}/4.625)^{1/2}r_{s,a}$ for a NFW subhalo and it is calculated for each parameter grid $(\rho_{s,i},r_{s,i},c_{t,i})$.
For the lower bound parameter $V_{50}$, we have two choices: $V_{50}=18\,\text{km/s}$, motivated by conventional theory of reionization~\cite{1996ApJ...465..608T,2000ApJ...542..535G,2006MNRAS.371..401H,2008MNRAS.390..920O}, and $V_{50}=10.5\,\text{km/s}$, based on the result of more resent analysis~\cite{2019MNRAS.488.4585G}. For $\sigma$, we adopt $\sigma=2.5\ \mathrm{km}\ \mathrm{s}^{-1}$, following Ref.~\cite{2019MNRAS.488.4585G}. For classical dSphs, we adopt $V_{50} = 25\ \text{km/s}$ according to \cite{2014ApJ...795L..13H}. Here we assume $\sigma=0\ \mathrm{km}\ \mathrm{s}^{-1}$ for simplicity. In this case $P_\text{form}(V_\text{peak})$ is equivalent to a step function $\Theta(\Vpeak-V_{50})$.




Using these quantities, the probability density distribution of the three profile parameters is then given by
\begin{equation}
    \pi_\text{sat.}(r_s,\rho_s,r_t) \propto \frac{\dd[3]{N_\text{sh}}}{\dd{r_s}\dd{\rho_s}\dd{r_t}} P_\text{form}(\Vpeak),
\end{equation}
where $\pi_\text{sat.}$ should be properly normalized to be a probability density distribution function. For the discrete sample points generated above, it is realized as 
\begin{equation}
    \pi_\text{sat.,i} = \frac{w_i P_\text{form}(V_{\text{peak},i})}{\sum_i w_i P_\text{form}(V_{\text{peak},i})}.
\end{equation}
Finally we smooth these samples $\qty{\pi_{\text{sat.},i}}$ to reconstruct a prior function $\pi_\text{sat.}(r_s,\rho_s,r_t)$ by using weighted kernel density estimation implemented in \texttt{scipy}~\cite{2020SciPy-NMeth}.

\subsubsection{SHMR prior}
\begin{figure}[tbp]
\centering
\smalltableandfigurefontsize
\begin{tikzpicture}[node distance=1.8cm]
\tikzstyle{dm} = [rectangle, rounded corners, minimum width=1.5cm, minimum height=0.5cm,text centered, draw=black, fill=gray!30]
\tikzstyle{star} = [diamond, minimum width=1.5cm, minimum height=0.5cm, text centered, draw=black]
\tikzstyle{evo} = [thick,->,>=stealth]
\tikzstyle{rel} = [thick,<->,>=stealth]
\node (dmz) [dm] {$m(z)$};
\node (dm0) [dm,right of=dmz,xshift=2.5cm,very thick] {m(0)};
\node (starz) [star,below of=dmz] {$m_\ast$};
\node (star0) [star,below of=dm0,very thick] {$m_\ast$};
\draw [evo] (dmz) -- node[anchor=south] {semi-analytic}  (dm0);
\draw [evo] (dmz) -- node[anchor=north] {model}  (dm0);
\draw [evo] (starz) -- node[anchor=south] {const.} (star0);
\draw [rel] (dmz) -- node[anchor=east] {SHMR} (starz);
\end{tikzpicture}
\caption{\tableandfigurefontsize A schematic picture to illustrate how to construct our SHMR prior. Horizontal one-side arrows indicate time evolution. Shapes with bold edges are values at present ($z=0$), appeared in the definition of the SHMR prior in \cref{eq:shmr_prior}.}
\label{fig:shmr_picture}
\end{figure}
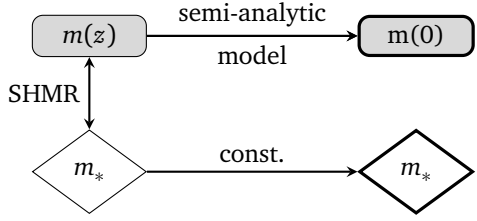

\begin{figure}
    \tableandfigurefontsize
    \centering
    \includegraphics[width=0.5\hsize]{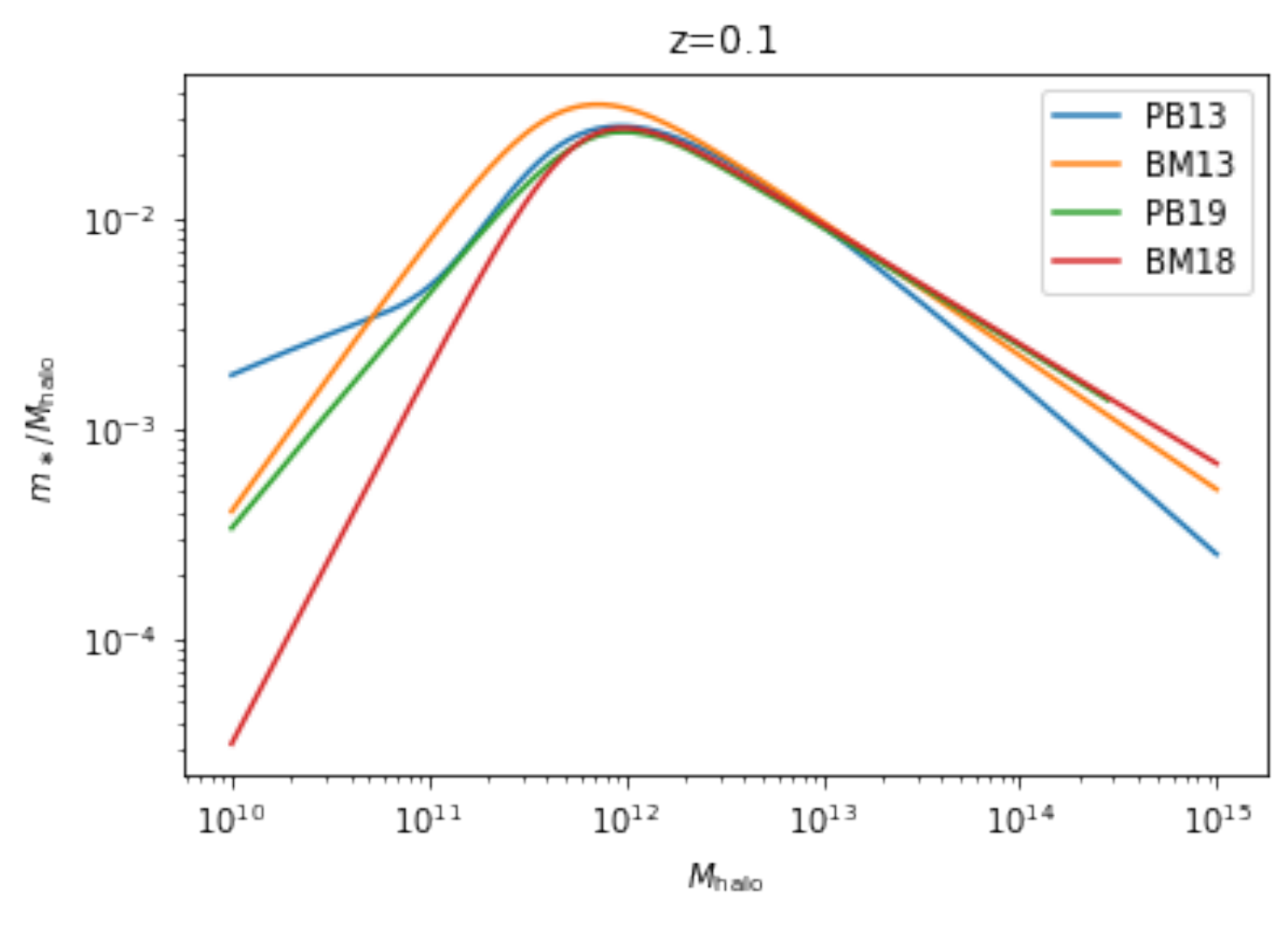}
    \caption{\tableandfigurefontsize
    SHMR function $m_\ast(M_h,z)$ normalized by the halo mass $M_h$. Here we fix $z=0.1$ for illustration purpose.}
    \label{fig:shmrs}
\end{figure}

In addition to the satellite prior, we use another prior motivated by the stellar-to-halo mass relation (SHMR) 
(see \cite{2018ARA&A..56..435W} for a review). This relation is obtained by fitting the structure formation model by using observed cosmological data sets with a simple assumption; the larger halo becomes, the more stars it hosts. Stellar mass $m_\ast$ is then written as a monotonic function of halo mass $m_{z}$ for given redshift $z$. We identify $m(z)$ with the subhalo mass at the accretion time and relate it to the current subhalo mass using the semi-analytic approach mentioned in \cref{subsubsec:satellite_prior}.
Here, for the simplicity, we assume that the stellar mass does not change after accretion and identify $m_\ast$ as the current stellar mass.\footnote{Basically, a galaxy is hosted at the center of a dark matter halo which is more massive than stellar mass, and the tidal force strips mainly the dominant component of the halo mass. Therefore stellar mass is not significantly changed until the most of halo mass is striped and halo mass and galaxy mass become comparable. However, once the halo mass becomes comparable, it cannot bound the galaxy within the halo and the galaxy gets collapsed, thus we cannot find such galaxies anymore.} The schematic figure of this procedure is shown in \cref{fig:shmr_picture}.

In order to check model dependence, we adopt four SHMR models~\cite{2013ApJ...770...57B,2013MNRAS.428.3121M,2019MNRAS.488.3143B,2018MNRAS.477.1822M} (hereafter {\pbi}, {\bmi}, {\pbii} and {\bmii}, respectively). These model have the following features:
\begin{itemize}
    \item \pbi: calibrated by the Bolshoi simulation, using a multi-power law fitting function for the SHMR and fitting the SHMR intrinsic parameters with other systematic parameters.
    \item \bmi: calibrated by the Millennium simulation, using a double-power law fitting function for the SHMR and simply fitting SHMR parameters.
    \item \pbii: updated data sets from \pbi, cosmological models calibrated by the Bolshoi-Planck simulation, using a double-power law fitting function for the SHMR and simply fitting SHMR parameters.
    \item \bmii: updated data sets from \bmi, cosmological models assuming double-power law for the evolution of baryon conversion efficiency calibrated by an independent simulation, using double-power law for the SHMR and simply fitting SHMR parameters.
\end{itemize}
The probability density distribution of stellar mass $m_\ast$ is then written as follows:
\begin{equation}
\label{eq:shmr_prior}
    P(m_\ast|m(z)) = \mathcal{N}[m_\ast;m_\ast(m(z),z),\delta m_\ast]
\end{equation}
where $m_\ast(M_h,z)$ is expected stellar mass for given halo mass $M_h$ at accretion redshift $z$. $\delta m_\ast$ denotes the model uncertainty of each SHMR model. We show the function $m_\ast(M_h,z)$ for each model in \cref{fig:shmrs}.

In terms of the Bayesian statistics, we can compare credibility of a model (model 1) to a reference model (model 0) by using Bayes factor $\mathrm{BF}$, defined as the ratio of Bayesian evidences $\mathcal{E}$:
\begin{equation}
    \mathrm{BF} = \frac{\mathcal{E}_1}{\mathcal{E}_0} = \frac{\int\dd{\Theta_1}\mathcal{L}_1(\Theta_1)\pi_1(\Theta_1)}{\int\dd{\Theta_0}\mathcal{L}_0(\Theta_0)\pi_0(\Theta_0)}.
\end{equation}
Here the minus logarithm of the Bayesian evidence $-\ln\mathcal{E}$ is approximated by the widely applicable Bayesian information criterion (WBIC)~\cite{watanabe2012}:
\begin{gather}
    -\ln\mathcal{E} \simeq \mathrm{WBIC} = -\frac{\int\dd{\Theta}\qty(\ln\mathcal{L}(\Theta))\mathcal{L}(\Theta)^\beta\pi(\Theta)}{\int\dd{\Theta}\mathcal{L}(\Theta)^\beta\pi(\Theta)},\\
    \beta = 1/\ln(\mathrm{\#data}).
\end{gather}
In this work, we calculate WBIC of each model using the MCMC algorithm and evaluate the Bayes factor according to
\begin{equation}\label{eq:bayesfactor_from_wbic}
    \ln{\mathrm{BF} = \ln{\mathcal{E}_1} - \ln{\mathcal{E}_0 \simeq -\mathrm{WBIC}_1 + \mathrm{WBIC}_0.}}
\end{equation}
Here $\mathrm{BF}>1$ or $\ln\mathrm{BF}>0$ means that model 1 is more credible than model 0. According to Ref.~\cite{Jeffreys:1939xee}, there is a scale for interpreting $\ln\mathrm{BF}$ into the strength of evidence as follows: \emph{Decisive} for $\ln\mathrm{BF}\gtrsim 4$, \emph{very strong} for $3\lesssim\ln\mathrm{BF}\lesssim 4$, \emph{strong} for $2\lesssim\ln\mathrm{BF}\lesssim 3$, \emph{substantial} for $1\lesssim\ln\mathrm{BF}\lesssim 2$, and \emph{barely worth mentioning} for $0\lesssim\ln\mathrm{BF}\lesssim 1$.

\subsection{Data}
\label{subsec:data}

\begin{table}[htbp]
    \tableandfigurefontsize
    \centering
    \caption{\tableandfigurefontsize The half-light radius, distance from Earth, stellar mass and reference list for the 27 ultra-faint dSphs analyzed in this paper. We derive the half-light radius and its error based on the value in \ando. The values of distance and stellar mass are from \ando\ and Ref.~\cite{2012AJ....144....4M}, respectively (see text for more details). The last column shows the source of kinematical data set.}
    \begin{tabular}{lrrrr}
\toprule
           Name & $\log_{10}R_e/[\mathrm{pc}]$ & $D_\ast [\text{kpc}]$ &  $M_\ast\ [M_\odot]$ &                                          Refs. \\
\midrule
     Aquarius 2 &                $2.09\pm0.08$ &                 $108$ & ${4.7}\times 10^{3}$ &                     \cite{2016MNRAS.463..712T} \\
     Bo\"otes 1 &                $2.20\pm0.02$ &                  $66$ & ${2.9}\times 10^{4}$ &                     \cite{2011ApJ...736..146K} \\
     Bo\"otes 2 &                $1.52\pm0.07$ &                  $42$ & ${1.0}\times 10^{3}$ & \cite{2009ApJ...690..453K,2016ApJ...817...41J} \\
CanesVenatici 1 &                $2.53\pm0.02$ &                 $218$ & ${2.3}\times 10^{5}$ &                     \cite{2007ApJ...670..313S} \\
CanesVenatici 2 &                $1.73\pm0.09$ &                 $160$ & ${7.9}\times 10^{3}$ &                     \cite{2007ApJ...670..313S} \\
       Carina 2 &                $1.87\pm0.05$ &                  $36$ & ${5.4}\times 10^{3}$ &                     \cite{2018ApJ...857..145L} \\
  ComaBerenices &                $1.76\pm0.03$ &                  $44$ & ${3.7}\times 10^{3}$ &                     \cite{2007ApJ...670..313S} \\
        Draco 2 &                $1.12\pm0.18$ &                  $20$ & ${1.0}\times 10^{3}$ &                     \cite{2016MNRAS.458L..59M} \\
     Eridanus 2 &                $2.20\pm0.05$ &                 $380$ & ${6.5}\times 10^{4}$ &                     \cite{2017ApJ...838....8L} \\
         Grus 1 &                $1.27\pm0.46$ &                 $120$ & ${2.1}\times 10^{3}$ &                     \cite{2016ApJ...819...53W} \\
       Hercules &                $2.08\pm0.04$ &                 $132$ & ${3.7}\times 10^{4}$ &                     \cite{2007ApJ...670..313S} \\
   Horologium 1 &                $1.49\pm0.10$ &                  $79$ & ${2.2}\times 10^{3}$ &                     \cite{2015ApJ...811...62K} \\
       Hydrus 1 &                $1.73\pm0.03$ &                  $28$ & ${6.5}\times 10^{3}$ &                     \cite{2018MNRAS.479.5343K} \\
          Leo 4 &                $2.01\pm0.05$ &                 $154$ & ${1.9}\times 10^{4}$ & \cite{2007ApJ...670..313S,2021ApJ...920...92J} \\
          Leo T &                $2.13\pm0.05$ &                 $417$ & ${1.4}\times 10^{5}$ &                     \cite{2007ApJ...670..313S} \\
          Leo 5 &                $1.57\pm0.18$ &                 $178$ & ${1.1}\times 10^{4}$ &                     \cite{2021ApJ...920...92J} \\
      Pegasus 3 &                $1.62\pm0.16$ &                 $215$ & ${3.6}\times 10^{3}$ &                     \cite{2016ApJ...833...16K} \\
       Pisces 2 &                $1.68\pm0.07$ &                 $182$ & ${8.6}\times 10^{3}$ &                     \cite{2015ApJ...810...56K} \\
    Reticulum 2 &                $1.49\pm0.02$ &                  $30$ & ${3.0}\times 10^{3}$ &                     \cite{2015ApJ...808...95S} \\
        Segue 1 &                $1.30\pm0.06$ &                  $23$ & ${3.4}\times 10^{2}$ &                     \cite{2011ApJ...733...46S} \\
        Segue 2 &                $1.53\pm0.04$ &                  $35$ & ${8.6}\times 10^{2}$ &                     \cite{2013ApJ...770...16K} \\
   Triangulum 2 &                $1.10\pm0.13$ &                  $30$ & ${4.5}\times 10^{2}$ &                     \cite{2017ApJ...838...83K} \\
       Tucana 2 &                $2.21\pm0.07$ &                  $57$ & ${2.8}\times 10^{3}$ &                     \cite{2016ApJ...819...53W} \\
       Tucana 3 &                $1.64\pm0.06$ &                  $25$ & ${7.9}\times 10^{2}$ &                     \cite{2017ApJ...838...11S} \\
    UrsaMajor 1 &                $2.18\pm0.02$ &                  $97$ & ${1.4}\times 10^{4}$ &                     \cite{2007ApJ...670..313S} \\
    UrsaMajor 2 &                $1.93\pm0.02$ &                  $32$ & ${4.1}\times 10^{3}$ &                     \cite{2007ApJ...670..313S} \\
      Willman 1 &                $1.30\pm0.05$ &                  $38$ & ${1.0}\times 10^{3}$ &                     \cite{2011AJ....142..128W} \\
\bottomrule
\end{tabular}

    \label{tab:dsph_table_ufd}
\end{table}

\begin{table}[htbp]
    \tableandfigurefontsize
    \centering
    \caption{\tableandfigurefontsize Same as \cref{tab:dsph_table_ufd}, but for classical dSphs.}
    \begin{tabular}{lrrrr}
\toprule
     Name & $\log_{10}R_e/[\mathrm{pc}]$ & $D_\ast [\text{kpc}]$ &  $M_\ast\ [M_\odot]$ &                      Refs. \\
\midrule
   Carina &              $2.392\pm0.005$ &                 $105$ & ${3.8}\times 10^{5}$ & \cite{2009AJ....137.3100W} \\
    Draco &              $2.256\pm0.005$ &                  $76$ & ${2.9}\times 10^{5}$ & \cite{2015MNRAS.448.2717W} \\
   Fornax &              $2.849\pm0.003$ &                 $147$ & ${2.0}\times 10^{7}$ & \cite{2009AJ....137.3100W} \\
    Leo 1 &              $2.353\pm0.004$ &                 $254$ & ${5.5}\times 10^{6}$ & \cite{2008ApJ...675..201M} \\
    Leo 2 &              $2.217\pm0.005$ &                 $233$ & ${7.4}\times 10^{5}$ & \cite{2017ApJ...836..202S} \\
 Sculptor &              $2.359\pm0.004$ &                  $86$ & ${2.3}\times 10^{6}$ & \cite{2009AJ....137.3100W} \\
Sextans 1 &              $2.538\pm0.004$ &                  $86$ & ${4.4}\times 10^{5}$ & \cite{2009AJ....137.3100W} \\
UrsaMinor &              $2.434\pm0.006$ &                  $76$ & ${2.9}\times 10^{5}$ & \cite{2018AJ....156..257S} \\
\bottomrule
\end{tabular}

    \label{tab:dsph_table_classical}
\end{table}

\begin{table}[htbp]
    \tableandfigurefontsize
    \centering
    \caption{\tableandfigurefontsize Scanning region of each parameter.}
    \begin{tabular}{crrc}
        \toprule
        parameter & min. & max. \\
        \midrule
        $\log_{10}R_e/[\text{pc}]$ & 1.0 & 3.5 \\
        $\log_{10}r_s/[\text{pc}]$ & 0.0 & 5.0 \\
        $\log_{10}\rho_s/[M_\odot\text{pc}^{-3}]$ & $-4.0$ & 4.0 \\
        $\log_{10}r_t/[\text{pc}]$ & 0.0 & 5.0 \\
        $-\log_{10}(1-\beta_\text{ani})$ & $-1.0$ & $1.0$\\
        $v_\text{dSph}/[\text{km}\,\text{s}^{-1}]$ & $-1000$ & 1000 & \\
        \bottomrule
    \end{tabular}
    \label{tab:param_list}
\end{table}

We analyze the dSphs listed in \cref{tab:dsph_table_ufd,tab:dsph_table_classical} according to \ando, where we show the half-light radius, distance, and stellar mass of each dSph. The half-light radius and distances are from \ando\ and also we use the values in Ref.~\cite{2012AJ....144....4M} for the stellar masses. For dSphs without stellar mass values in Ref.~\cite{2012AJ....144....4M}, we calculate their stellar masses from apparent magnitudes and distances, assuming $M/L=1$ according to Ref.~\cite{2012AJ....144....4M}.
The last column in \cref{tab:dsph_table_ufd,tab:dsph_table_classical} indicates references of the kinematical data set. In general, kinematical data set includes member stars and foreground stars. For data sets having membership flag, we extract stars identified as members. For those containing membership probability $P_M$, we choose member-like stars ($P_M > 0.95$). For the other data sets having no membership information, we adopt the selection criteria illustrated and described in the reference. In addition, we remove member stars identified as binary stars in order to avoid accidental increase of the velocity dispersion.


\subsection{Analysis}
Based on the likelihood and priors defined above, we can calculate the posterior probability density distribution (or simply posterior) $P(\Theta|D)$ by using the Bayes' theorem:
\begin{equation}\label{eq:bayes}
    P(\Theta|D) = \frac{\mathcal{L}(\Theta)\pi(\Theta)}{\int\dd{\Theta}\mathcal{L}(\Theta)\pi(\Theta)},
\end{equation}
where
\begin{equation}
    \pi = \begin{cases}
        \pi_\text{photo.} & \text{(without any cosmological priors)} \\
        \pi_\text{photo.}\pi_\text{sat.} & \text{(satellite prior only)} \\
        \pi_\text{photo.}\pi_\text{sat.+SHMR} & \text{(satellite \& SHMR prior)}
    \end{cases}
\end{equation}
Here, as mentioned in \cref{subsec:priors}, the satellite prior $\pi_\text{sat.}$ is selected from two candidates $\text{sat.}_{10.5}$ and $\text{sat.}_{18}$, and the SHMR model for $\pi_\text{sat.+SHMR}$ is chosen from {\pbi}, {\bmi}, {\pbii} and {\bmii}.

Instead of calculating \cref{eq:bayes} straightforwardly, we obtain samples from the posterior by using the Markov Chain Monte-Carlo methods. In this paper, we use the Affine invariant ensemble sampler implemented in \texttt{emcee}~\cite{2013PASP..125..306F}. We scan the parameter region as shown in \cref{tab:param_list}. For $R_e$, $r_s$, $\rho_s$ and $r_t$ we adopt the logarithmic scale, reflecting that they are positive. The range for the anisotropy $\beta_{ani}$ is set to include both of radial and tangential cases. Since $v_\text{dSph}$ is strongly constrained by the likelihood function, we choose its limits large enough to include the estimated value.

\section{Results}
\label{sec:results}

\begin{figure}[tbp!]
    \centering
    \tableandfigurefontsize
    \includegraphics[width=0.45\hsize]{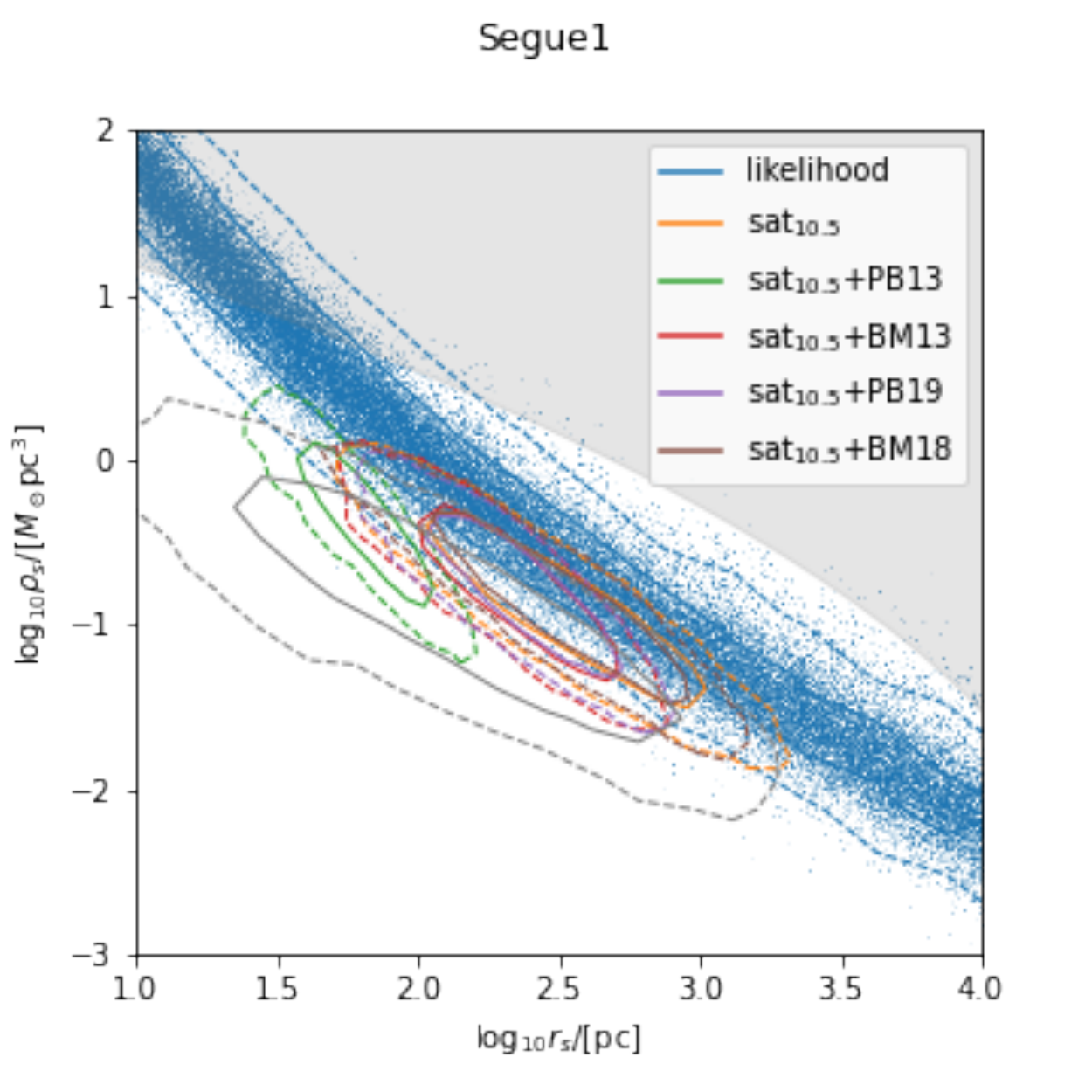}
    \includegraphics[width=0.45\hsize]{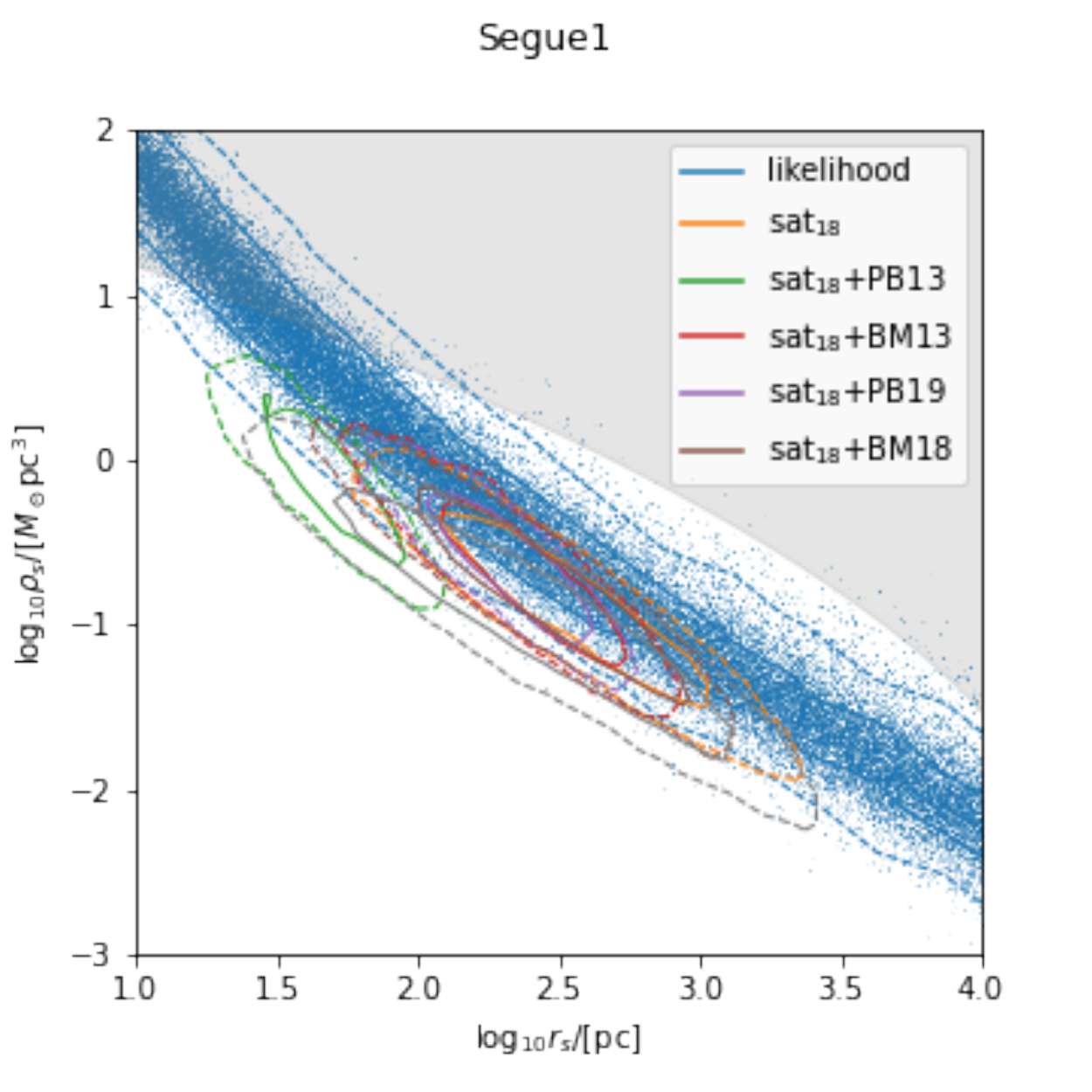}
    \caption{\tableandfigurefontsize Posterior probability density function projected onto $r_s$-$\rho_s$ plain for the case of Segue 1. The left and right panels show result assuming the $\mathrm{sat.}_{10.5}$ and $\mathrm{sat.}_{18}$ model, respectively. Blue dots are distributed according to likelihood only analysis (without any cosmological priors), while colored contours show posteriors with cosmological priors. Gray shaded area shows the cosmological constraint adopted in Ref.~\cite{Geringer-Sameth:2014yza}. For the other dSphs, see \cref{fig:posteriors_v50_105,fig:posteriors_v50_180}.}
    \label{fig:posterior_segue1}
\end{figure}

\begin{figure}[tbp!]
    \centering
    \tableandfigurefontsize
    \begin{minipage}{0.45\hsize}
    \includegraphics[width=\hsize]{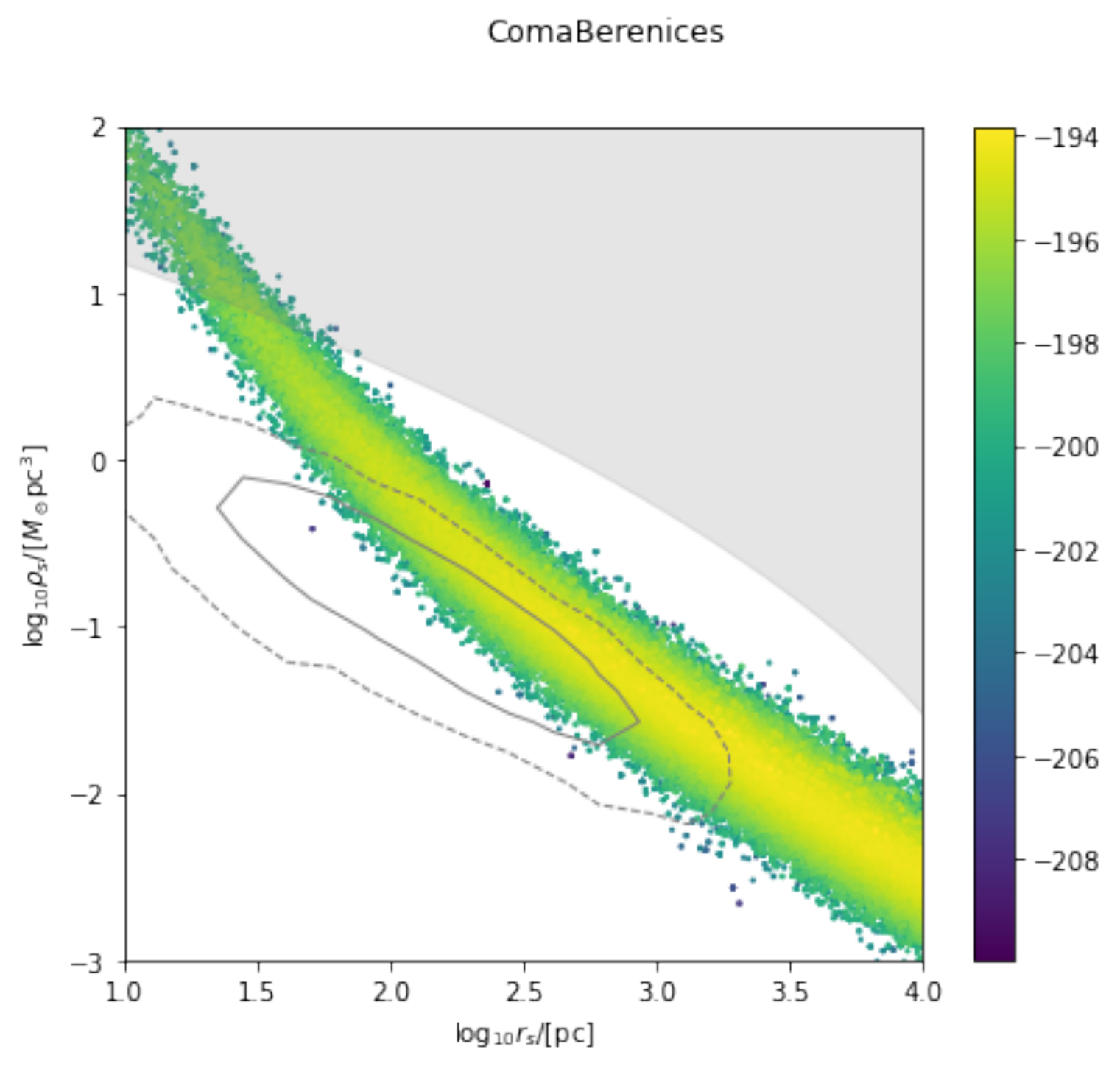}
    \end{minipage}
    \begin{minipage}{0.45\hsize}
    \includegraphics[width=\hsize]{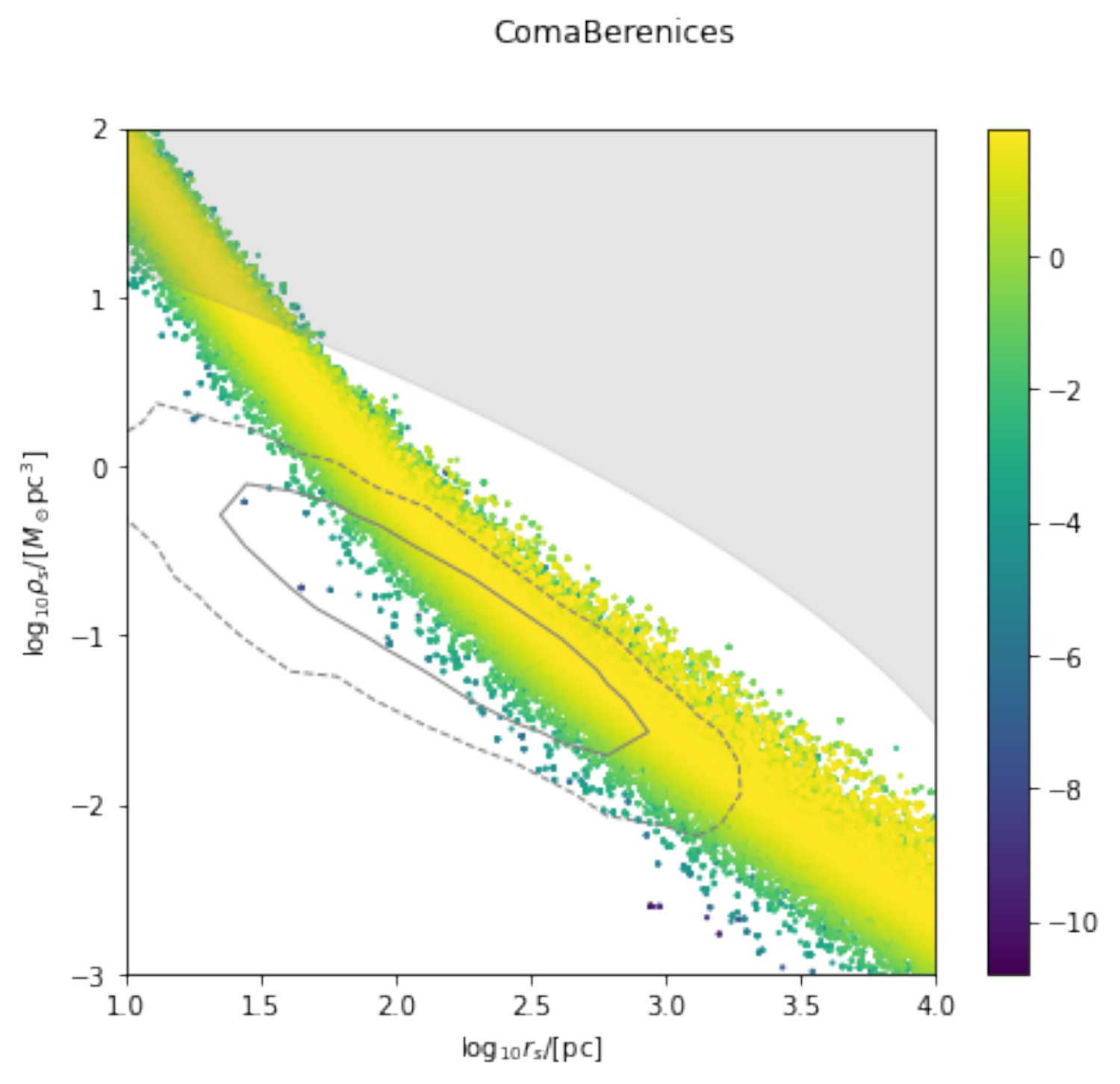}
    \end{minipage}
    \caption{\tableandfigurefontsize Results of likelihood only (without any cosmological priors) analyses. The left panel is for the $R$-dependent likelihood function, (\cref{eq:likelihood}), while the right panel is for the $R$-independent likelihood (see text), respectively. Color of heat map corresponds to the value of profiled likelihood function $\mathcal{L}_\text{prof}(r_s,\rho_s)\equiv\max_{r_t}\mathcal{L}(r_s,\rho_s,r_t)$. Gray contours denote the $1\sigma$ and $2\sigma$ regions of the satellite prior and gray shaded region shows the rough cosmological prior adopted in Ref.~\cite{Geringer-Sameth:2014yza}. For illustration purpose, we show only the Coma Berenices case.}
    \label{fig:posterior_degen}
\end{figure}

\begin{sidewaystable}[p]
    \centering
    \tableandfigurefontsize
    \caption{\smalltableandfigurefontsize
    Median and 1$\sigma$ interval of the estimated $J$-factors. $\mathrm{sat.}_{10.5}$ and $\mathrm{sat.}_{18}$ denote the satellite prior assuming $V_{50}=10.5$ and $18\ \mathrm{km/s}$, respectively.}
    \renewcommand{\arraystretch}{1.2}
    \begin{tabular}{l|rrr|rr|rr|rr|rr}
\toprule
{} & \multicolumn{3}{l}{w/o SHMR} & \multicolumn{2}{l}{\pbi} & \multicolumn{2}{l}{\bmi} & \multicolumn{2}{l}{\pbii} & \multicolumn{2}{l}{\bmii} \\
{} &                       flat &     $\mathrm{sat.}_{10.5}$ &       $\mathrm{sat.}_{18}$ &     $\mathrm{sat.}_{10.5}$ &       $\mathrm{sat.}_{18}$ &     $\mathrm{sat.}_{10.5}$ &       $\mathrm{sat.}_{18}$ &     $\mathrm{sat.}_{10.5}$ &       $\mathrm{sat.}_{18}$ &     $\mathrm{sat.}_{10.5}$ &       $\mathrm{sat.}_{18}$ \\
\midrule
Aquarius2      &  ${18.25}_{-0.55}^{+0.61}$ &  ${17.62}_{-0.36}^{+0.37}$ &  ${17.84}_{-0.36}^{+0.35}$ &  ${17.21}_{-0.24}^{+0.26}$ &  ${17.31}_{-0.24}^{+0.26}$ &  ${17.57}_{-0.32}^{+0.31}$ &  ${17.73}_{-0.37}^{+0.34}$ &  ${17.71}_{-0.35}^{+0.31}$ &  ${17.80}_{-0.35}^{+0.31}$ &  ${17.92}_{-0.33}^{+0.34}$ &  ${17.91}_{-0.34}^{+0.33}$ \\
Bootes1        &  ${18.24}_{-0.26}^{+0.28}$ &  ${18.03}_{-0.21}^{+0.22}$ &  ${18.11}_{-0.21}^{+0.21}$ &  ${17.95}_{-0.18}^{+0.19}$ &  ${17.99}_{-0.17}^{+0.18}$ &  ${18.09}_{-0.21}^{+0.21}$ &  ${18.09}_{-0.21}^{+0.21}$ &  ${18.12}_{-0.20}^{+0.22}$ &  ${18.12}_{-0.20}^{+0.21}$ &  ${18.31}_{-0.18}^{+0.19}$ &  ${18.30}_{-0.17}^{+0.18}$ \\
Bootes2        &  ${16.64}_{-4.90}^{+2.80}$ &  ${17.69}_{-0.81}^{+0.65}$ &  ${18.41}_{-0.53}^{+0.46}$ &  ${17.48}_{-0.29}^{+0.28}$ &  ${17.53}_{-0.29}^{+0.32}$ &  ${17.64}_{-0.49}^{+0.53}$ &  ${17.84}_{-0.67}^{+0.69}$ &  ${17.74}_{-0.73}^{+0.58}$ &  ${18.33}_{-0.42}^{+0.39}$ &  ${18.33}_{-0.38}^{+0.41}$ &  ${18.35}_{-0.41}^{+0.44}$ \\
CanesVenatici1 &  ${17.59}_{-0.19}^{+0.32}$ &  ${17.42}_{-0.12}^{+0.13}$ &  ${17.45}_{-0.12}^{+0.12}$ &  ${17.41}_{-0.12}^{+0.13}$ &  ${17.45}_{-0.12}^{+0.14}$ &  ${17.45}_{-0.11}^{+0.11}$ &  ${17.44}_{-0.12}^{+0.12}$ &  ${17.47}_{-0.12}^{+0.12}$ &  ${17.46}_{-0.12}^{+0.13}$ &  ${17.46}_{-0.07}^{+0.10}$ &  ${17.46}_{-0.07}^{+0.10}$ \\
CanesVenatici2 &  ${17.94}_{-0.45}^{+0.50}$ &  ${17.40}_{-0.36}^{+0.37}$ &  ${17.58}_{-0.31}^{+0.30}$ &  ${16.98}_{-0.25}^{+0.23}$ &  ${17.08}_{-0.25}^{+0.23}$ &  ${17.38}_{-0.31}^{+0.27}$ &  ${17.48}_{-0.32}^{+0.27}$ &  ${17.47}_{-0.28}^{+0.27}$ &  ${17.51}_{-0.28}^{+0.27}$ &  ${17.68}_{-0.31}^{+0.30}$ &  ${17.68}_{-0.30}^{+0.30}$ \\
Carina2        &  ${18.38}_{-0.55}^{+0.56}$ &  ${18.06}_{-0.37}^{+0.41}$ &  ${18.35}_{-0.36}^{+0.37}$ &  ${18.00}_{-0.26}^{+0.25}$ &  ${18.06}_{-0.25}^{+0.26}$ &  ${18.15}_{-0.32}^{+0.33}$ &  ${18.18}_{-0.36}^{+0.39}$ &  ${18.17}_{-0.42}^{+0.39}$ &  ${18.39}_{-0.35}^{+0.33}$ &  ${18.53}_{-0.27}^{+0.30}$ &  ${18.56}_{-0.27}^{+0.30}$ \\
ComaBerenices  &  ${18.95}_{-0.36}^{+0.34}$ &  ${18.58}_{-0.29}^{+0.28}$ &  ${18.70}_{-0.26}^{+0.26}$ &  ${18.13}_{-0.23}^{+0.23}$ &  ${18.19}_{-0.23}^{+0.24}$ &  ${18.49}_{-0.27}^{+0.25}$ &  ${18.60}_{-0.26}^{+0.24}$ &  ${18.55}_{-0.27}^{+0.25}$ &  ${18.63}_{-0.24}^{+0.23}$ &  ${18.71}_{-0.26}^{+0.25}$ &  ${18.71}_{-0.27}^{+0.26}$ \\
Draco2         &  ${16.80}_{-4.77}^{+2.54}$ &  ${18.30}_{-0.78}^{+0.62}$ &  ${18.93}_{-0.49}^{+0.43}$ &  ${18.12}_{-0.29}^{+0.29}$ &  ${18.15}_{-0.27}^{+0.29}$ &  ${18.28}_{-0.48}^{+0.49}$ &  ${18.40}_{-0.61}^{+0.69}$ &  ${18.35}_{-0.70}^{+0.54}$ &  ${18.89}_{-0.43}^{+0.37}$ &  ${18.90}_{-0.35}^{+0.37}$ &  ${18.93}_{-0.35}^{+0.38}$ \\
Eridanus2      &  ${17.29}_{-0.36}^{+0.38}$ &  ${16.91}_{-0.24}^{+0.25}$ &  ${16.98}_{-0.23}^{+0.24}$ &  ${16.76}_{-0.19}^{+0.20}$ &  ${16.81}_{-0.19}^{+0.19}$ &  ${16.97}_{-0.23}^{+0.22}$ &  ${16.96}_{-0.22}^{+0.22}$ &  ${17.03}_{-0.21}^{+0.21}$ &  ${17.03}_{-0.23}^{+0.23}$ &  ${17.15}_{-0.16}^{+0.14}$ &  ${17.14}_{-0.15}^{+0.14}$ \\
Grus1          &  ${17.42}_{-0.89}^{+0.88}$ &  ${17.09}_{-0.48}^{+0.46}$ &  ${17.44}_{-0.44}^{+0.44}$ &  ${16.77}_{-0.25}^{+0.26}$ &  ${16.85}_{-0.26}^{+0.28}$ &  ${17.05}_{-0.41}^{+0.42}$ &  ${17.26}_{-0.50}^{+0.46}$ &  ${17.15}_{-0.48}^{+0.41}$ &  ${17.43}_{-0.39}^{+0.36}$ &  ${17.53}_{-0.36}^{+0.37}$ &  ${17.51}_{-0.35}^{+0.39}$ \\
Hercules       &  ${17.89}_{-0.37}^{+0.37}$ &  ${17.51}_{-0.27}^{+0.30}$ &  ${17.69}_{-0.27}^{+0.28}$ &  ${17.43}_{-0.21}^{+0.21}$ &  ${17.48}_{-0.23}^{+0.22}$ &  ${17.62}_{-0.27}^{+0.26}$ &  ${17.61}_{-0.25}^{+0.26}$ &  ${17.69}_{-0.29}^{+0.26}$ &  ${17.69}_{-0.28}^{+0.27}$ &  ${17.93}_{-0.20}^{+0.18}$ &  ${17.95}_{-0.21}^{+0.18}$ \\
Horologium1    &  ${19.08}_{-0.63}^{+0.68}$ &  ${18.11}_{-0.40}^{+0.41}$ &  ${18.26}_{-0.37}^{+0.33}$ &  ${17.53}_{-0.23}^{+0.24}$ &  ${17.59}_{-0.24}^{+0.25}$ &  ${17.95}_{-0.34}^{+0.31}$ &  ${18.18}_{-0.36}^{+0.30}$ &  ${18.05}_{-0.34}^{+0.31}$ &  ${18.18}_{-0.30}^{+0.28}$ &  ${18.25}_{-0.36}^{+0.35}$ &  ${18.29}_{-0.36}^{+0.34}$ \\
Hydrus1        &  ${18.52}_{-0.32}^{+0.36}$ &  ${18.33}_{-0.27}^{+0.29}$ &  ${18.50}_{-0.28}^{+0.31}$ &  ${18.25}_{-0.19}^{+0.21}$ &  ${18.30}_{-0.21}^{+0.21}$ &  ${18.37}_{-0.26}^{+0.26}$ &  ${18.38}_{-0.26}^{+0.31}$ &  ${18.42}_{-0.31}^{+0.29}$ &  ${18.51}_{-0.27}^{+0.26}$ &  ${18.69}_{-0.22}^{+0.25}$ &  ${18.68}_{-0.22}^{+0.25}$ \\
Leo4           &  ${15.63}_{-5.06}^{+1.85}$ &  ${16.72}_{-0.61}^{+0.53}$ &  ${17.17}_{-0.46}^{+0.47}$ &  ${16.96}_{-0.27}^{+0.27}$ &  ${17.01}_{-0.29}^{+0.27}$ &  ${17.09}_{-0.37}^{+0.39}$ &  ${17.15}_{-0.41}^{+0.37}$ &  ${16.95}_{-0.50}^{+0.53}$ &  ${17.18}_{-0.47}^{+0.43}$ &  ${17.70}_{-0.27}^{+0.28}$ &  ${17.69}_{-0.28}^{+0.27}$ \\
Leo5           &  ${17.18}_{-0.82}^{+0.78}$ &  ${16.92}_{-0.48}^{+0.45}$ &  ${17.28}_{-0.43}^{+0.36}$ &  ${16.73}_{-0.26}^{+0.27}$ &  ${16.87}_{-0.26}^{+0.26}$ &  ${17.07}_{-0.34}^{+0.34}$ &  ${17.16}_{-0.39}^{+0.37}$ &  ${17.17}_{-0.43}^{+0.36}$ &  ${17.29}_{-0.37}^{+0.32}$ &  ${17.56}_{-0.30}^{+0.30}$ &  ${17.51}_{-0.31}^{+0.33}$ \\
LeoT           &  ${17.61}_{-0.44}^{+0.43}$ &  ${16.95}_{-0.31}^{+0.31}$ &  ${17.01}_{-0.29}^{+0.27}$ &  ${16.82}_{-0.22}^{+0.22}$ &  ${16.87}_{-0.23}^{+0.23}$ &  ${17.07}_{-0.29}^{+0.26}$ &  ${17.09}_{-0.28}^{+0.24}$ &  ${17.12}_{-0.30}^{+0.26}$ &  ${17.12}_{-0.32}^{+0.27}$ &  ${17.06}_{-0.12}^{+0.10}$ &  ${17.05}_{-0.14}^{+0.10}$ \\
Pegasus3       &  ${17.82}_{-2.10}^{+1.01}$ &  ${16.72}_{-0.72}^{+0.60}$ &  ${17.23}_{-0.51}^{+0.44}$ &  ${16.41}_{-0.28}^{+0.29}$ &  ${16.45}_{-0.31}^{+0.31}$ &  ${16.78}_{-0.51}^{+0.46}$ &  ${17.00}_{-0.66}^{+0.50}$ &  ${16.88}_{-0.74}^{+0.47}$ &  ${17.17}_{-0.39}^{+0.35}$ &  ${17.35}_{-0.42}^{+0.39}$ &  ${17.35}_{-0.39}^{+0.39}$ \\
Pisces2        &  ${17.24}_{-1.17}^{+0.91}$ &  ${16.74}_{-0.56}^{+0.55}$ &  ${17.15}_{-0.48}^{+0.46}$ &  ${16.70}_{-0.27}^{+0.27}$ &  ${16.72}_{-0.30}^{+0.30}$ &  ${16.87}_{-0.38}^{+0.42}$ &  ${16.94}_{-0.45}^{+0.47}$ &  ${16.93}_{-0.60}^{+0.48}$ &  ${17.15}_{-0.46}^{+0.42}$ &  ${17.47}_{-0.33}^{+0.33}$ &  ${17.50}_{-0.34}^{+0.34}$ \\
Reticulum2     &  ${18.98}_{-0.36}^{+0.37}$ &  ${18.65}_{-0.30}^{+0.30}$ &  ${18.80}_{-0.28}^{+0.27}$ &  ${18.31}_{-0.20}^{+0.21}$ &  ${18.35}_{-0.20}^{+0.21}$ &  ${18.57}_{-0.28}^{+0.27}$ &  ${18.70}_{-0.32}^{+0.29}$ &  ${18.68}_{-0.28}^{+0.25}$ &  ${18.75}_{-0.26}^{+0.25}$ &  ${18.81}_{-0.27}^{+0.27}$ &  ${18.82}_{-0.27}^{+0.26}$ \\
Segue1         &  ${19.74}_{-0.41}^{+0.39}$ &  ${19.20}_{-0.39}^{+0.33}$ &  ${19.32}_{-0.33}^{+0.28}$ &  ${18.22}_{-0.31}^{+0.29}$ &  ${18.36}_{-0.33}^{+0.31}$ &  ${18.95}_{-0.37}^{+0.30}$ &  ${19.23}_{-0.30}^{+0.27}$ &  ${18.98}_{-0.34}^{+0.30}$ &  ${19.21}_{-0.29}^{+0.25}$ &  ${19.20}_{-0.35}^{+0.31}$ &  ${19.23}_{-0.36}^{+0.30}$ \\
Segue2         &  ${18.02}_{-2.06}^{+0.70}$ &  ${18.01}_{-0.50}^{+0.46}$ &  ${18.42}_{-0.41}^{+0.37}$ &  ${17.66}_{-0.26}^{+0.26}$ &  ${17.72}_{-0.28}^{+0.27}$ &  ${17.93}_{-0.44}^{+0.40}$ &  ${18.11}_{-0.60}^{+0.44}$ &  ${18.11}_{-0.51}^{+0.37}$ &  ${18.37}_{-0.37}^{+0.30}$ &  ${18.38}_{-0.30}^{+0.32}$ &  ${18.39}_{-0.34}^{+0.35}$ \\
Triangulum2    &  ${14.36}_{-3.86}^{+2.91}$ &  ${17.75}_{-0.86}^{+0.67}$ &  ${18.54}_{-0.49}^{+0.43}$ &  ${17.56}_{-0.29}^{+0.29}$ &  ${17.63}_{-0.28}^{+0.27}$ &  ${17.68}_{-0.49}^{+0.55}$ &  ${17.96}_{-0.68}^{+0.65}$ &  ${17.77}_{-0.70}^{+0.57}$ &  ${18.44}_{-0.42}^{+0.36}$ &  ${18.43}_{-0.38}^{+0.37}$ &  ${18.39}_{-0.36}^{+0.43}$ \\
Tucana2        &  ${18.14}_{-0.52}^{+0.58}$ &  ${17.87}_{-0.36}^{+0.38}$ &  ${18.16}_{-0.35}^{+0.36}$ &  ${17.57}_{-0.25}^{+0.26}$ &  ${17.68}_{-0.23}^{+0.25}$ &  ${17.84}_{-0.30}^{+0.33}$ &  ${18.02}_{-0.35}^{+0.36}$ &  ${17.96}_{-0.35}^{+0.33}$ &  ${18.10}_{-0.32}^{+0.32}$ &  ${18.20}_{-0.30}^{+0.32}$ &  ${18.19}_{-0.29}^{+0.32}$ \\
Tucana3        &  ${15.71}_{-4.35}^{+1.79}$ &  ${17.52}_{-0.56}^{+0.45}$ &  ${18.06}_{-0.33}^{+0.34}$ &  ${17.73}_{-0.27}^{+0.25}$ &  ${17.80}_{-0.25}^{+0.26}$ &  ${17.63}_{-0.29}^{+0.35}$ &  ${17.65}_{-0.30}^{+0.35}$ &  ${17.53}_{-0.38}^{+0.46}$ &  ${18.05}_{-0.50}^{+0.43}$ &  ${18.29}_{-0.24}^{+0.27}$ &  ${18.25}_{-0.23}^{+0.27}$ \\
UrsaMajor1     &  ${18.66}_{-0.29}^{+0.30}$ &  ${18.30}_{-0.22}^{+0.22}$ &  ${18.34}_{-0.20}^{+0.20}$ &  ${18.03}_{-0.19}^{+0.21}$ &  ${18.09}_{-0.20}^{+0.19}$ &  ${18.26}_{-0.20}^{+0.20}$ &  ${18.27}_{-0.19}^{+0.20}$ &  ${18.30}_{-0.20}^{+0.20}$ &  ${18.30}_{-0.19}^{+0.20}$ &  ${18.39}_{-0.19}^{+0.19}$ &  ${18.38}_{-0.19}^{+0.20}$ \\
UrsaMajor2     &  ${19.54}_{-0.38}^{+0.38}$ &  ${19.01}_{-0.28}^{+0.29}$ &  ${19.10}_{-0.24}^{+0.25}$ &  ${18.56}_{-0.24}^{+0.23}$ &  ${18.63}_{-0.24}^{+0.24}$ &  ${18.91}_{-0.25}^{+0.26}$ &  ${19.05}_{-0.26}^{+0.25}$ &  ${18.98}_{-0.25}^{+0.25}$ &  ${19.04}_{-0.24}^{+0.25}$ &  ${19.12}_{-0.27}^{+0.26}$ &  ${19.13}_{-0.27}^{+0.27}$ \\
Willman1       &  ${19.50}_{-0.44}^{+0.44}$ &  ${18.85}_{-0.38}^{+0.34}$ &  ${18.95}_{-0.34}^{+0.29}$ &  ${18.04}_{-0.28}^{+0.27}$ &  ${18.14}_{-0.30}^{+0.27}$ &  ${18.59}_{-0.32}^{+0.30}$ &  ${18.86}_{-0.32}^{+0.29}$ &  ${18.70}_{-0.33}^{+0.29}$ &  ${18.84}_{-0.28}^{+0.26}$ &  ${18.86}_{-0.33}^{+0.32}$ &  ${18.92}_{-0.35}^{+0.31}$ \\
\bottomrule
\end{tabular}

    \label{tab:j_table_ufd}
\end{sidewaystable}

\begin{table}[tbp!]
    \centering
    \tableandfigurefontsize
    \caption{\tableandfigurefontsize
    Same as \cref{tab:j_table_ufd}, but for classical dSphs.}
    \renewcommand{\arraystretch}{1.2}
    \begin{tabular}{l|rr|rrrrrrrrrrrrrrrrr}
\toprule
{} & \multicolumn{2}{l}{w/o SHMR} &                       \pbi &                       \bmi &                      \pbii &                      \bmii \\
{} &                       flat &            $\mathrm{sat.}$ &            $\mathrm{sat.}$ &            $\mathrm{sat.}$ &            $\mathrm{sat.}$ &            $\mathrm{sat.}$ \\
\midrule
Carina    &  ${17.86}_{-0.07}^{+0.09}$ &  ${17.86}_{-0.06}^{+0.07}$ &  ${17.85}_{-0.06}^{+0.06}$ &  ${17.85}_{-0.06}^{+0.06}$ &  ${17.86}_{-0.06}^{+0.07}$ &  ${17.87}_{-0.06}^{+0.06}$ \\
Draco     &  ${18.92}_{-0.06}^{+0.06}$ &  ${18.89}_{-0.06}^{+0.06}$ &  ${18.85}_{-0.06}^{+0.06}$ &  ${18.85}_{-0.06}^{+0.06}$ &  ${18.85}_{-0.06}^{+0.06}$ &  ${18.88}_{-0.06}^{+0.06}$ \\
Fornax    &  ${17.93}_{-0.08}^{+0.20}$ &  ${18.03}_{-0.10}^{+0.11}$ &  ${18.02}_{-0.08}^{+0.10}$ &  ${18.00}_{-0.07}^{+0.09}$ &  ${18.02}_{-0.08}^{+0.09}$ &  ${17.96}_{-0.06}^{+0.07}$ \\
Leo1      &  ${17.80}_{-0.14}^{+0.22}$ &  ${17.71}_{-0.09}^{+0.10}$ &  ${17.73}_{-0.08}^{+0.08}$ &  ${17.73}_{-0.09}^{+0.11}$ &  ${17.74}_{-0.09}^{+0.12}$ &  ${17.78}_{-0.11}^{+0.14}$ \\
Leo2      &  ${17.82}_{-0.20}^{+0.25}$ &  ${17.70}_{-0.14}^{+0.16}$ &  ${17.64}_{-0.11}^{+0.13}$ &  ${17.69}_{-0.13}^{+0.15}$ &  ${17.71}_{-0.14}^{+0.15}$ &  ${17.73}_{-0.14}^{+0.17}$ \\
Sculptor  &  ${18.56}_{-0.05}^{+0.07}$ &  ${18.55}_{-0.04}^{+0.04}$ &  ${18.55}_{-0.04}^{+0.04}$ &  ${18.55}_{-0.04}^{+0.04}$ &  ${18.55}_{-0.04}^{+0.05}$ &  ${18.56}_{-0.04}^{+0.04}$ \\
Sextans1  &  ${18.09}_{-0.16}^{+0.40}$ &  ${18.12}_{-0.13}^{+0.15}$ &  ${18.09}_{-0.11}^{+0.14}$ &  ${18.09}_{-0.12}^{+0.14}$ &  ${18.12}_{-0.12}^{+0.15}$ &  ${18.19}_{-0.13}^{+0.15}$ \\
UrsaMinor &  ${18.47}_{-0.09}^{+0.13}$ &  ${18.46}_{-0.08}^{+0.09}$ &  ${18.50}_{-0.08}^{+0.09}$ &  ${18.46}_{-0.08}^{+0.09}$ &  ${18.46}_{-0.07}^{+0.08}$ &  ${18.47}_{-0.08}^{+0.09}$ \\
\bottomrule
\end{tabular}

    \label{tab:j_table_classical}
\end{table}

\begin{sidewaysfigure}[p]
    \centering
    \tableandfigurefontsize
    \includegraphics[width=0.8\hsize]{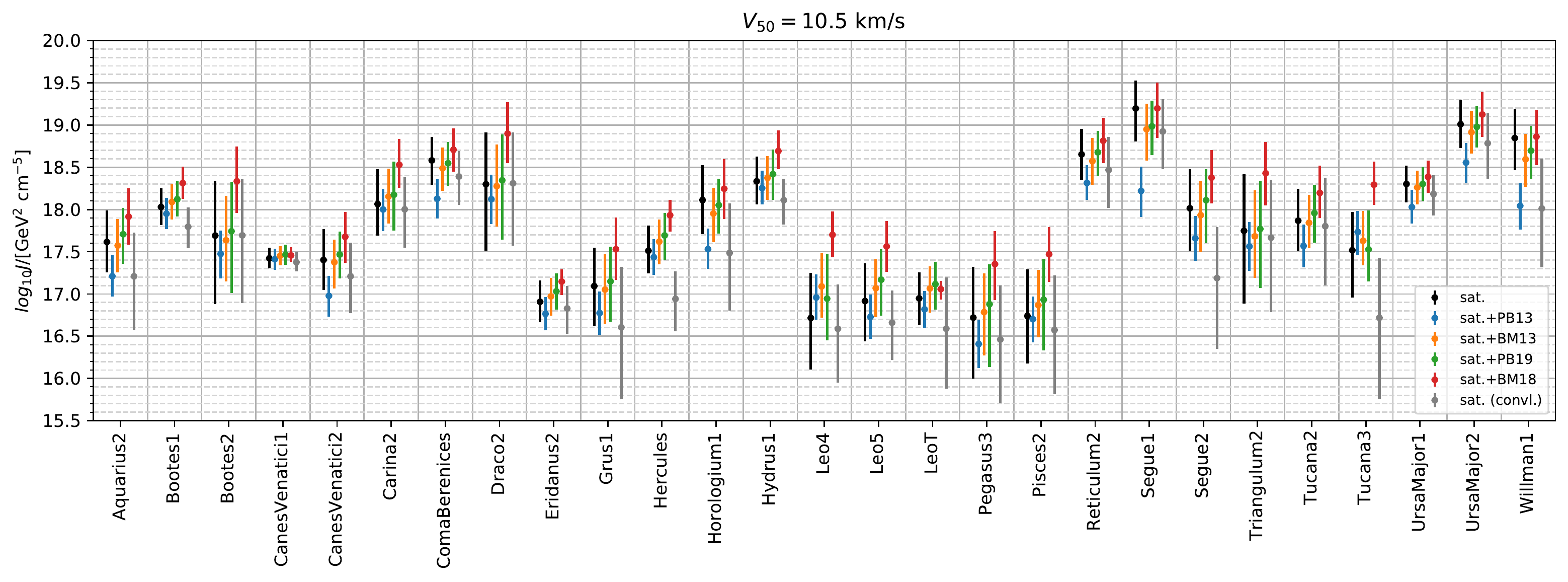}
    \includegraphics[width=0.8\hsize]{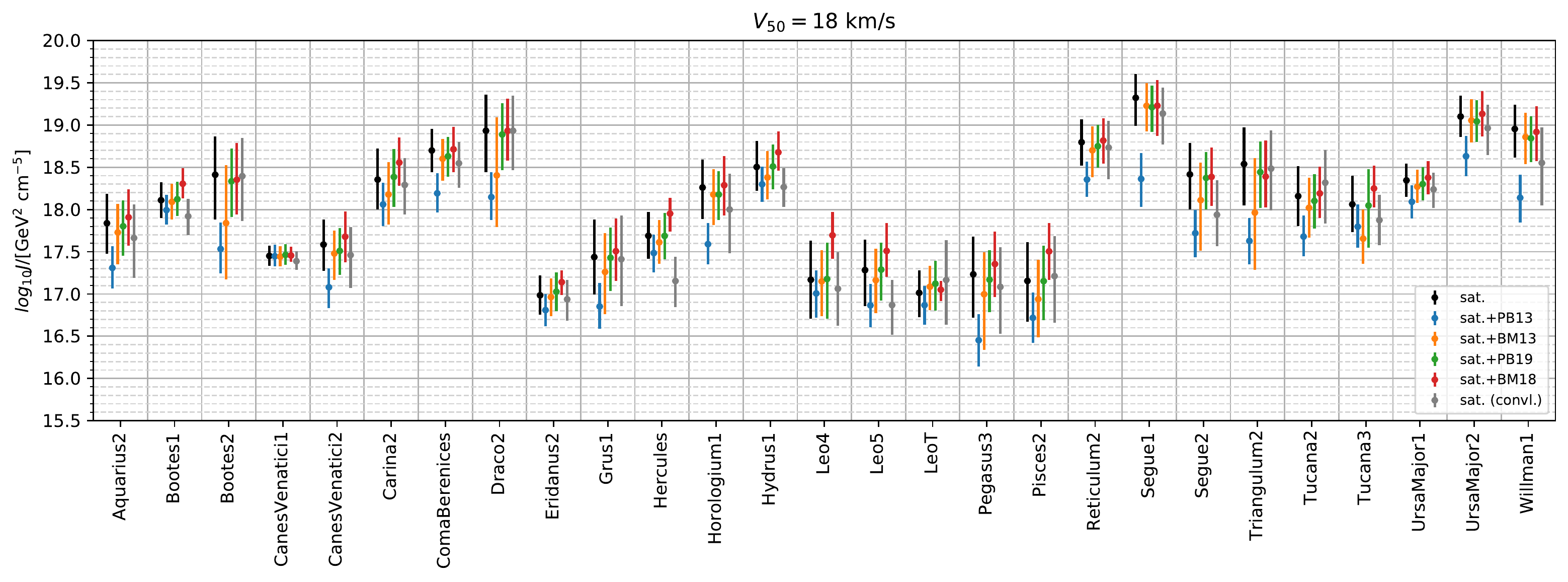}
    \caption{\tableandfigurefontsize
    Estimated $J$-factors of UFDs in \cref{tab:j_table_ufd} (dots) and their 68\% credible intervals (error bars). Black bars are the results with satellite prior only analysis, while gray ones show the results of conventional analysis~\cite{Ando:2020yyk}. Blue, orange, green and red bars correspond to analyses with the satellite prior and \pbi, \bmi, \pbii\ and \bmii\ priors, respectively. The upper panel shows results when $V_{50}=10.5\ \mathrm{km/s}$, while the lower panel is for $V_{50}=18\ \mathrm{km/s}$.}
    \label{fig:j_table_ufd}
\end{sidewaysfigure}

\begin{figure}[tbp!]
    \centering
    \tableandfigurefontsize
    \includegraphics[width=0.8\hsize]{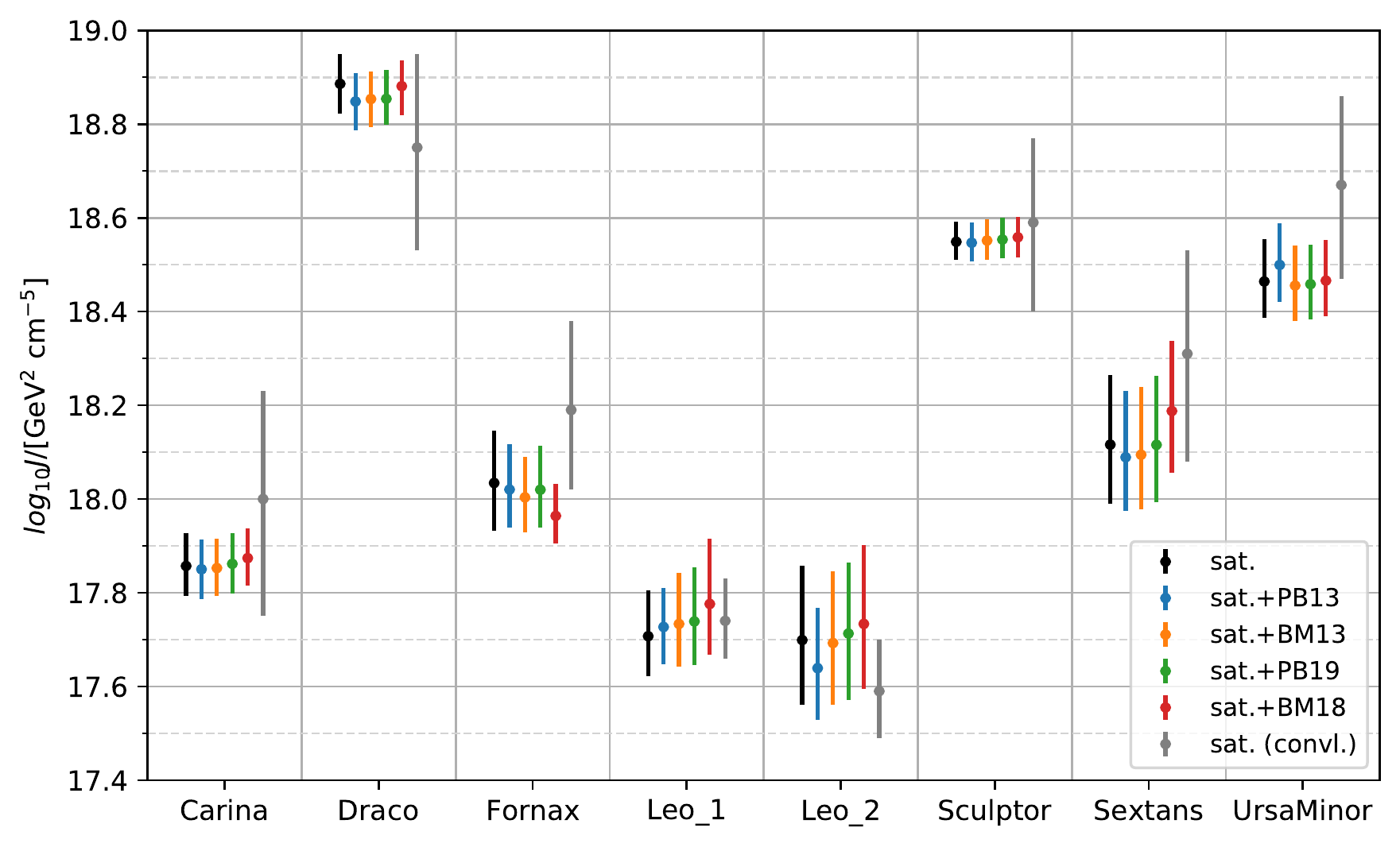}
    \caption{\tableandfigurefontsize
    Same figure as \cref{fig:j_table_ufd} but for the classical dSphs listed in \cref{tab:j_table_classical}.}
    \label{fig:j_table_classical}
\end{figure}

\begin{table}[tbp!]
    \centering
    \tableandfigurefontsize
    \caption{\tableandfigurefontsize
    The natural logarithm of Bayes factors of each model calculated according to \cref{eq:bayesfactor_from_wbic}. Column 1 shows the Bayes factor of $\mathrm{sat}_{18}$ to a reference model $\mathrm{sat}_{10.5}$ for each dSph. Columns 2-5 shows the Bayes factors of the satellite prior and SHMR analyses to the satellite prior only analysis $\mathrm{sat}_{10.5}$ as a reference, so as Columns 6-9 not for $\mathrm{sat}_{10.5}$ but $\mathrm{sat}_{18}$ cases.
    By definition, positive (negative) values mean that the corresponding model is more (less) credible than the reference model. }
    \begin{tabular}{lr|rrrr|rrrr}
\toprule
{} & $\mathrm{sat.}_{18}$/$\mathrm{sat.}_{10.5}$ & \multicolumn{4}{l}{$\qty(\mathrm{sat.}_{10.5}+\mathrm{SHMR})/\mathrm{sat.}_{10.5}$} & \multicolumn{4}{l}{$\qty(\mathrm{sat.}_{18}+\mathrm{SHMR})/\mathrm{sat.}_{18}$} \\
{} &                                    w/o SHMR &                   \pbi &  \bmi & \pbii & \bmii &                 \pbi &  \bmi & \pbii & \bmii \\
\midrule
Aquarius2      &                                        0.77 &                  -1.16 &  0.17 &  0.34 &  1.13 &                -2.16 & -0.35 &  0.05 &  0.29 \\
Bootes1        &                                       -0.01 &                   0.34 &  0.20 &  0.17 &  0.09 &                 0.12 &  0.10 &  0.21 &  0.11 \\
Bootes2        &                                       -0.09 &                   0.05 &  0.06 & -0.05 & -0.16 &                 0.07 &  0.04 &  0.01 & -0.01 \\
CanesVenatici1 &                                        0.49 &                   0.36 &  0.49 &  0.07 &  0.34 &                 0.33 & -0.01 & -0.31 & -0.03 \\
CanesVenatici2 &                                        1.29 &                  -0.70 &  0.64 &  0.92 &  2.08 &                -2.61 & -0.70 &  0.15 &  0.71 \\
Carina2        &                                       -0.14 &                   0.65 &  0.44 &  0.03 & -0.35 &                 0.76 &  0.30 &  0.29 & -0.19 \\
ComaBerenices  &                                        1.06 &                  -1.71 & -0.09 &  0.35 &  1.75 &                -3.07 & -0.52 &  0.07 &  0.64 \\
Draco2         &                                        0.16 &                   0.10 &  0.10 &  0.04 &  0.24 &                -0.07 & -0.09 & -0.05 & -0.01 \\
Eridanus2      &                                        0.79 &                  -0.22 &  0.76 &  0.94 &  1.53 &                -0.62 & -0.03 &  0.05 &  0.78 \\
Grus1          &                                       -0.30 &                   0.34 &  0.14 & -0.07 & -0.40 &                 0.27 &  0.19 &  0.10 & -0.07 \\
Hercules       &                                        0.88 &                   0.58 &  0.96 &  0.59 &  1.06 &                -0.04 & -0.06 & -0.07 &  0.15 \\
Horologium1    &                                        1.12 &                  -3.87 & -0.78 & -0.44 &  1.25 &                -4.13 & -0.58 & -0.03 &  0.25 \\
Hydrus1        &                                       -0.17 &                   0.30 &  0.23 & -0.20 & -0.89 &                 0.38 &  0.09 & -0.03 & -0.85 \\
Leo4           &                                       -0.16 &                   0.34 &  0.02 & -0.05 & -0.93 &                 0.44 &  0.15 &  0.01 & -0.72 \\
Leo5           &                                       -0.03 &                   0.34 &  0.51 &  0.04 &  0.24 &                -0.47 & -0.15 &  0.31 &  0.45 \\
LeoT           &                                        1.39 &                   0.65 &  1.85 &  1.43 &  0.84 &                -0.01 &  0.47 &  0.08 & -0.61 \\
Pegasus3       &                                        1.28 &                  -0.19 &  0.37 &  0.11 &  1.62 &                -1.65 & -0.98 & -0.06 &  0.25 \\
Pisces2        &                                        0.27 &                   0.49 &  0.26 & -0.04 & -0.07 &                 0.29 & -0.01 & -0.11 & -0.28 \\
Reticulum2     &                                        0.96 &                  -1.08 &  0.03 &  0.17 &  1.13 &                -1.85 & -0.82 & -0.12 &  0.13 \\
Segue1         &                                        1.89 &                  -2.63 & -1.00 & -0.27 &  1.36 &                -4.29 & -1.01 & -0.05 & -0.41 \\
Segue2         &                                        0.08 &                   0.11 &  0.12 &  0.21 &  0.26 &                -0.17 & -0.20 &  0.04 & -0.10 \\
Triangulum2    &                                       -0.65 &                  -0.01 &  0.14 & -0.02 & -0.52 &                 0.57 &  0.37 &  0.08 &  0.16 \\
Tucana2        &                                       -0.13 &                  -0.23 &  0.16 &  0.11 & -0.09 &                -0.54 &  0.12 &  0.16 & -0.06 \\
Tucana3        &                                       -2.75 &                   0.56 &  0.18 & -0.01 & -2.82 &                 3.39 &  2.68 & -0.08 &  0.08 \\
UrsaMajor1     &                                        1.06 &                  -5.26 & -0.16 &  0.61 &  1.84 &                -5.30 & -0.45 & -0.08 &  0.80 \\
UrsaMajor2     &                                        1.25 &                  -4.98 & -1.11 &  0.05 &  1.67 &                -5.93 & -0.63 & -0.21 &  0.37 \\
Willman1       &                                        2.07 &                  -3.05 & -1.05 & -0.62 &  1.71 &                -4.90 & -1.26 & -0.37 & -0.29 \\
\bottomrule
\end{tabular}

    \label{tab:wbic}
\end{table}

\Cref{fig:posterior_segue1} shows the posterior projected onto $r_s$-$\rho_s$ plane. The value of likelihood function is shown by blue dots and contours. Colored contours denote posteriors assuming the satellite and/or SHMR priors. The satellite prior itself is shown by gray contours and gray shaded area shows the cosmological constraint adopted in Ref.~\cite{Geringer-Sameth:2014yza}. For illustration purpose, we only show the result of Segue 1. For other dSphs, see \cref{fig:posteriors_v50_105,fig:posteriors_v50_180}.

In order to clarify the advantage of $R$-dependent analysis, we compare results of $R$-dependent and $R$-independent likelihood analyses in \cref{fig:posterior_degen}, where $R$-independent likelihood is defined similarly to \cref{eq:likelihood} but the velocity dispersion $\sigmalossq(R)$ is replaced by averaged dispersion $\overline{\sigmalossq}$. In \cref{fig:posterior_degen} the color of the heat map corresponds to the value of likelihood functions.

\Cref{tab:j_table_ufd,tab:j_table_classical} show the median values of $J$-factor posteriors with 68\% ($\sim 1\sigma$) credible intervals. Left three columns show results without SHMR priors. In particular, ``flat'' column denotes those without any cosmological priors (only with the likelihood and the photometry prior). Following columns are those with SHMR priors, \pbi, \bmi, \pbii, and \bmii, respectively. 
These results are also shown in \cref{fig:j_table_ufd}. In \cref{fig:j_table_ufd}, we also show the results of conventional analysis~\cite{Ando:2020yyk} as gray bars for comparison.

We show the Bayes factor of each model in \cref{tab:wbic}. Column 1 shows the Bayes factor of $\mathrm{sat}_{18}$ to a reference model $\mathrm{sat}_{10.5}$ for each dSphs. Columns 2-5 shows the Bayes factors of the satellite prior $\mathrm{sat}_{10.5}$ and SHMR analyses to the satellite prior only analysis as a reference. Columns 6-9 are same as Columns 2-5 but for $\mathrm{sat}_{18}$ cases. Here a positive (negative) value indicates that the corresponding model is more (less) credible than $\mathrm{sat}_{10.5}$.

\section{Discussion}
\subsection{Posterior}
For Bo\"otes 2, Draco 2, Leo 4, Pegesus 3, Pisces 2, Segue 2, Triangulum 2 and Tucana 3, their posterior distributions of $r_s$-$\rho_s$ without satellite priors (likelihood) are broadly distributed (\cref{fig:posteriors_v50_105,fig:posteriors_v50_180}). This is because observational errors of spectroscopic data set are too large to exclude small $r_s$-$\rho_s$ region (dSph without dark matter). In such a case, the GS15-like cut excludes heavier halo mass region but estimated $J$-factor is still distributed broadly towards the small $r_s$-$\rho_s$ region, thus the choice of scanning range of $r_s$ and $\rho_s$ strongly affects the result of estimation. This problem is solved by introducing the satellite prior because it excludes the small $r_s$-$\rho_s$ region based on the formation history of dSphs.

For the other ultra faint dSphs, posterior distribution becomes more ridgy thanks to a large amount of kinematical data. In contrast with those obtained in \ando, the likelihood edges becomes narrow towards the upper left (compact) or lower right (faint) regions, which indicates that the height of likelihood peak varies from the upper left to the lower right. This is thanks to the radial dependence of the velocity dispersion $\sigma(R)$; even though $\overline{\sigma}$ can be constant by varying $r_s$ and $\rho_s$ properly, $\sigma(R)$ cannot be kept to fit observed stellar velocity distribution at all radii.

\Cref{fig:posterior_degen} shows that introducing $R$-dependence in the likelihood function mitigate the degeneracy between $\rho_s$ and $r_s$ in \ando. Since certain combinations of $\rho_s$ and $r_s$ gives same value of mean $\sigma_\text{los}^2$, $R$-independent likelihood as used in \ando\ has a degeneracy problem. In contrast, the function $\sigma_\text{los}^2$ is not equivalent even for such a combination, hence it allows us to distinguish these parameter sets. Introducing $R$-dependence however causes another issue, namely, arbitrariness of anisotropy function $\beta(r)$, which is just assumed to be constant for simplicity in this study. In order to remove unexpected bias, this arbitrariness should be carefully treated in the further study as well as other arbitrariness such as the axisymmetrisity.

\subsection{$J$-factor and Bayes factor}

\Cref{fig:j_table_ufd} shows that, in the satellite prior only analysis, our estimates of log-$J$-factor are larger by  $\sim\order{0.1}$ than those estimated in \ando. This is because the $R$-dependence of our likelihood function weakens the $\rho_s$-$r_s$ degeneracy, as mentioned in the previous section, and excludes too compact (small $r_s$, large $\rho_s$) or faint (large $r_s$, small $\rho_s$) dark matter halo with small $J$-factor value.

\Cref{fig:j_table_ufd} also shows that SHMR priors decrease the uncertainty of $J$-factor by up to about 50\%, but estimated median values have SHMR model dependence and some estimations are not consistent with each other. For instance, the \pbi\ prior tends to predict smaller $J$-factor than other priors for dSphs with large $J$-factors such as Segue 1. Conversely, the \bmii\ prior gives larger $J$-factor than other priors for small $J$-factor dSphs such as Leo 4. These features come from the difference of SHMR models. As shown in \cref{fig:shmrs}, SHMR models have different slopes for small $M_h$ region around the mass scale of dSph halos. In particular, \pbi\ model has large $m_\ast/M_h$ ratio, while \bmii\ one smaller $m_\ast/M_h$ than others. Once $m_\ast$ is fixed by observations, large $m_\ast/M_h$ gives small $M_h$, and vice versa.
We note that Ref.~\cite{2008MNRAS.390.1453W} reported $M/L\sim1.6$, thus our estimates of stellar mass obtained by assuming $M/L=1$ are potentially smaller than actual values. However, this discrepancy has no significant effect on our estimation because of the scatter of SHMR models.

Bayes factors help us understand the model dependence of the estimated $J$-factors. \Cref{tab:j_table_ufd,tab:wbic} show that models whose estimate is deviated from the result of satellite prior only analysis tend to have small Bayes factors. For instance, the \pbi\ model shows $\ln\mathrm{BF}\lesssim -3$ for Segue 1 and Willman 1, and the \bmii\ shows $\ln\mathrm{BF}\lesssim -1$ for Leo 4. It means that, in terms of the Bayesian analysis, the results of \pbi\ for Segue 1 and Willman 1 are very strongly less reliable than those of the satellite prior only analysis, and the results of \bmii\ for Leo 4 are substantially less reliable, respectively. We can understand this feature through posteriors in \cref{fig:posteriors_v50_105,fig:posteriors_v50_180}. For these dSphs, posteriors obtained by \pbi\ or \bmii\ are significantly deviated from the contour of the satellite prior only analysis, which means that these SHMR models and the satellite prior are incompatible.
In contrast, models having comparable $J$-factors to the satellite prior only analysis have Bayes factors almost equal to or larger than the satellite prior only analysis. This tendency of the Bayes factors indicates that the estimated $J$-factor values with the satellite prior only analysis are stable even when considering SHMRs.

We can utilize this tendency in the opposite direction; not evaluating dark matter profiles by using SHMRs, but evaluating SHMRs by using dark matter profiles. The relation between $J$-factors and Bayes factors suggest some possibilities that there are some unknown biases in the observation of these dSphs or that some SHMR models having small Bayes factors are invalid for certain ultra faint dSphs. The latter possibility could originate from the difference of the construction of these models; the \pbi\ model predicts larger $m_\ast/M_h$ values than the others around small halo mass region, while those of the \bmii\ model is smaller than the others (see \cref{fig:shmrs}). In particular, Ref.~\cite{2019MNRAS.488.3143B} indicated that \pbi\ assumed a strong surface-brightness incompleteness correction for faint galaxies that is no longer observationally supported \cite{2016MNRAS.463.2746W}, which causes the overestimate of the SHMR around low halo mass region. For \bmii, Ref.~\cite{2018MNRAS.477.1822M} pointed out that the underestimate of the \bmii\ model around low-halo mass region occurs to compensate the overestimation of the number of massive galaxies caused by the Eddington bias.
Further investigation of these features would help us to improve and calibrate these SHMR models using dSph observation or reveal some unknown nature of dSphs.

Since the $J$-factor values of the ultra faint dSphs obtained in this work are not significantly different from conventional values, there are no significant updates for the current dark matter constraints of the indirect detection experiment. The detection sensitivity depends on the lower bounds of $J$-factors. Because $J$-factors of dSphs with the largest $J$-factors such as Segue 1 and Ursa Major 2 do not change significantly even when considering cosmological priors having largest Bayes factors, constraints on dark matter parameters do also not show significant difference. The constraints however could be updated when only we select a part of dSphs as detection targets, where $J$-factor lower bound of each dSph matters.

In contrast, from \cref{tab:j_table_classical,fig:j_table_classical}, the $J$-factor uncertainty of classical dSphs obtained in this work are reduced up to about 20\% of the results of \ando\ due to the consideration of $R$-dependence of the velocity dispersion. In particular, the Draco dSph, having one of the largest $J$-factors, has larger lower bound by about $0.25$ in logarithmic scale than \ando. Since indirect detection sensitivity reflects the lower bound of $J$-factor, the sensitivity might be stronger than the results of \ando. Here we should note that our results has implicit bias of dSph model construction. In other conventional works such as Ref.~\cite{Geringer-Sameth:2014yza} the uncertainty of dark matter profile is taken into account by introducing more general dark matter profile models and they indicate the deviation of dark matter profile from the simple NFW profile.
In this paper, however, we neglect the uncertainty of dark matter profile by fixing it to be the NFW profile and also the that of the anisotropy profile by assuming constant model, thus our results have an implicit bias based on the model construction. In order to calculate more reliable $J$-factor values, we need further investigation to implement the flexibility of dark matter profile into the cosmological prior analysis.


\section{Summary and conclusion}
\label{sec:summary}
In this paper, we utilized two cosmological priors (satellite and SHMR) and a likelihood function with radial dependence to obtain better constraints on the dark matter halo profile of dSphs through the kinematical fitting using spherical Jeans equation. We prepared some different setups for the cosmological priors and estimate the posterior probability density function and $J$-factor. We compared these models and showed that our $J$-factor estimates obtained by using the satellite prior are stable in terms of their Bayes factors even when considering another cosmological prior, the SHMR prior. 
Cosmological priors and the R-dependence of the likelihood mitigates the degeneracy between parameters and decrease the uncertainty of $J$-factor values upto about 50\% for ultrafaint dSphs and about 20\% for classical dSphs. These estimates would be updated by introducing the flexibility of dSph models (e.g. anitotropy, halo profile and non-sphericity).

\section*{Acknowledgements}
This research made use of 
Astropy,\footnote{http://www.astropy.org} a community-developed core Python package for Astronomy~\cite{astropypaperi,astropypaperii}, 
NumPy,\footnote{https://numpy.org/} the fundamental package for scientific computing with Python~\cite{harris2020array},
and pandas,\footnote{https://pandas.pydata.org/} a fast, powerful, flexible and easy to use open source data analysis and manipulation tool, built on top of the Python programming language~\cite{jeff_reback_2021_4681666,mckinney-proc-scipy-2010}.
This work was supported by NEXT KAKENHI Grant Numbers, JP20H05850 (S.H. and S.A.), JP20H05861 (S.A.), JSPS KAKENHI Grant Numbers, JP20H01895, JP21K13909 and JP21H05447 (K.H.).
Kavli IPMU is supported by World Premier International Research Centre Initiative (WPI), MEXT, Japan.

\appendix

\section{Posteriors}
Posterior probability density distribution projected onto $r_s$-$\rho_s$ plane are shown in \cref{fig:posteriors_v50_105,fig:posteriors_v50_180,fig:posteriors_classical}. Here \cref{fig:posteriors_v50_105,fig:posteriors_v50_180} are for ultra faint dSphs, while \cref{fig:posteriors_classical} is for classical dSphs.

\begin{figure}
    \centering
    \tableandfigurefontsize
    \includegraphics[width=\posteriorfigwidth]{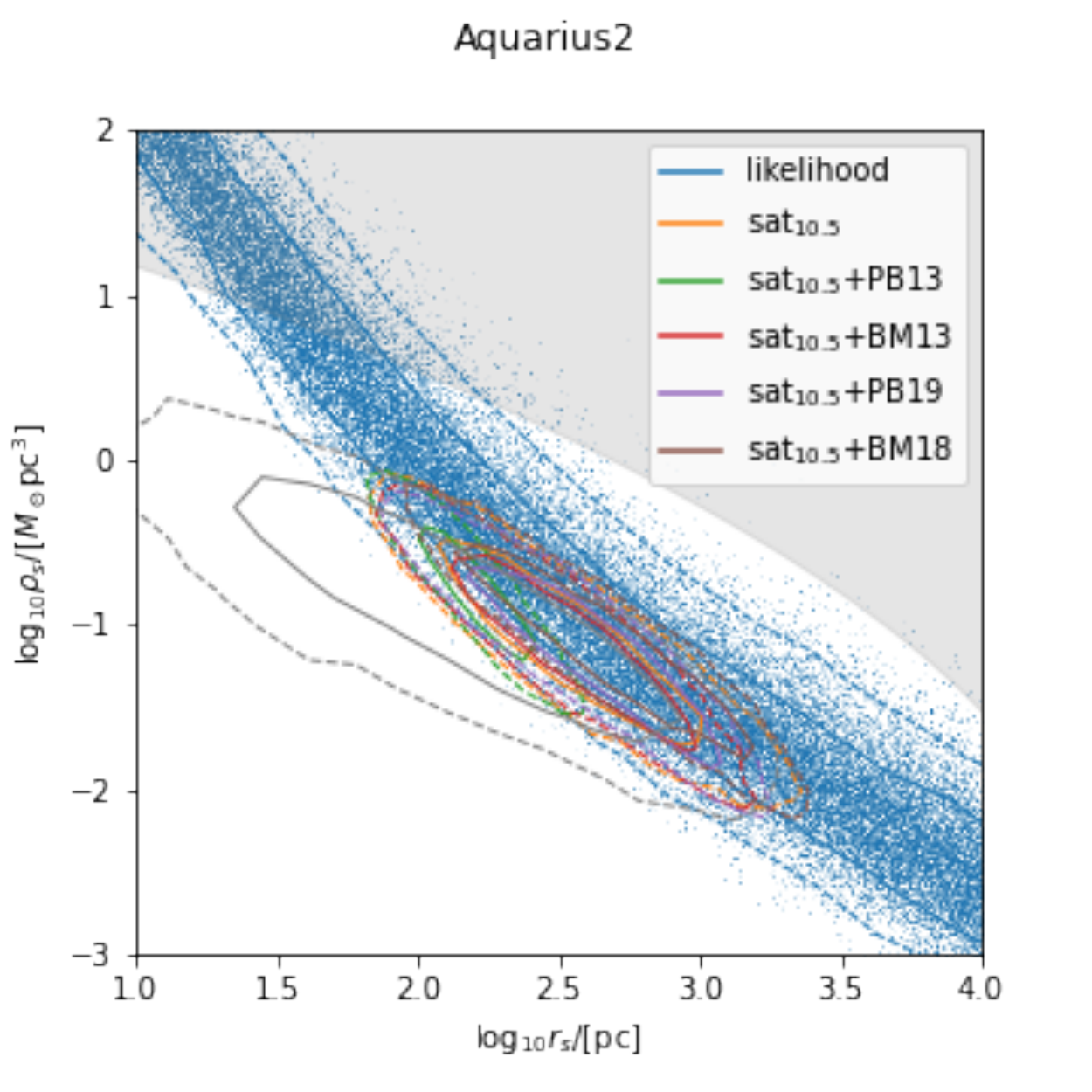}
    \includegraphics[width=\posteriorfigwidth]{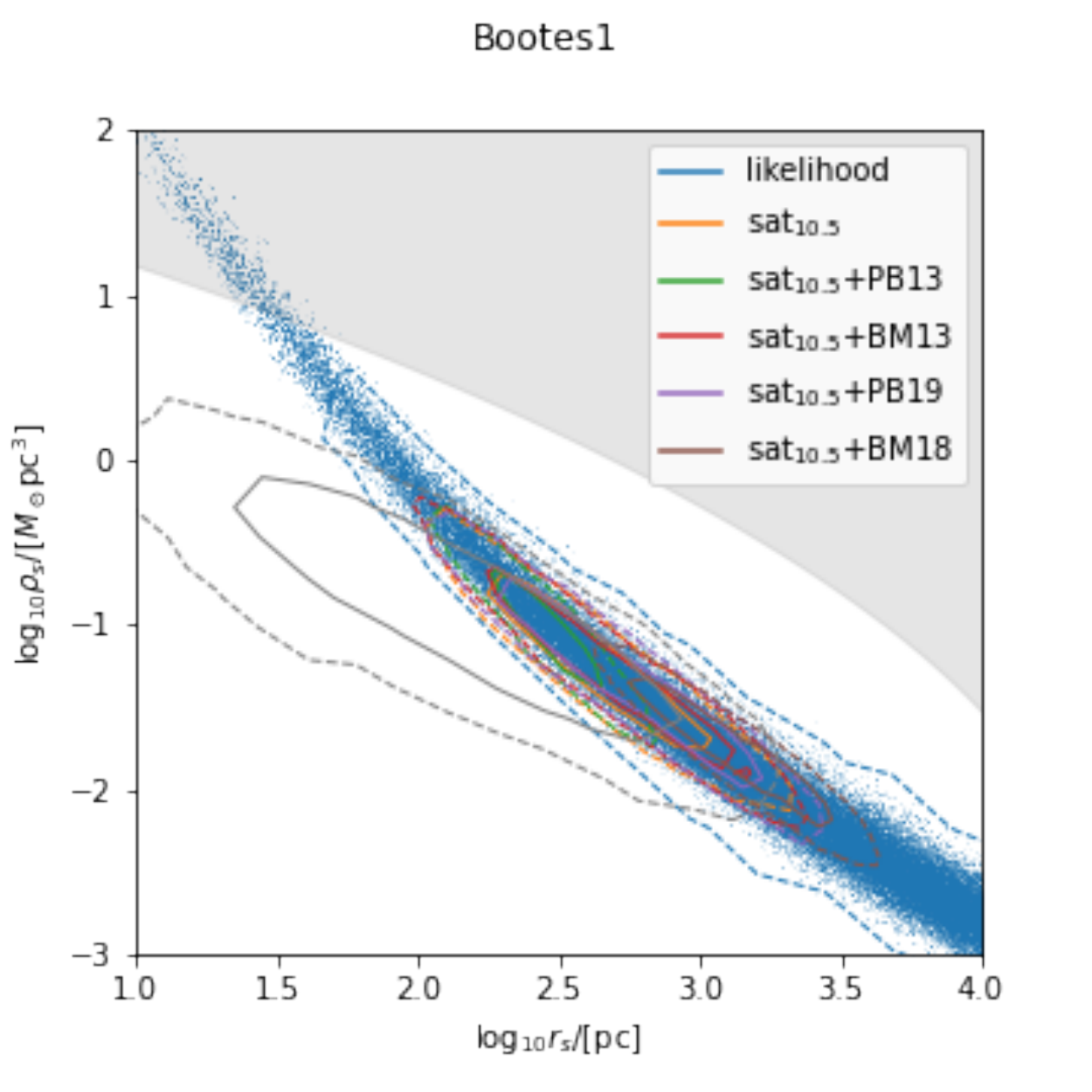}
    \includegraphics[width=\posteriorfigwidth]{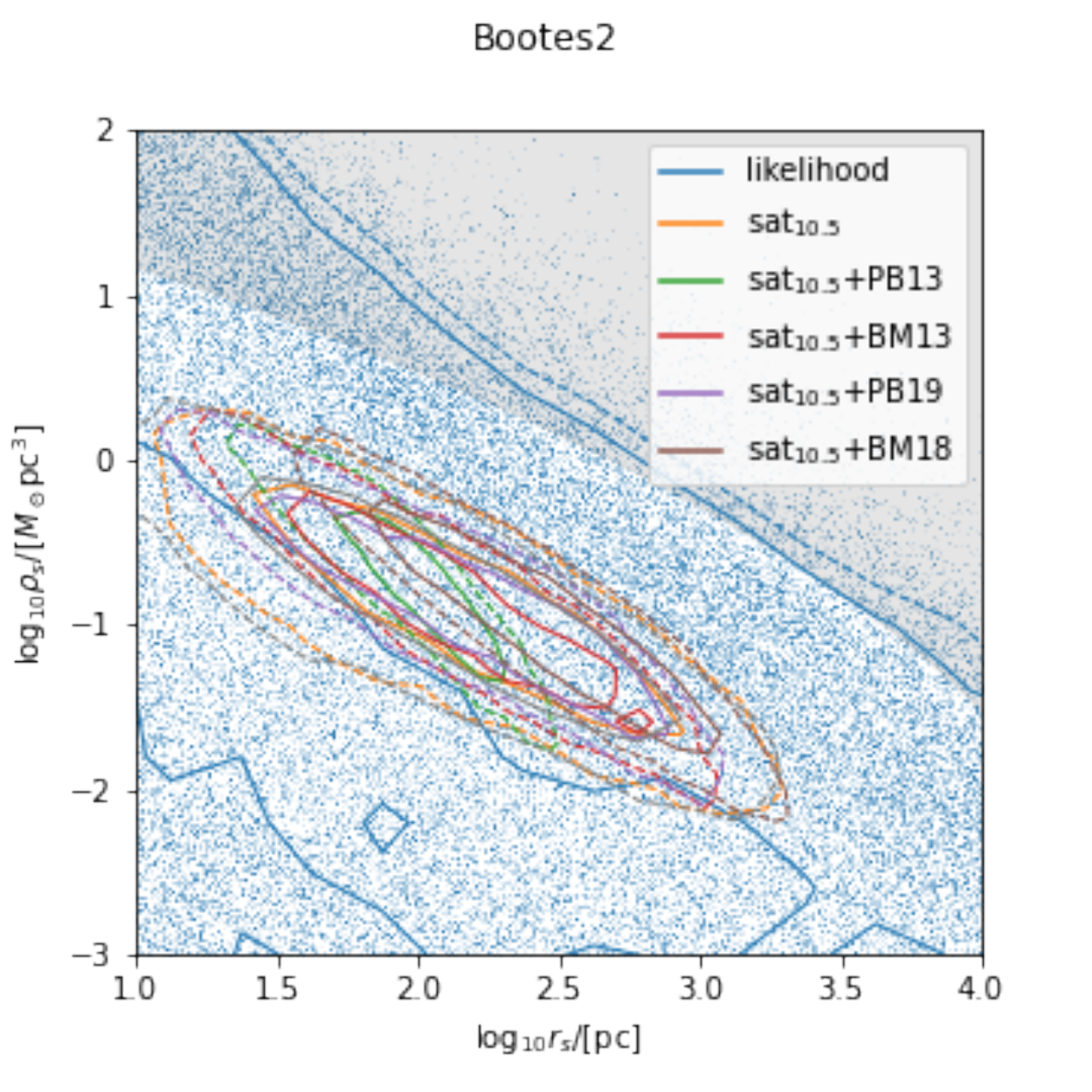}
    \includegraphics[width=\posteriorfigwidth]{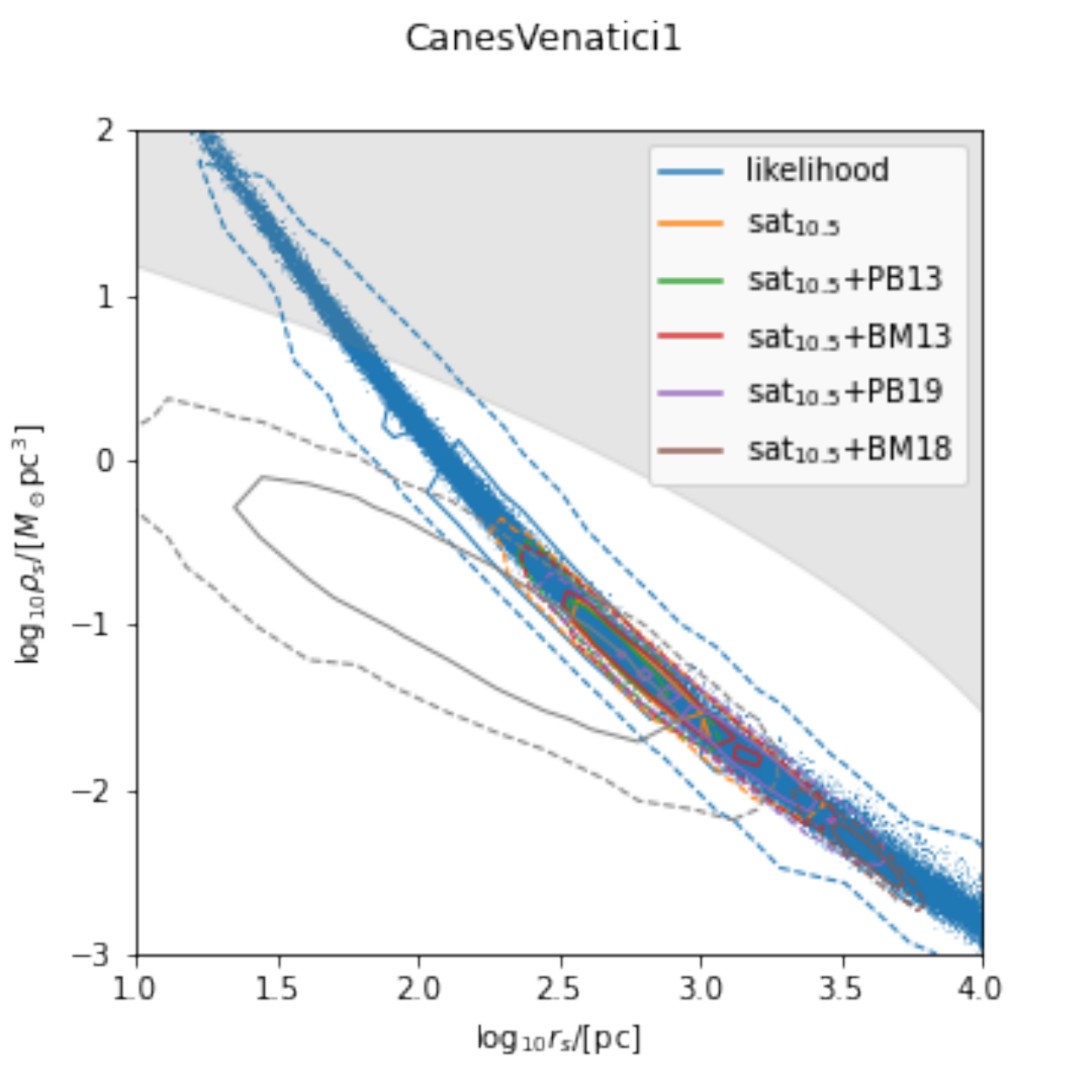}
    \includegraphics[width=\posteriorfigwidth]{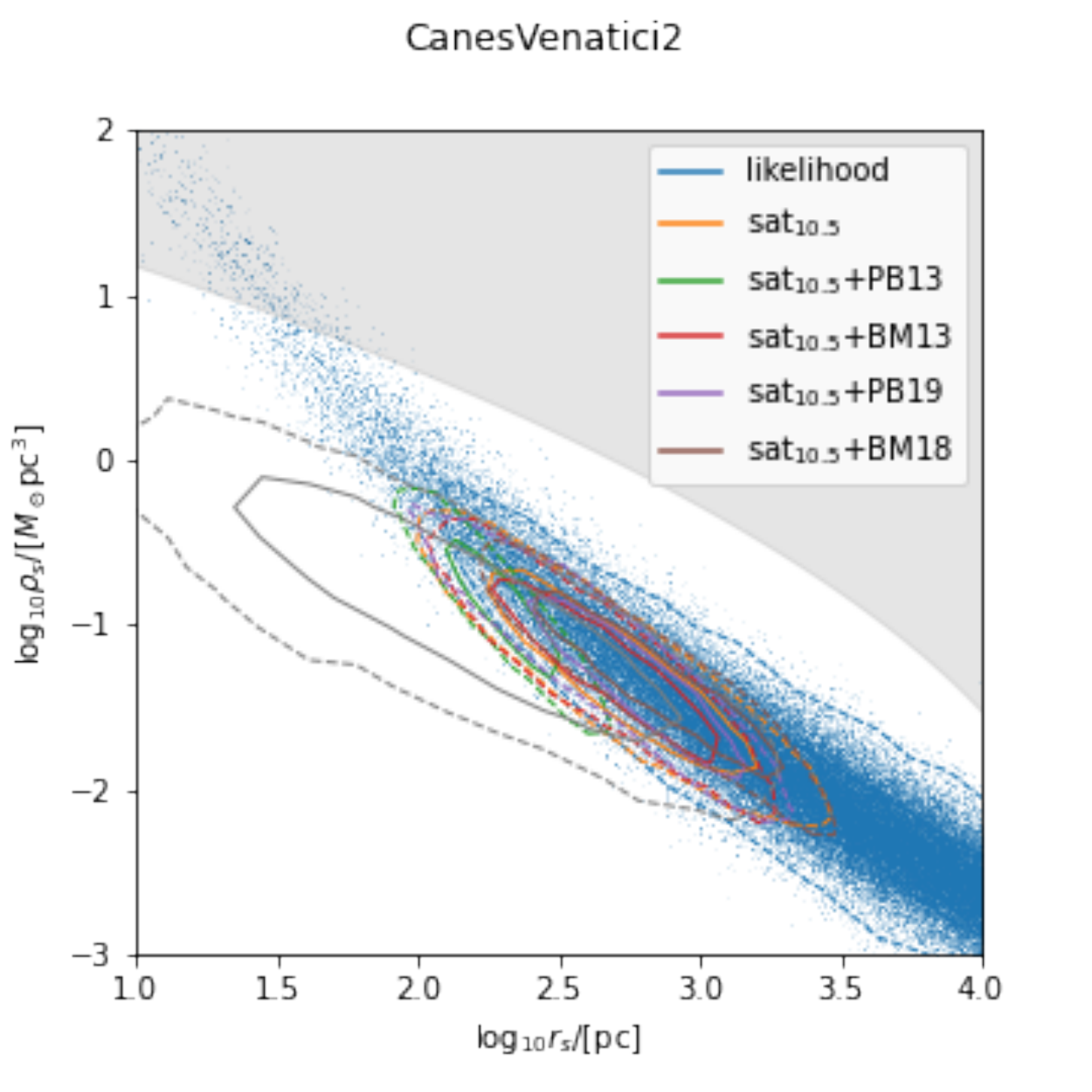}
    \includegraphics[width=\posteriorfigwidth]{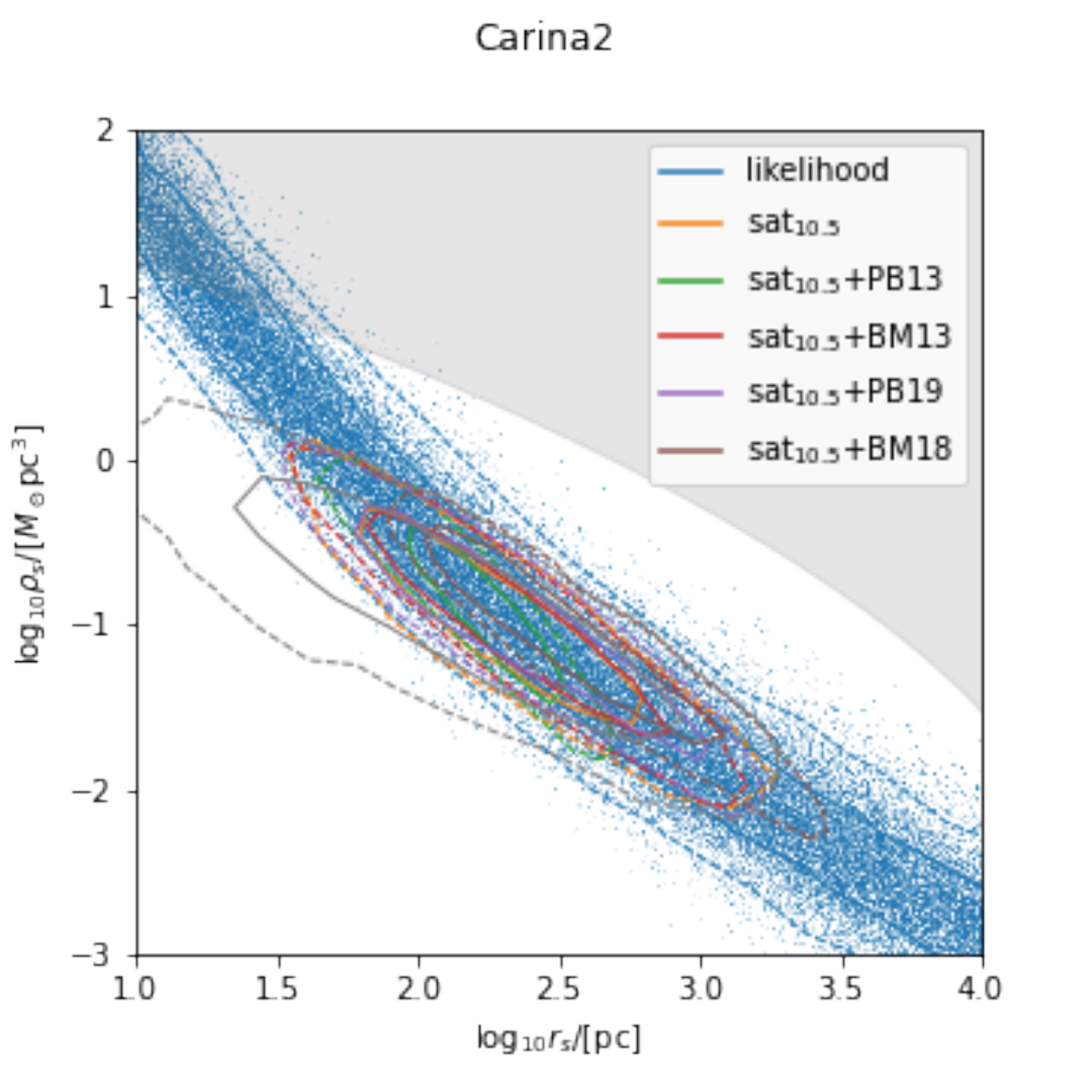}
    \includegraphics[width=\posteriorfigwidth]{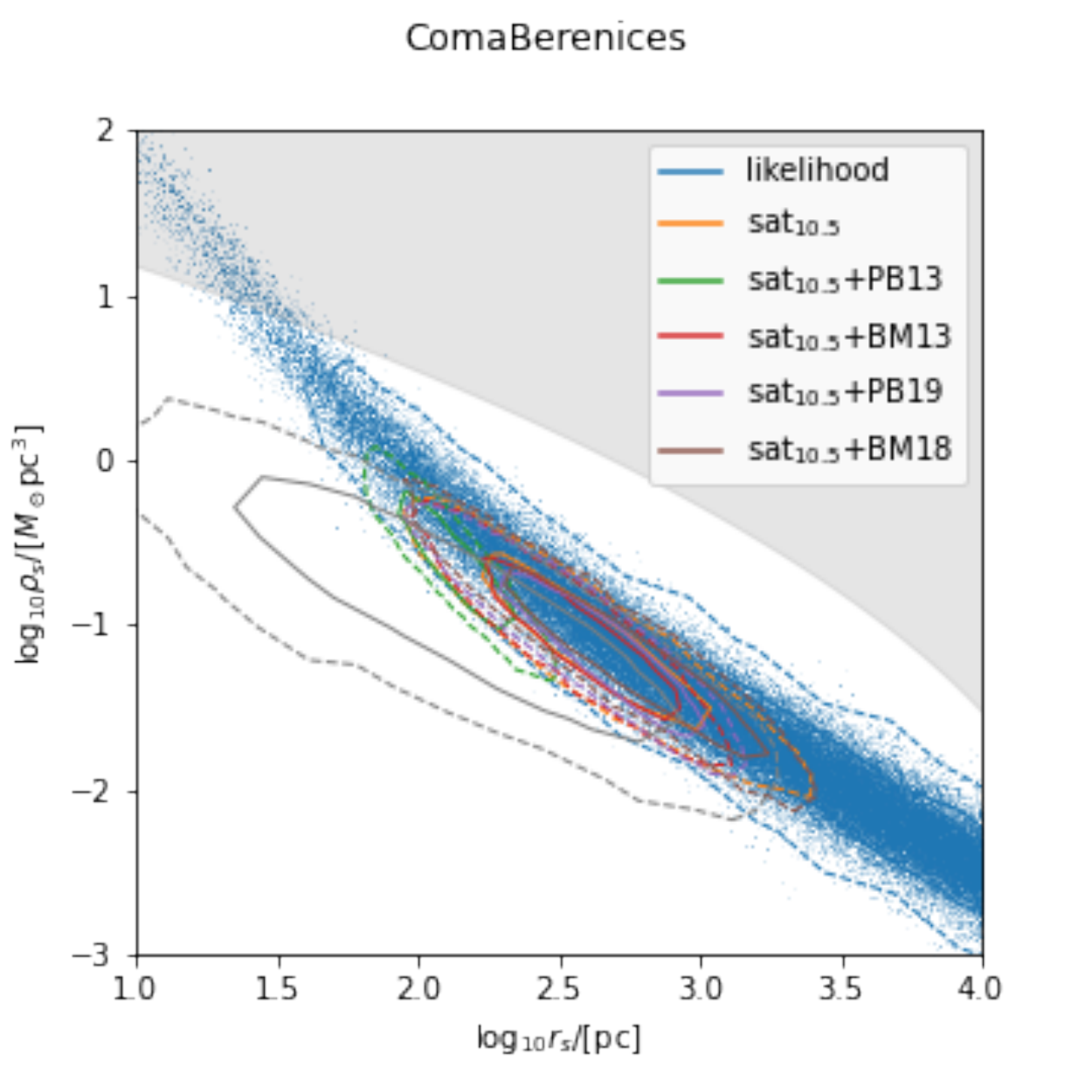}
    \includegraphics[width=\posteriorfigwidth]{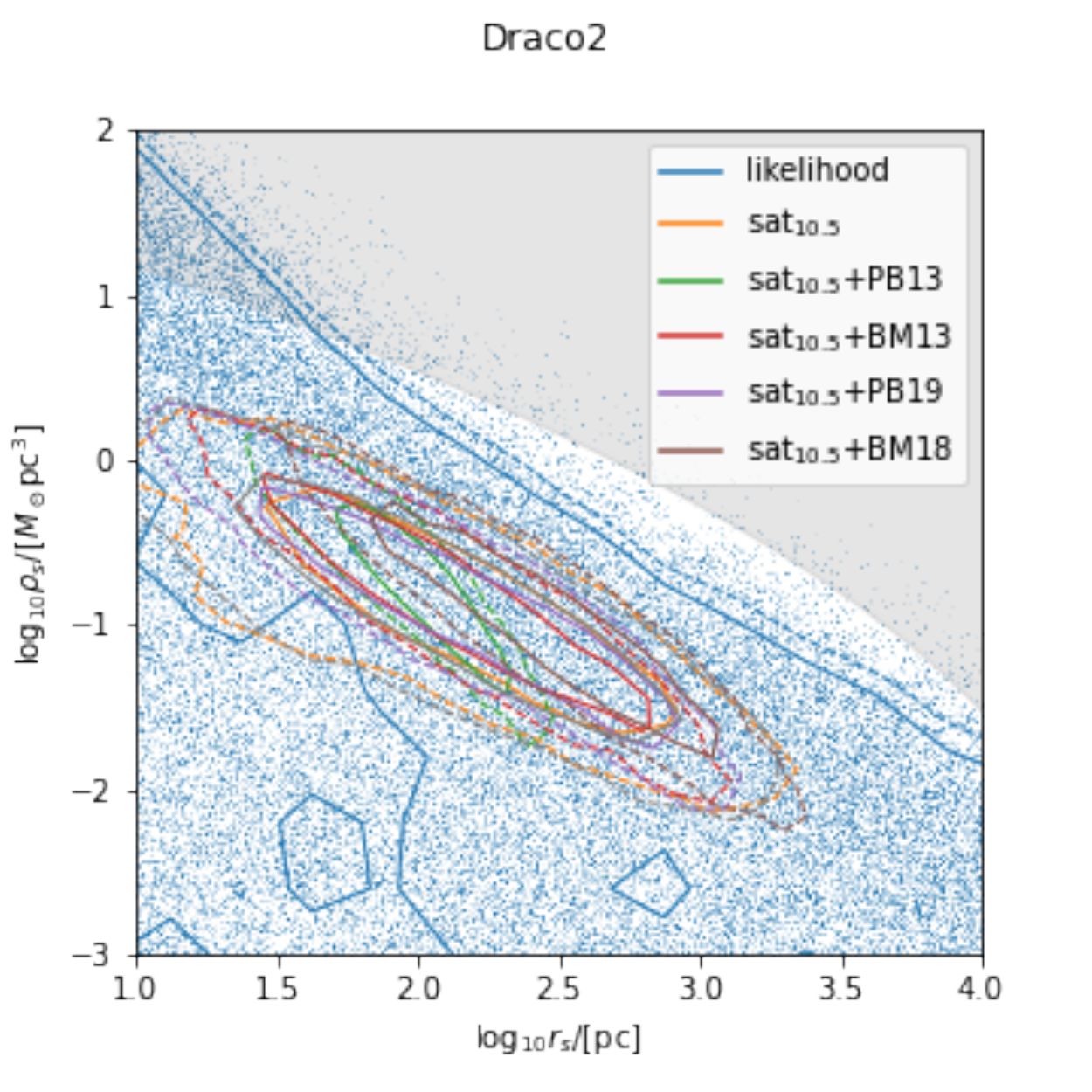}
    \includegraphics[width=\posteriorfigwidth]{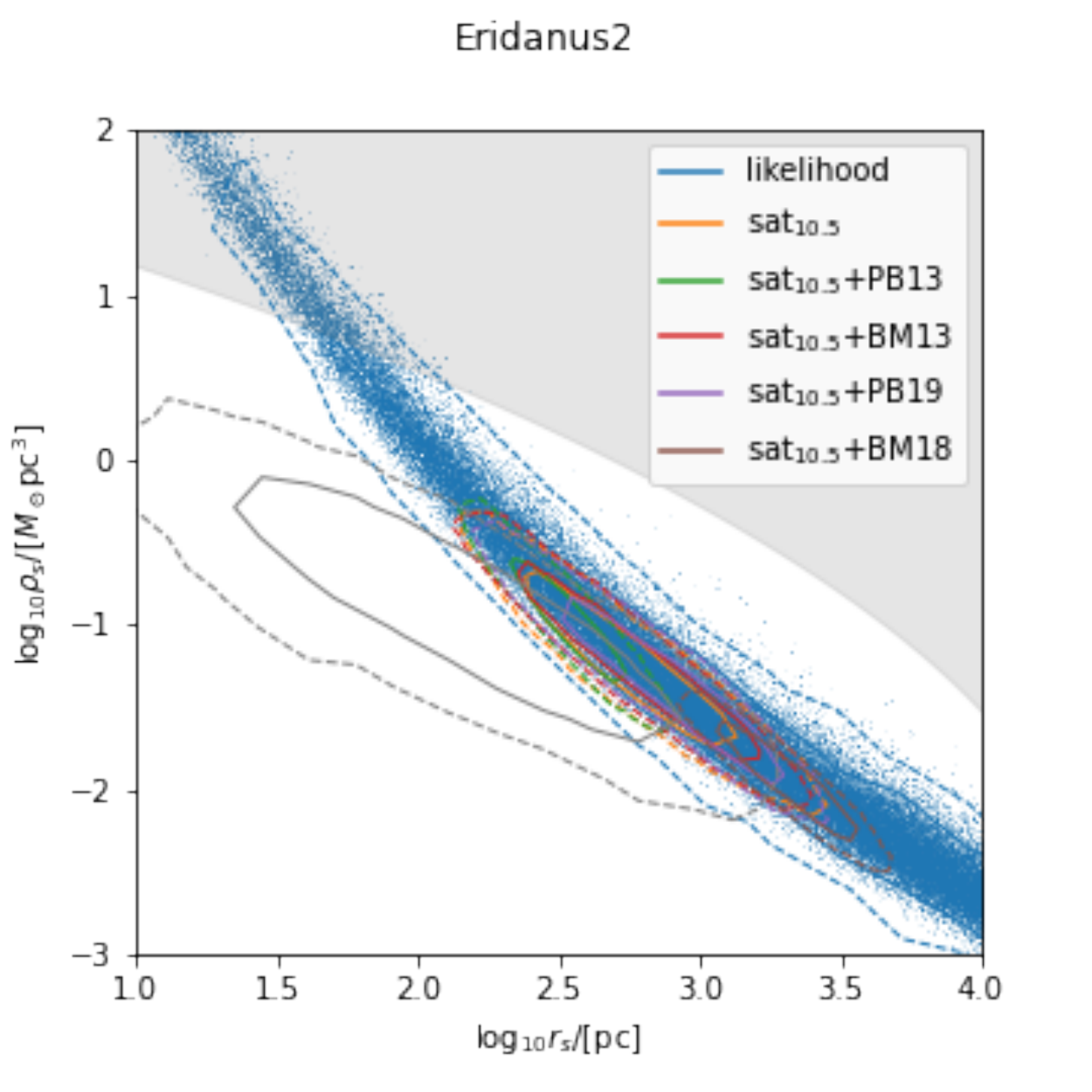}
    \includegraphics[width=\posteriorfigwidth]{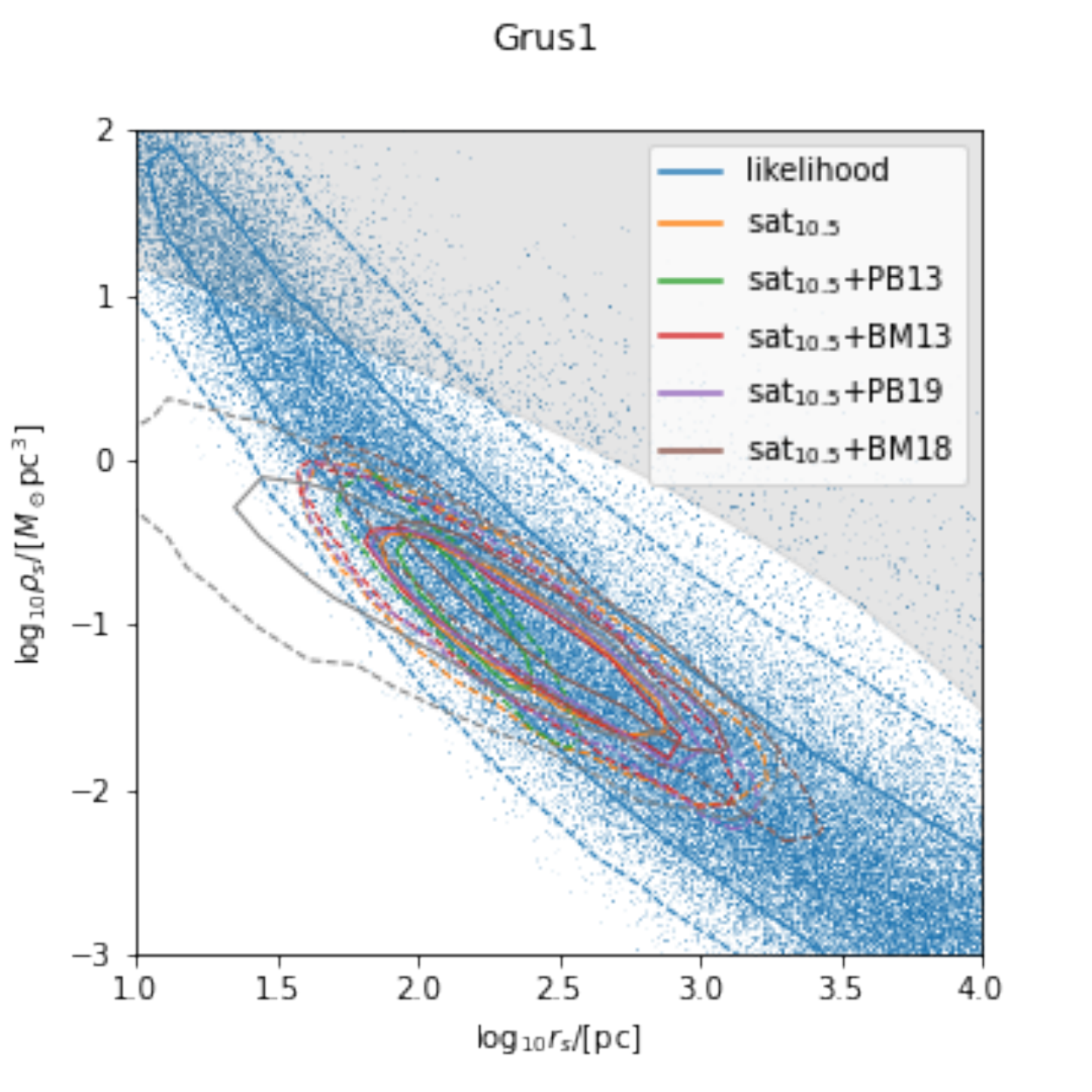}
    \includegraphics[width=\posteriorfigwidth]{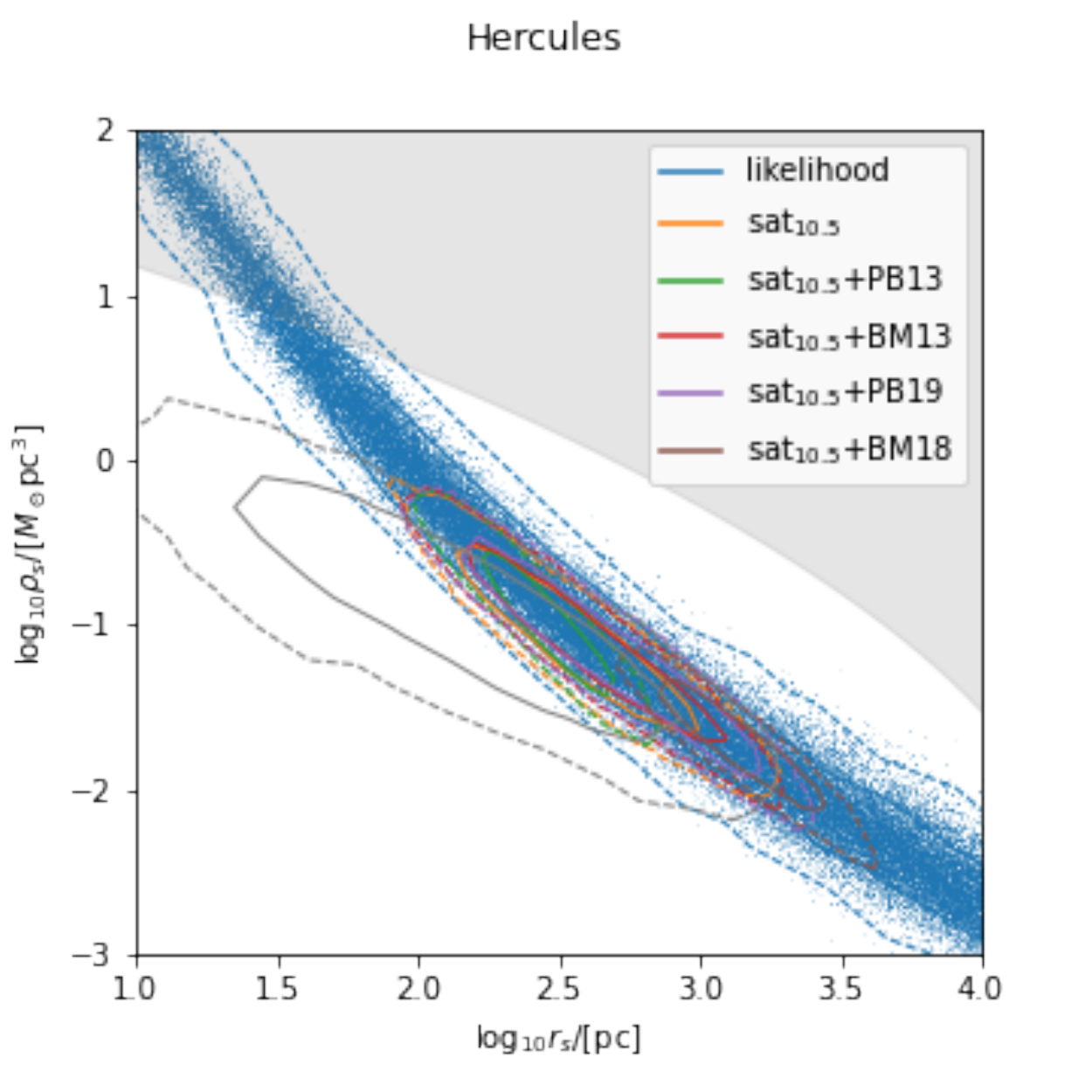}
    \includegraphics[width=\posteriorfigwidth]{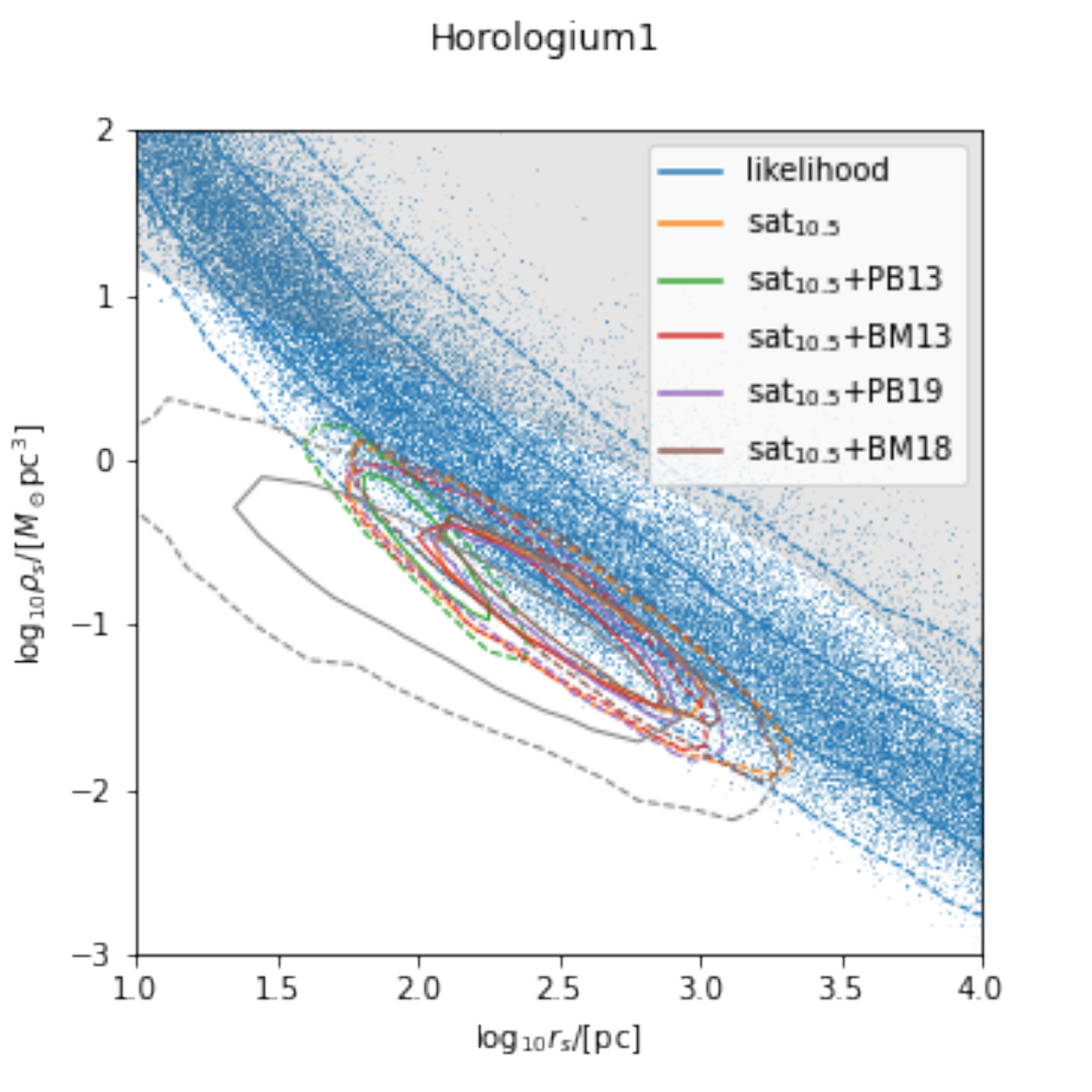}
    \caption{\tableandfigurefontsize
    Posterior probability density distribution projected onto $r_s$-$\rho_s$ plain with $\mathrm{sat.}_{10.5}$ prior. The gray shaded area shows the cosmological constraint adopted in Ref.~\cite{Geringer-Sameth:2014yza}. The gray contours shows the probability density distribution of is the satellite prior $\mathrm{sat.}_{10.5}$. The blue dots and contours illustrate the shape of the likelihood function (flat prior). Colored contours shows the posterior probability density distribution assuming our cosmological priors.}
    \label{fig:posteriors_v50_105}
\end{figure}

\begin{figure}
    \ContinuedFloat
    \captionsetup{list=off,format=cont}
    \centering
    \tableandfigurefontsize
    \includegraphics[width=\posteriorfigwidth]{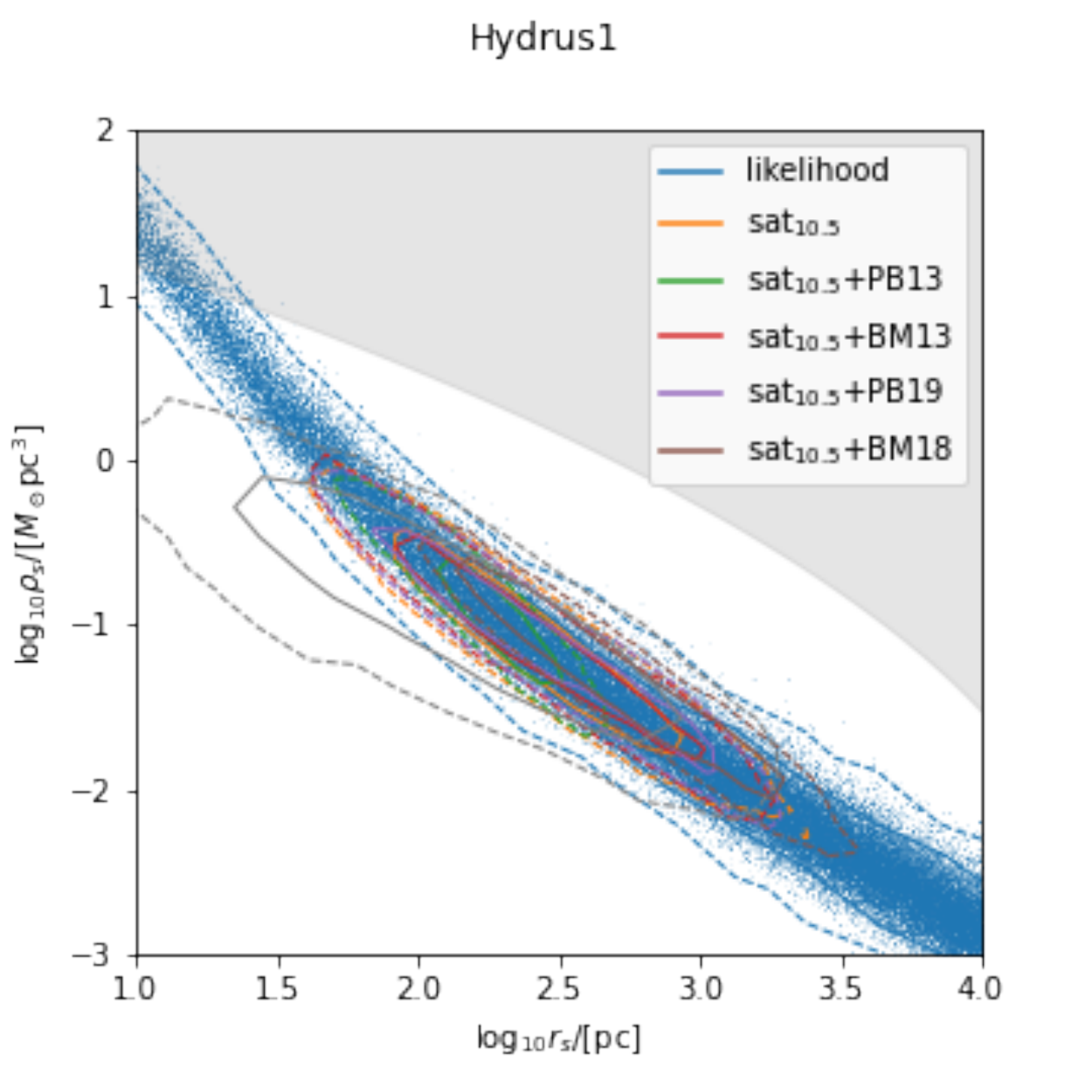}
    \includegraphics[width=\posteriorfigwidth]{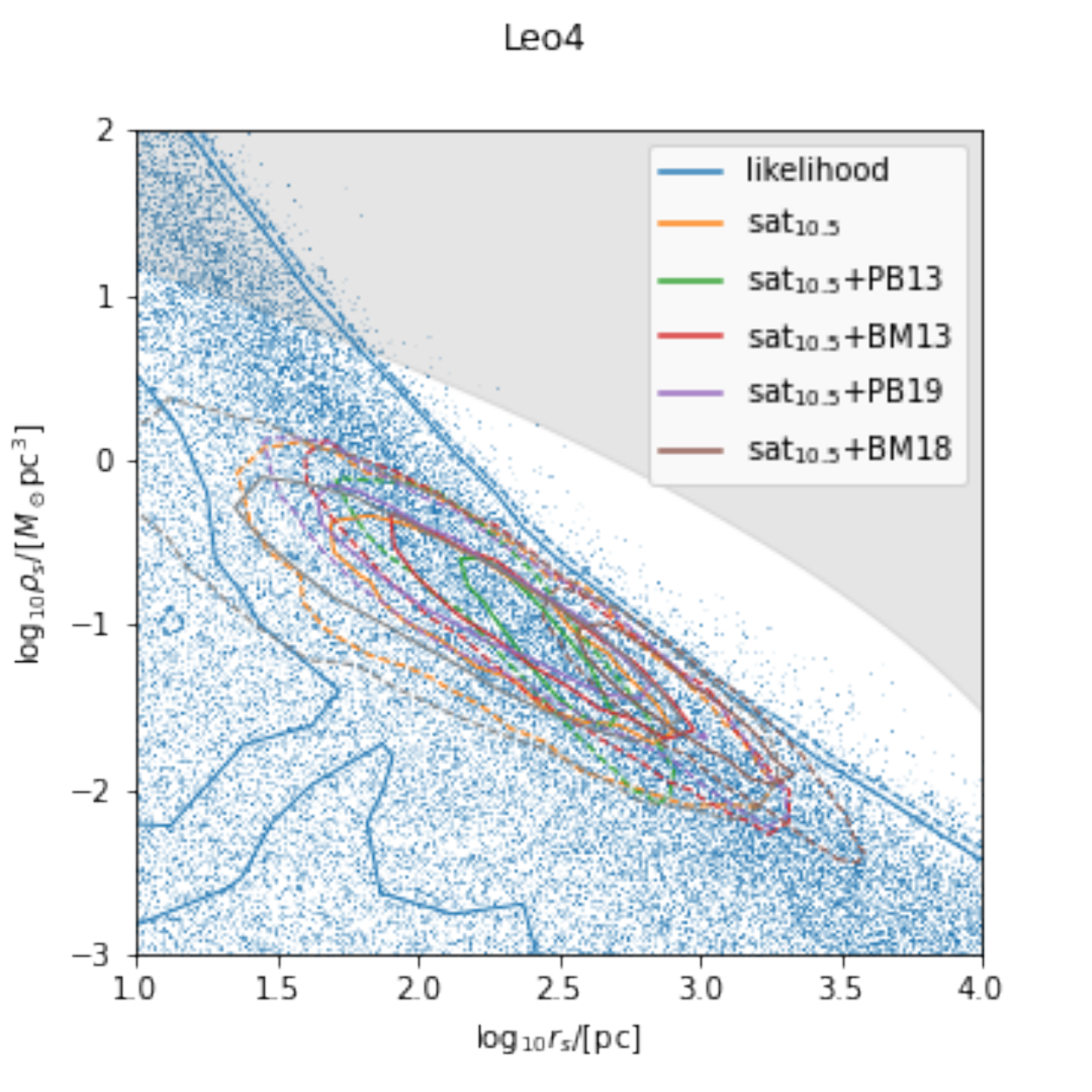}
    \includegraphics[width=\posteriorfigwidth]{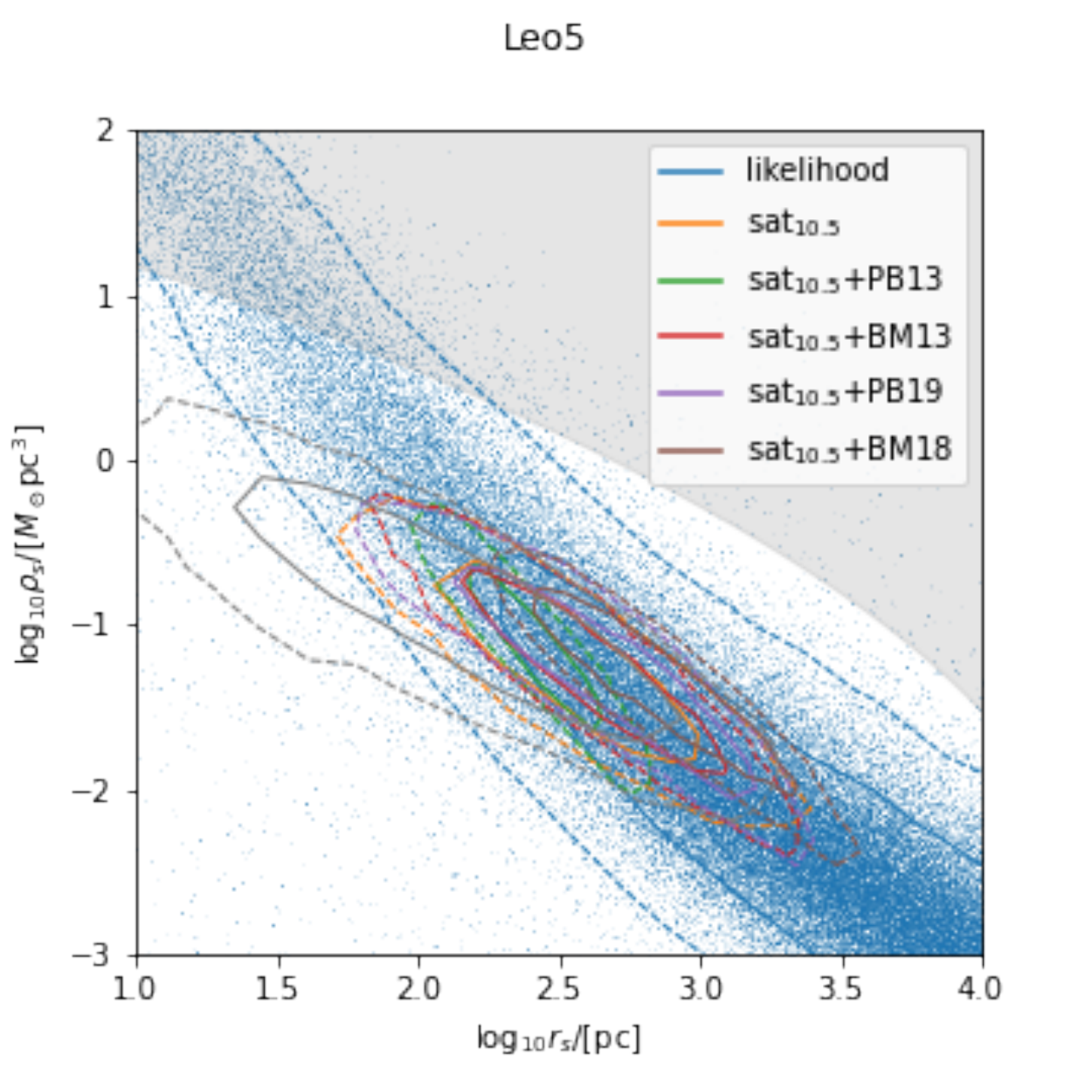}
    \includegraphics[width=\posteriorfigwidth]{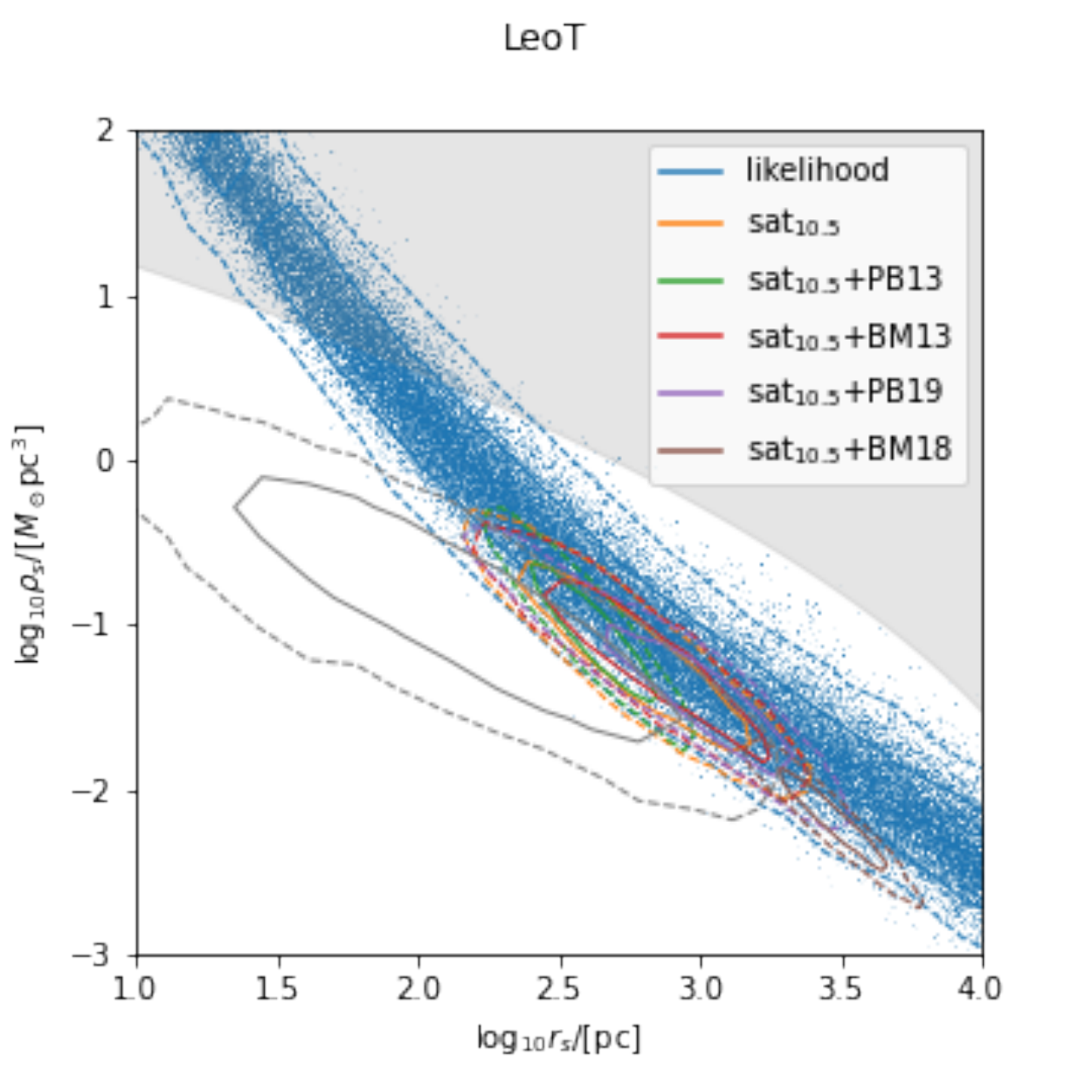}
    \includegraphics[width=\posteriorfigwidth]{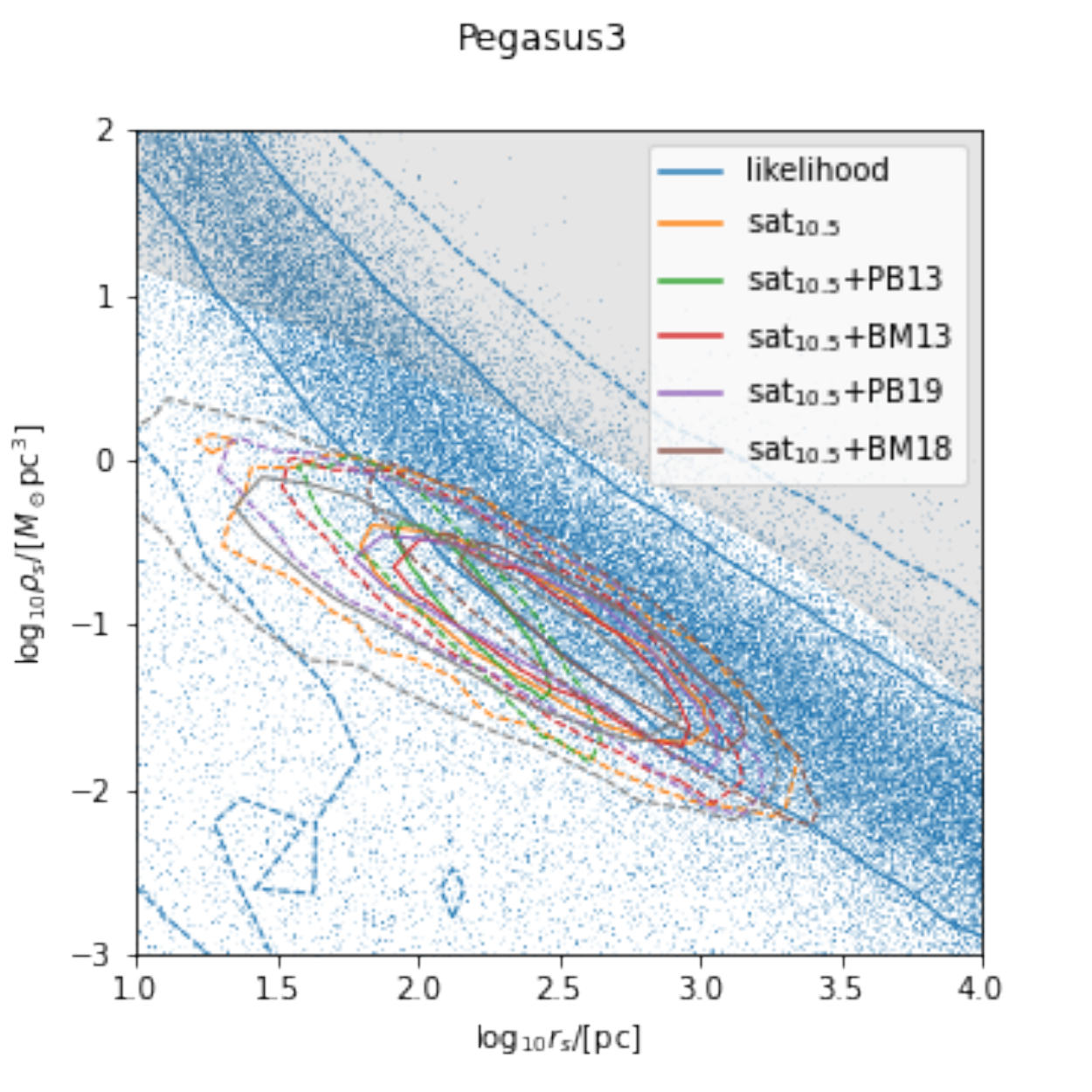}
    \includegraphics[width=\posteriorfigwidth]{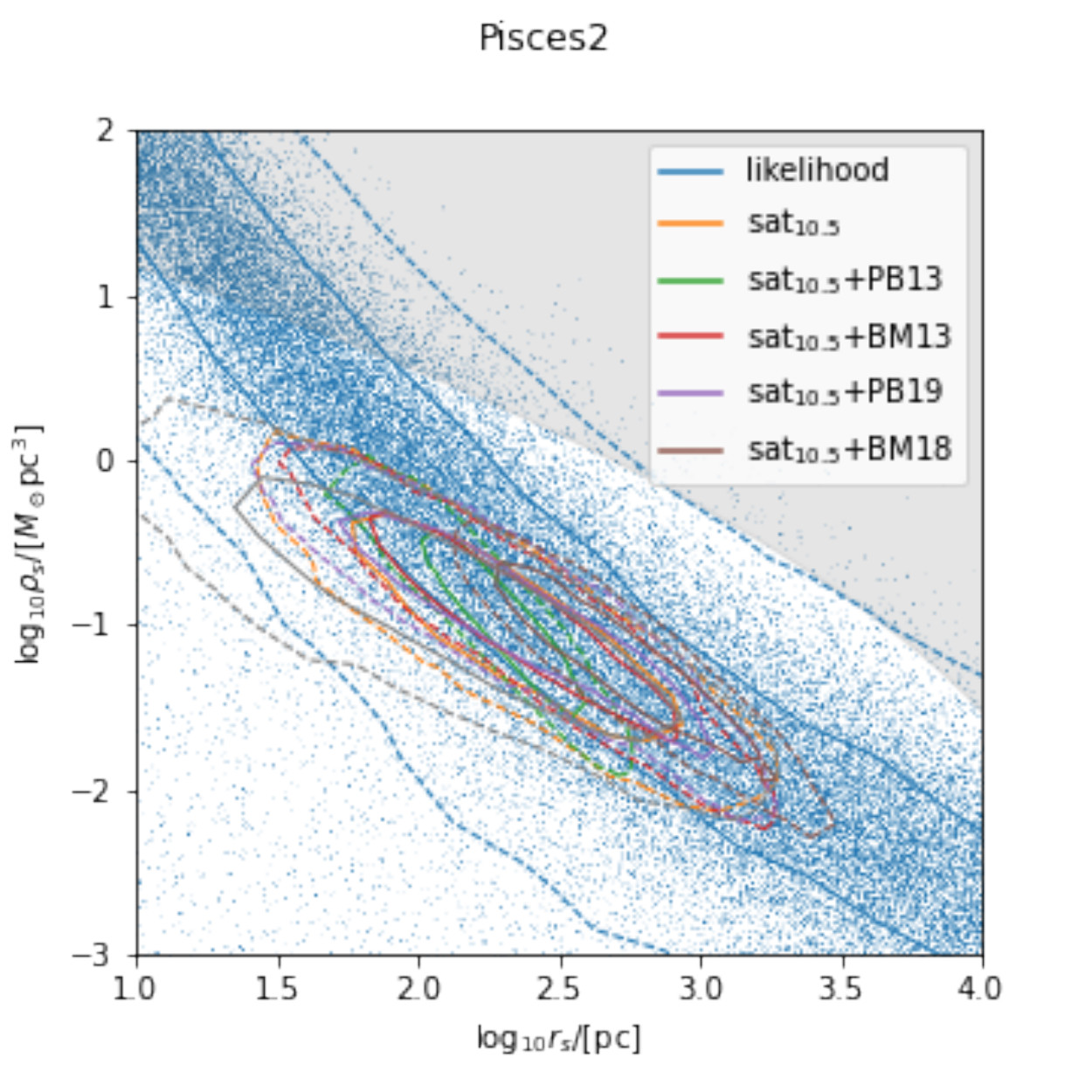}
    \includegraphics[width=\posteriorfigwidth]{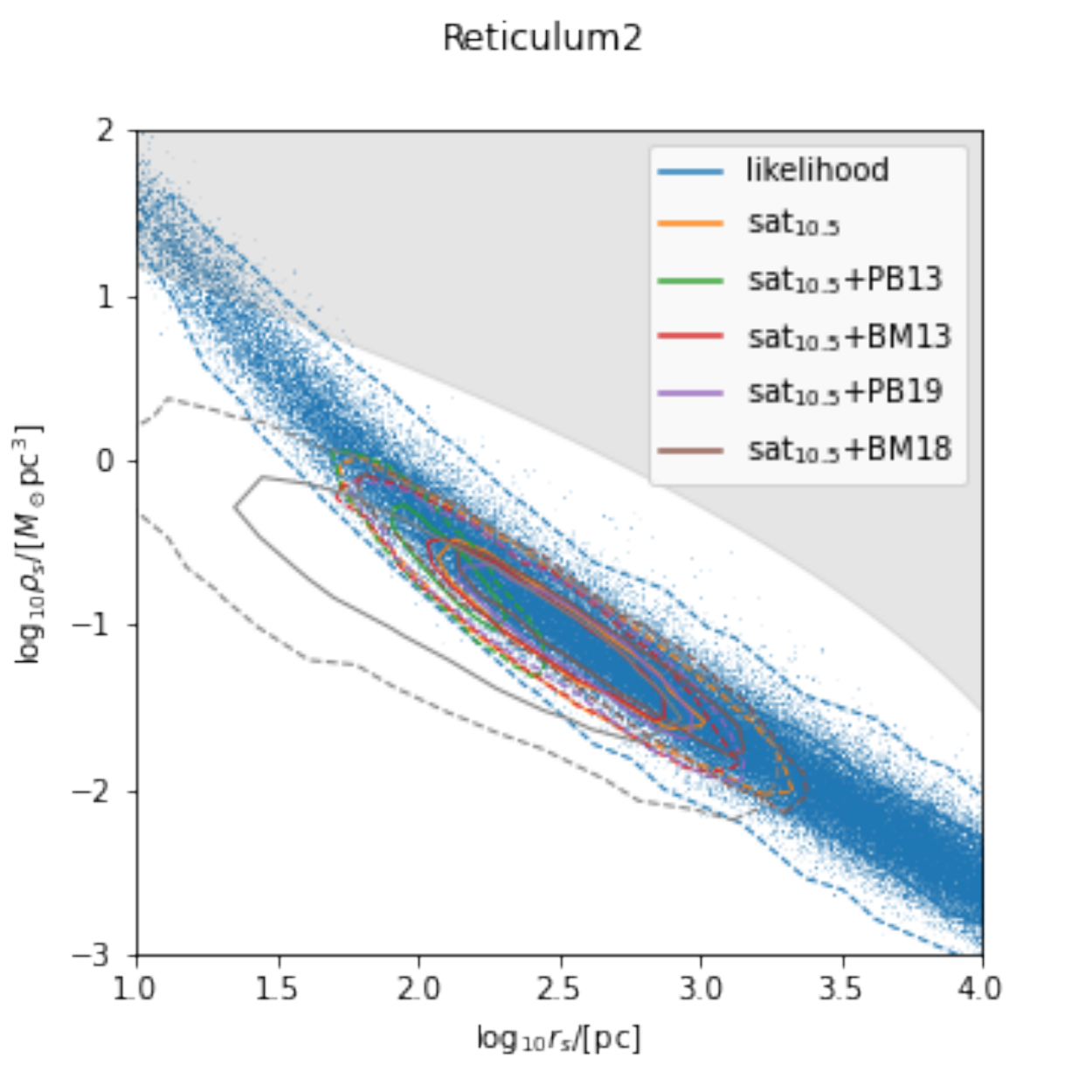}
    \includegraphics[width=\posteriorfigwidth]{fig/Segue1_v50_105.pdf}
    \includegraphics[width=\posteriorfigwidth]{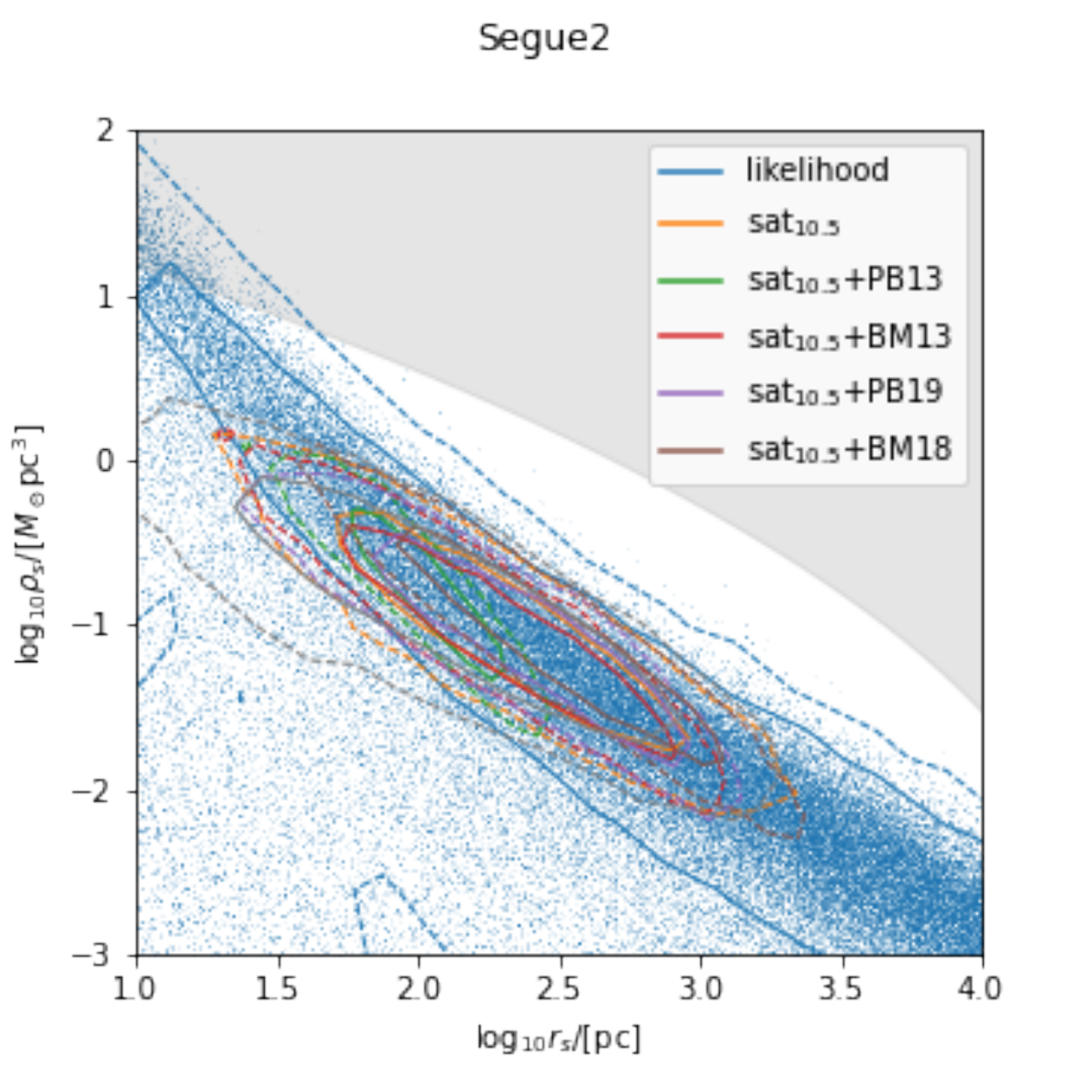}
    \includegraphics[width=\posteriorfigwidth]{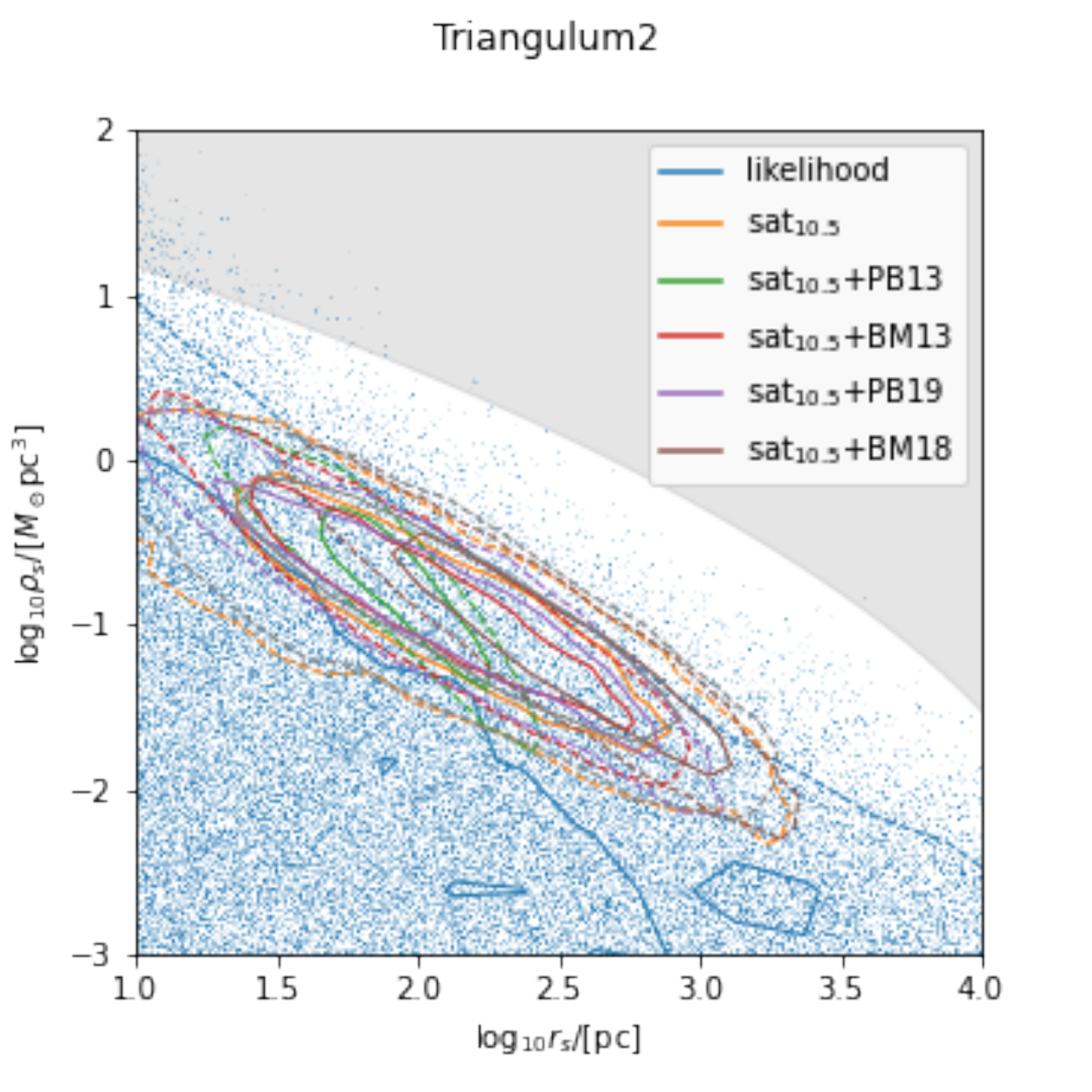}
    \includegraphics[width=\posteriorfigwidth]{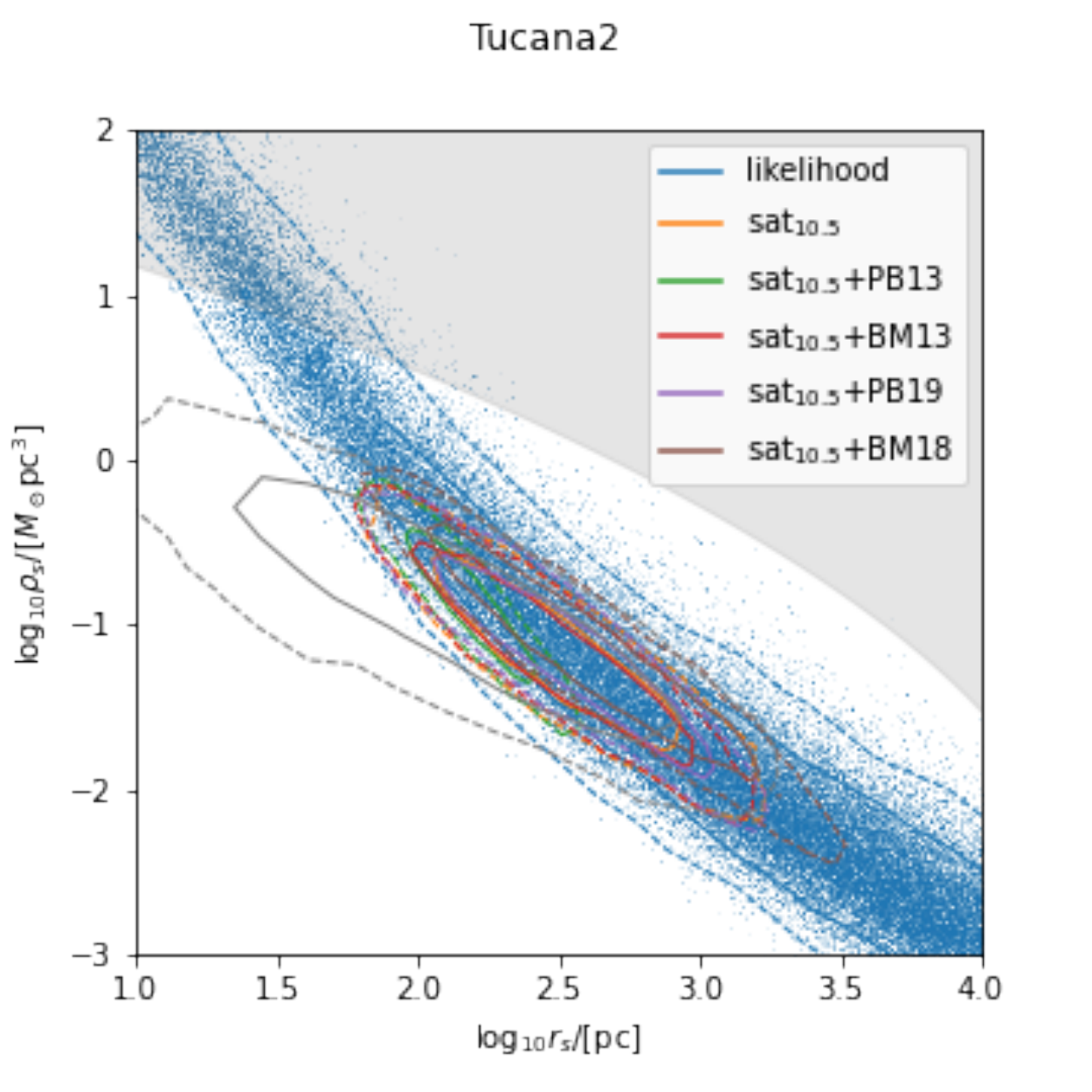}
    \includegraphics[width=\posteriorfigwidth]{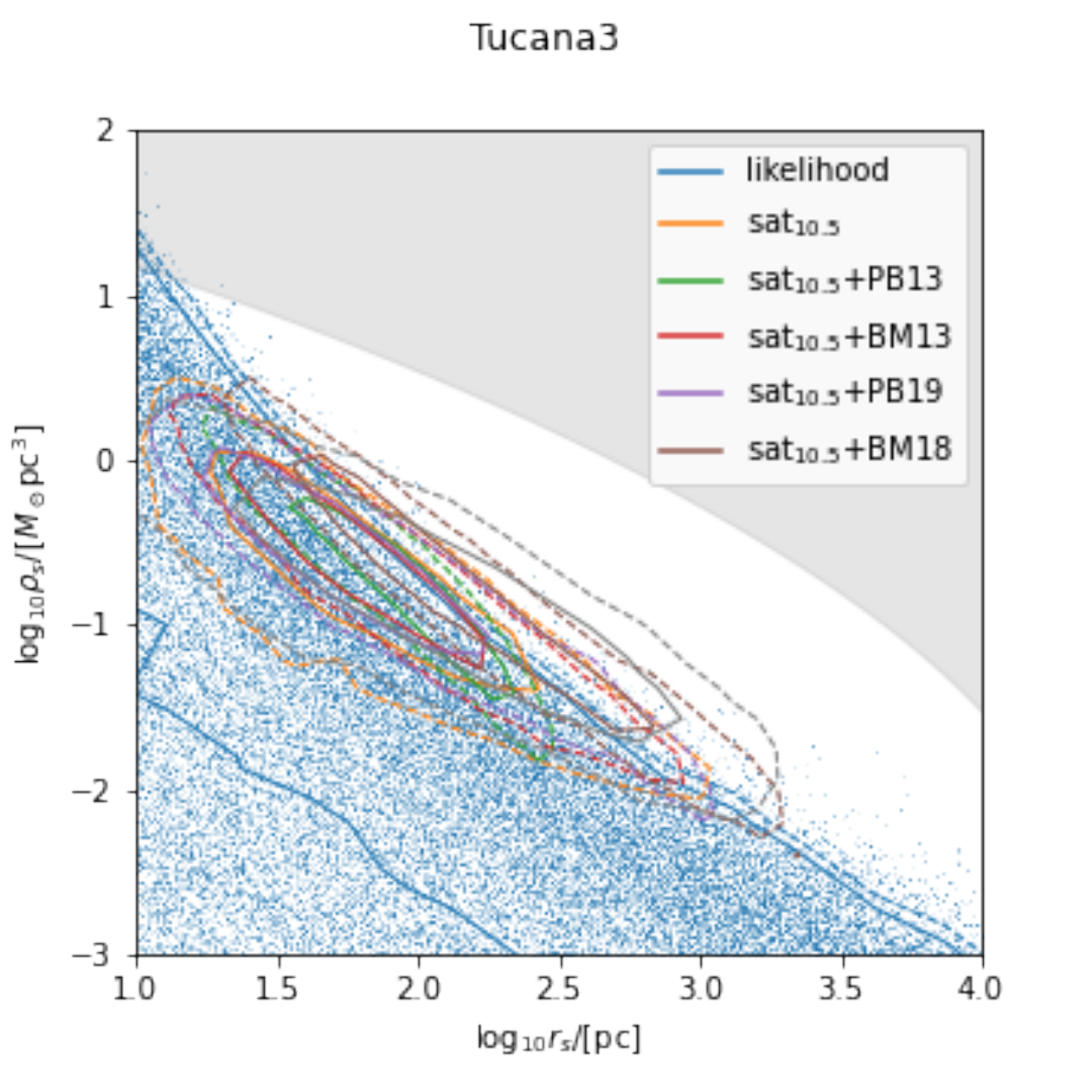}
    \includegraphics[width=\posteriorfigwidth]{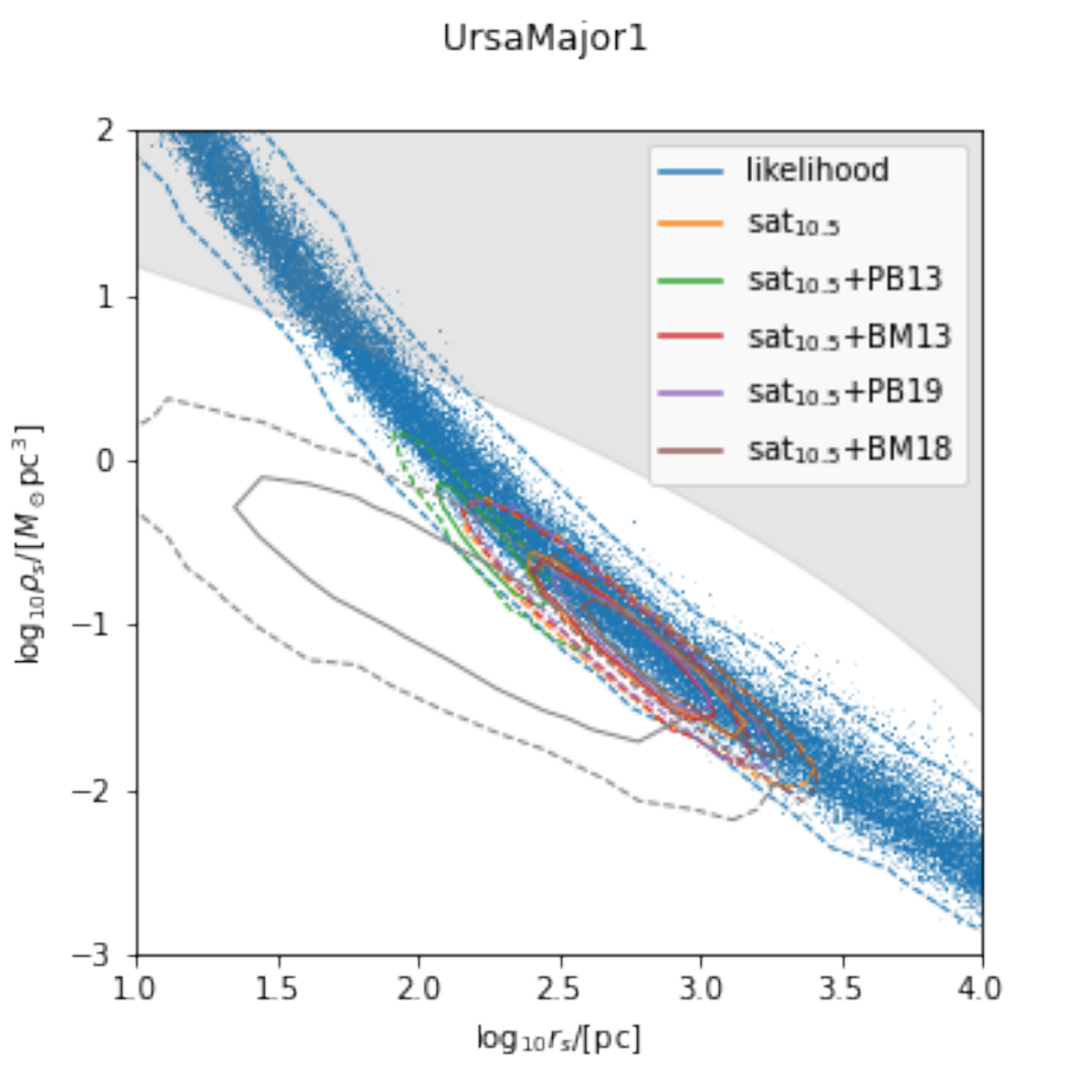}
    \includegraphics[width=\posteriorfigwidth]{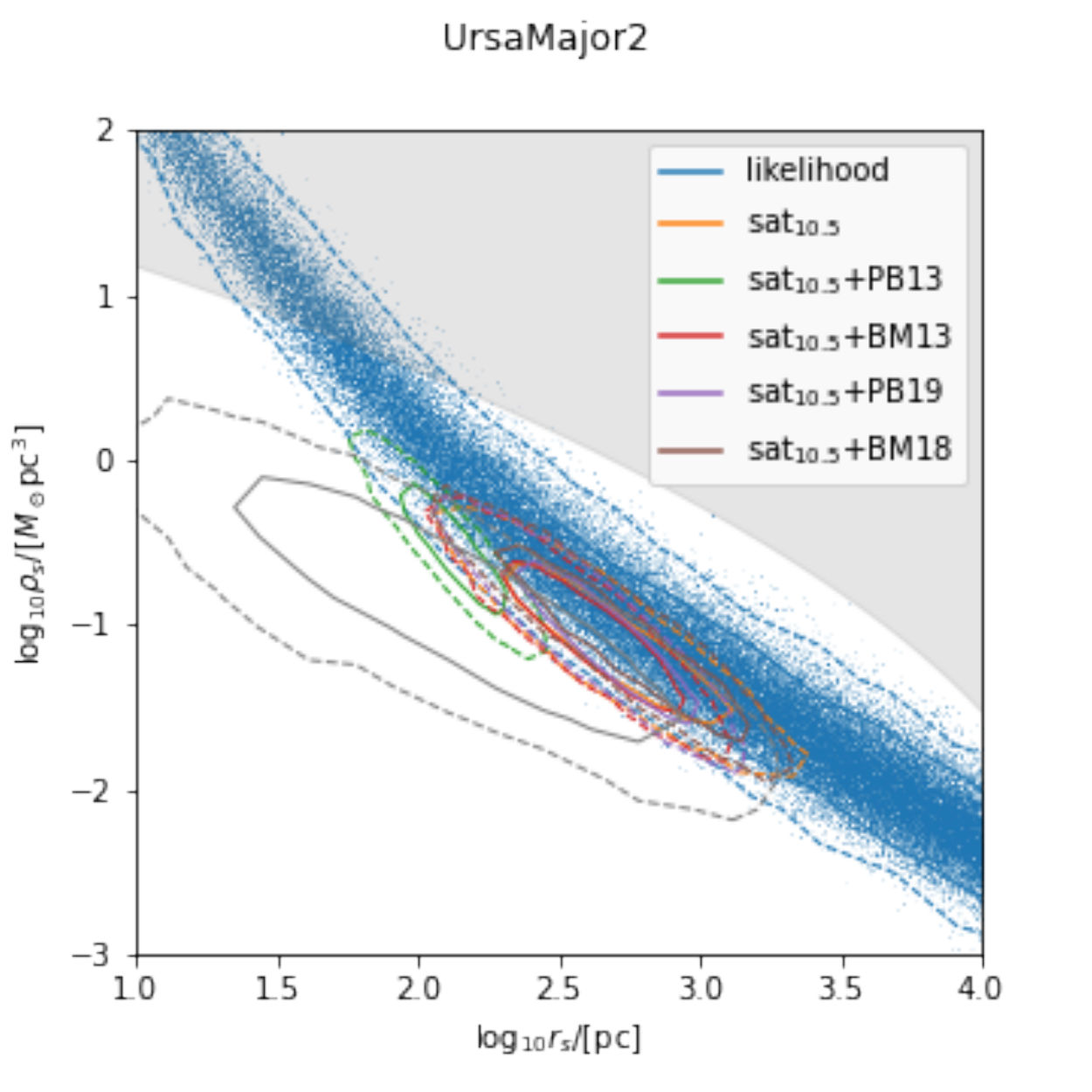}
    \includegraphics[width=\posteriorfigwidth]{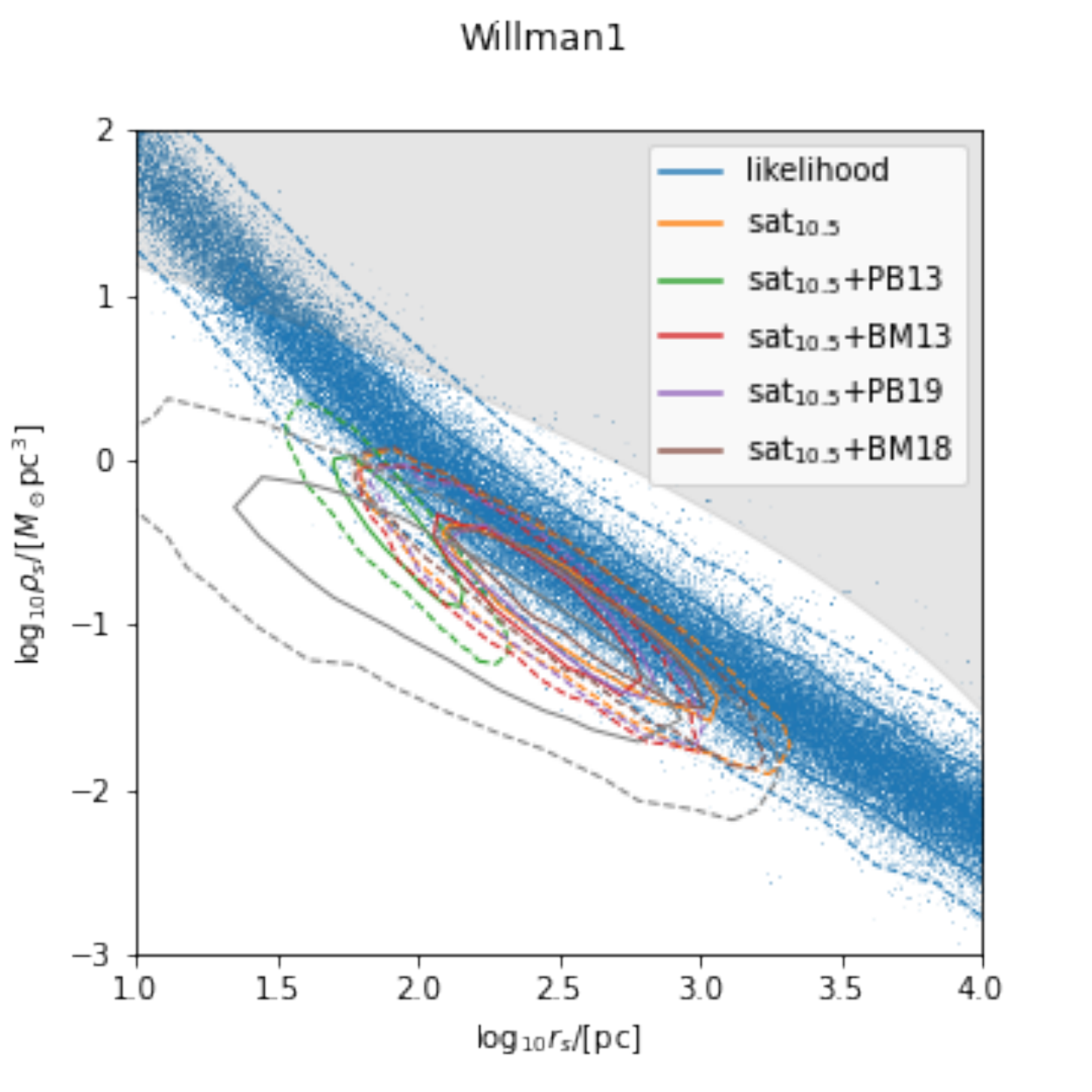}
    \caption{\tableandfigurefontsize}
\end{figure}

\begin{figure}
    \centering
    \tableandfigurefontsize
    \includegraphics[width=\posteriorfigwidth]{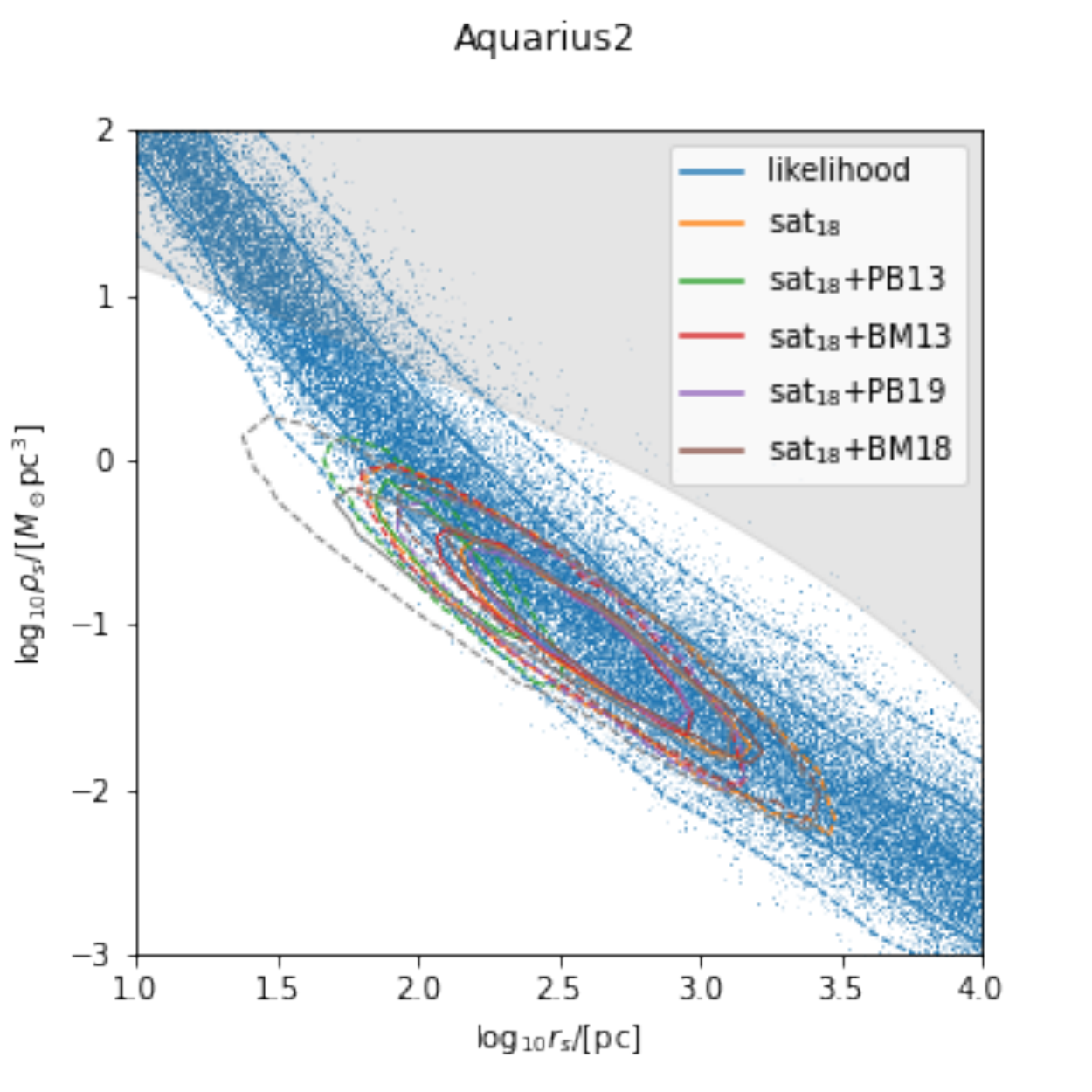}
    \includegraphics[width=\posteriorfigwidth]{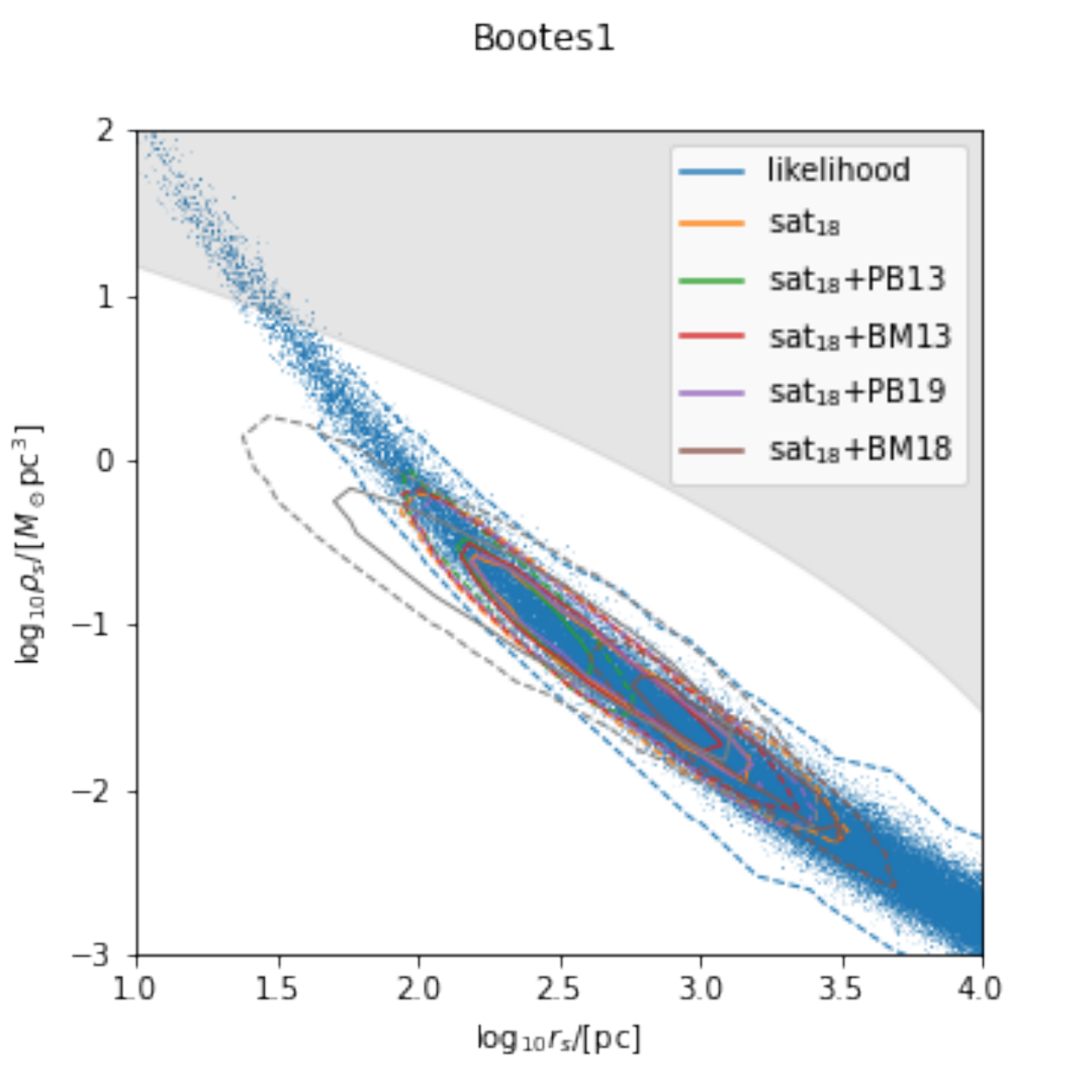}
    \includegraphics[width=\posteriorfigwidth]{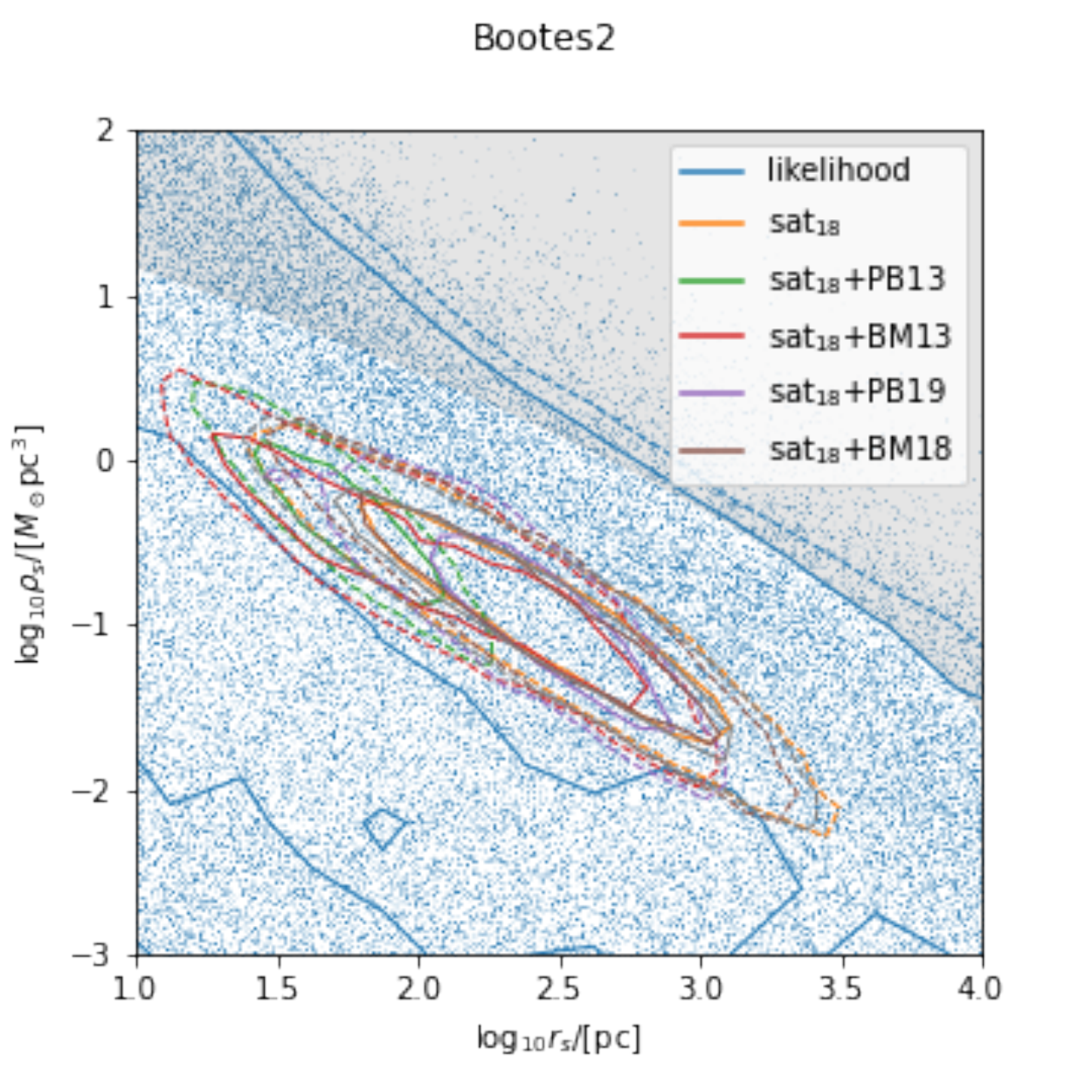}
    \includegraphics[width=\posteriorfigwidth]{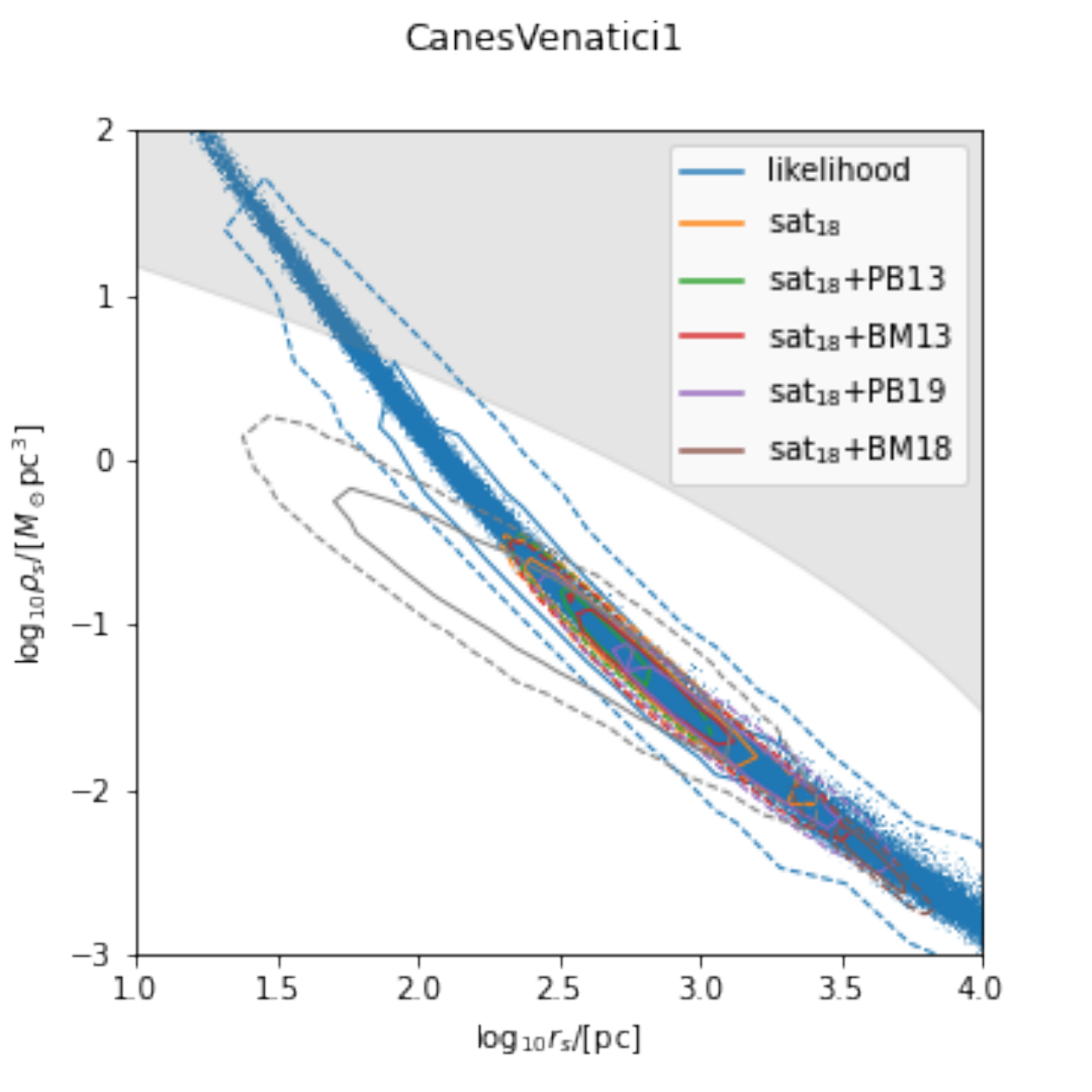}
    \includegraphics[width=\posteriorfigwidth]{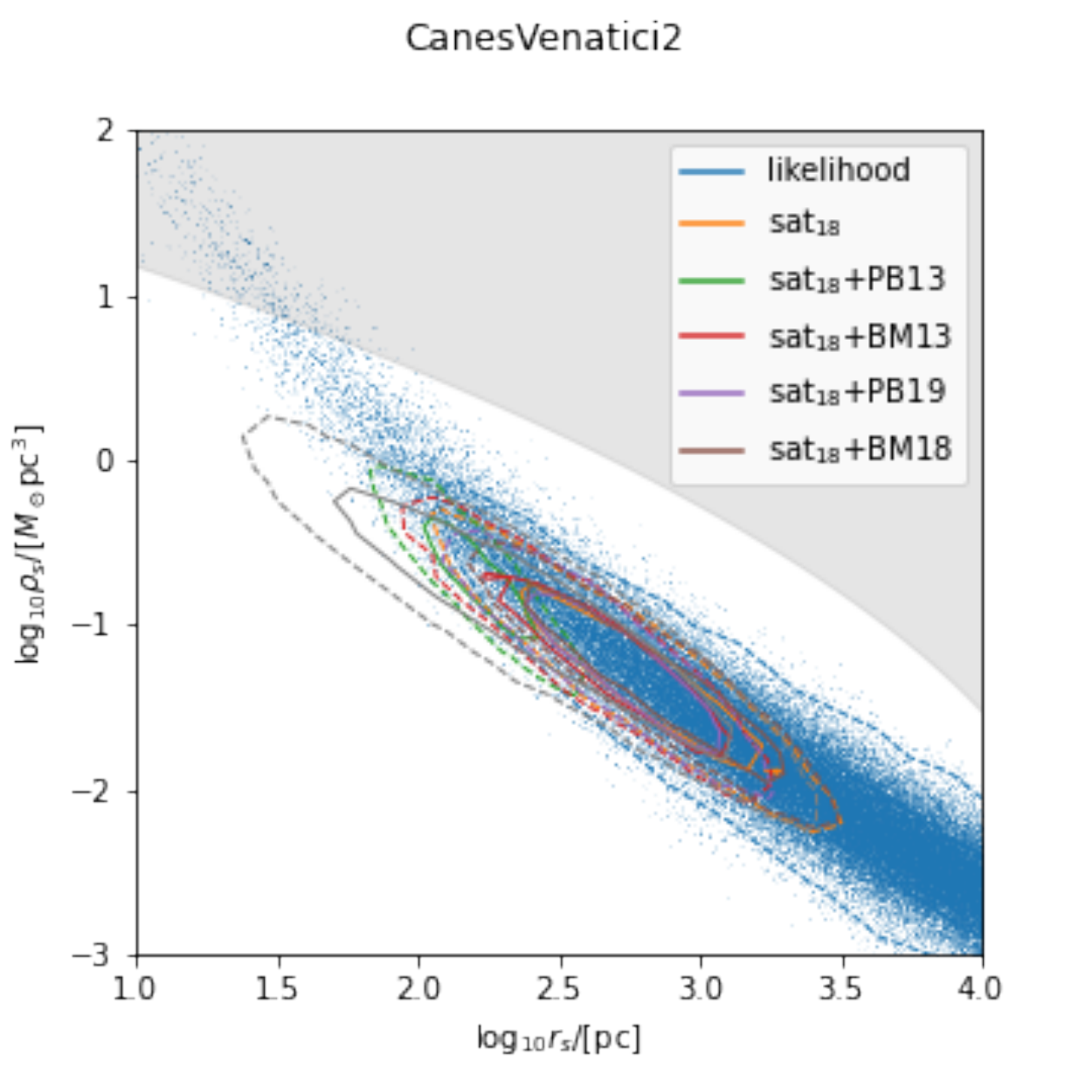}
    \includegraphics[width=\posteriorfigwidth]{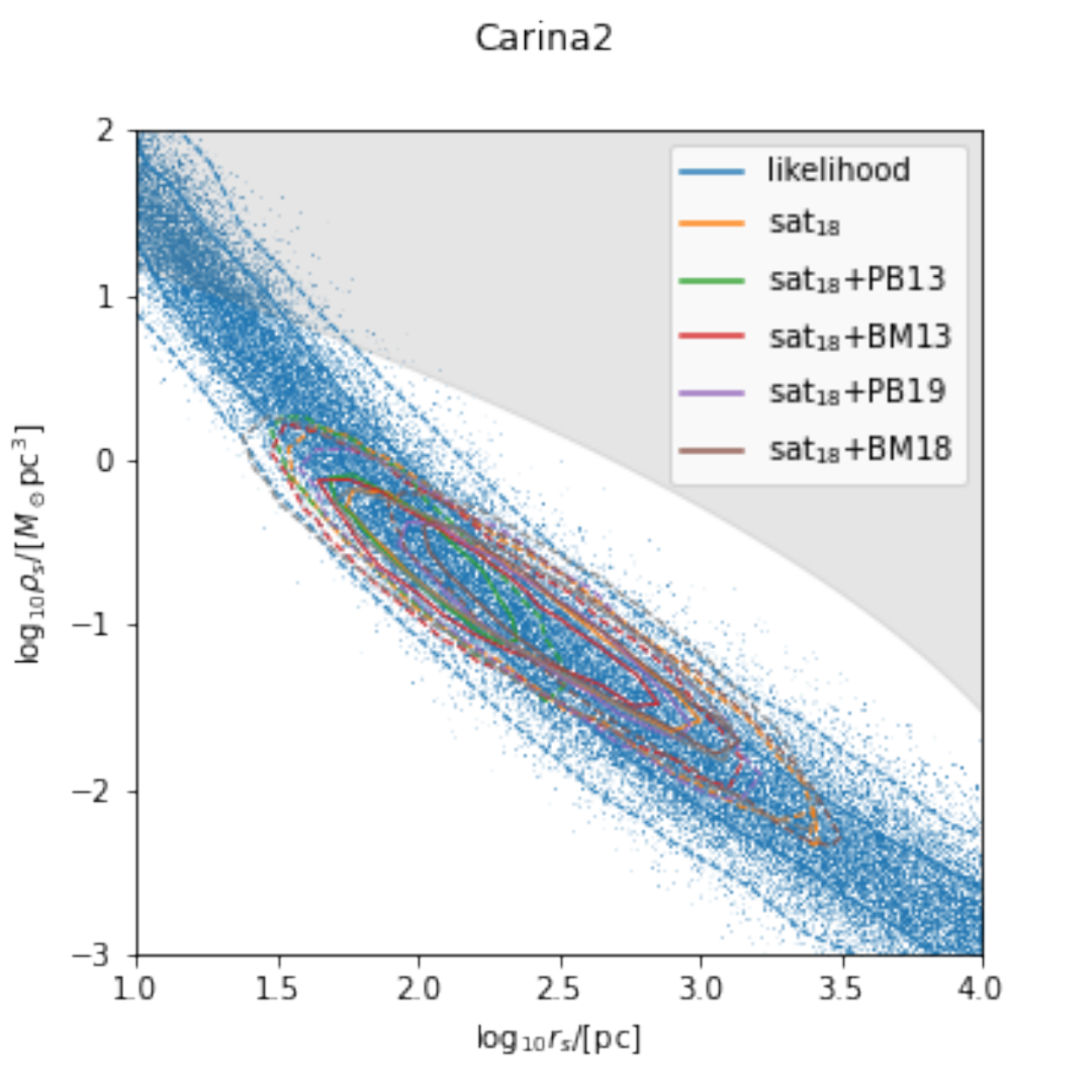}
    \includegraphics[width=\posteriorfigwidth]{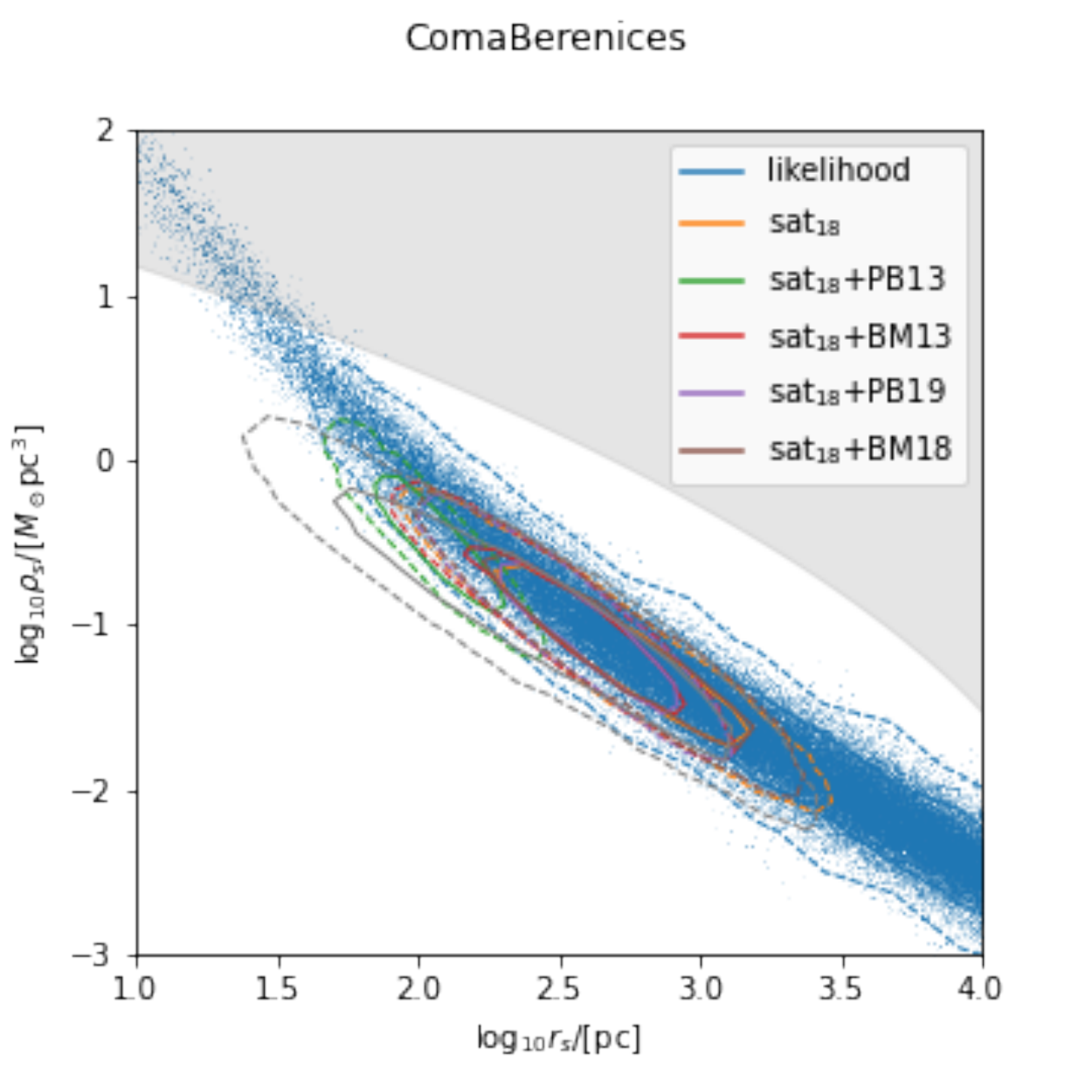}
    \includegraphics[width=\posteriorfigwidth]{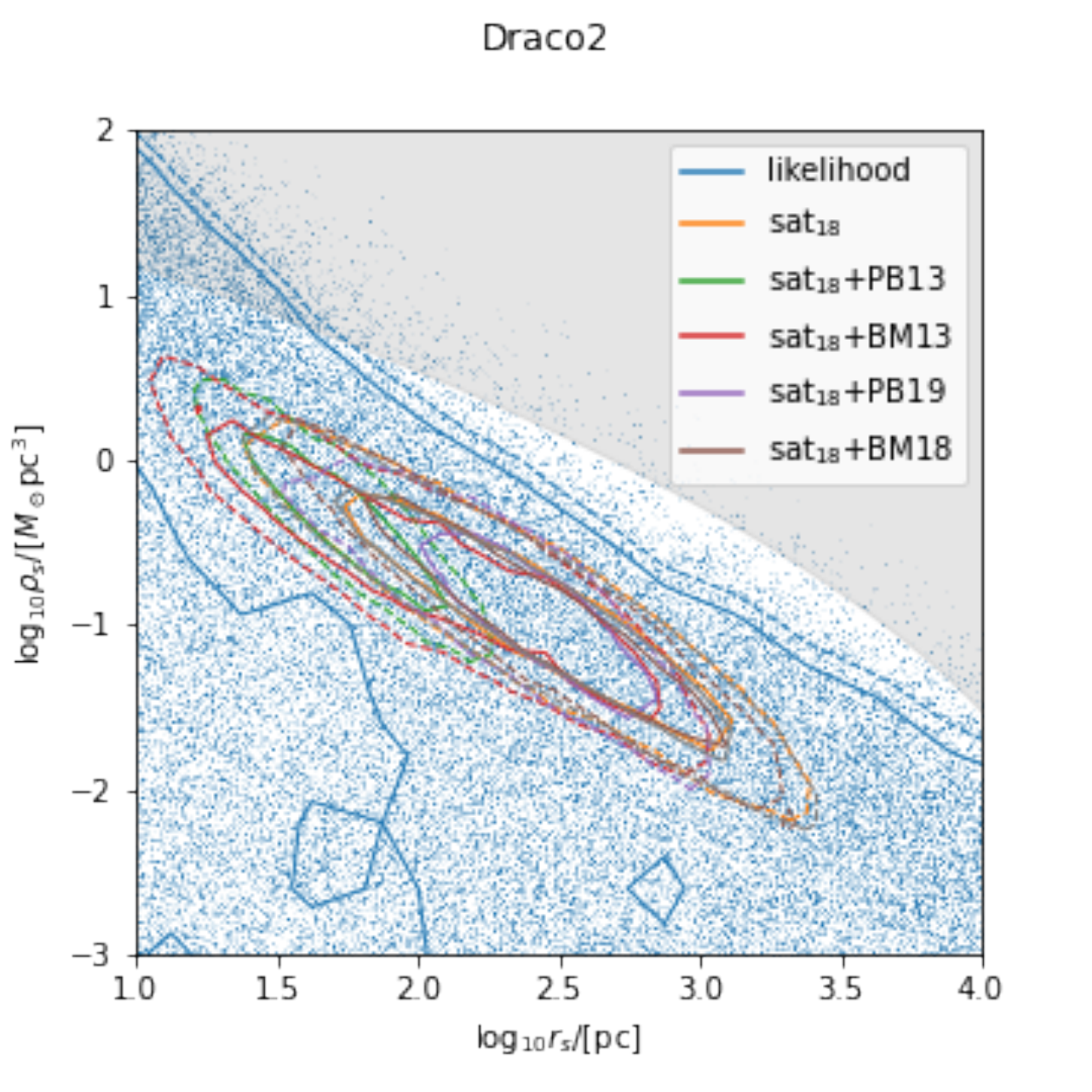}
    \includegraphics[width=\posteriorfigwidth]{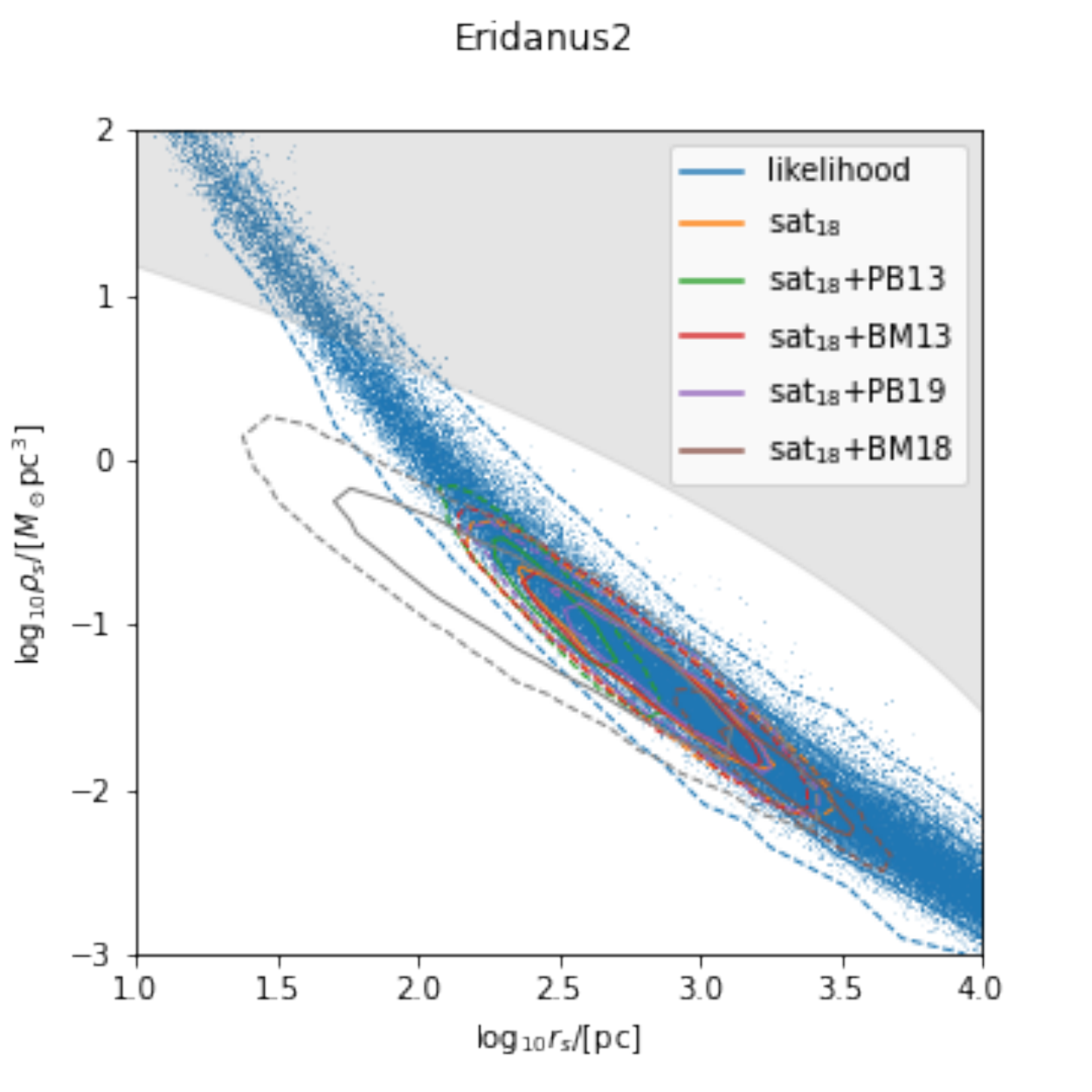}
    \includegraphics[width=\posteriorfigwidth]{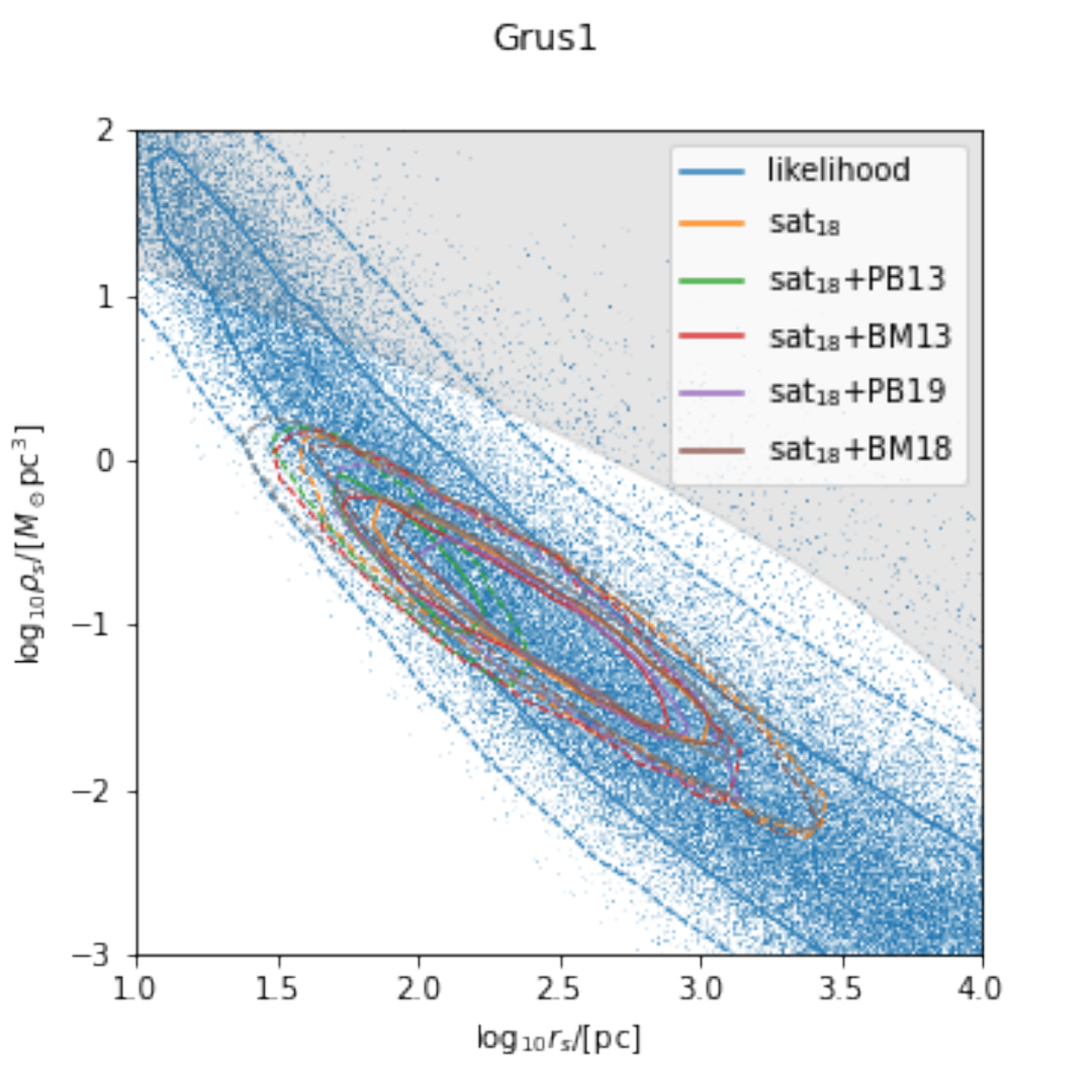}
    \includegraphics[width=\posteriorfigwidth]{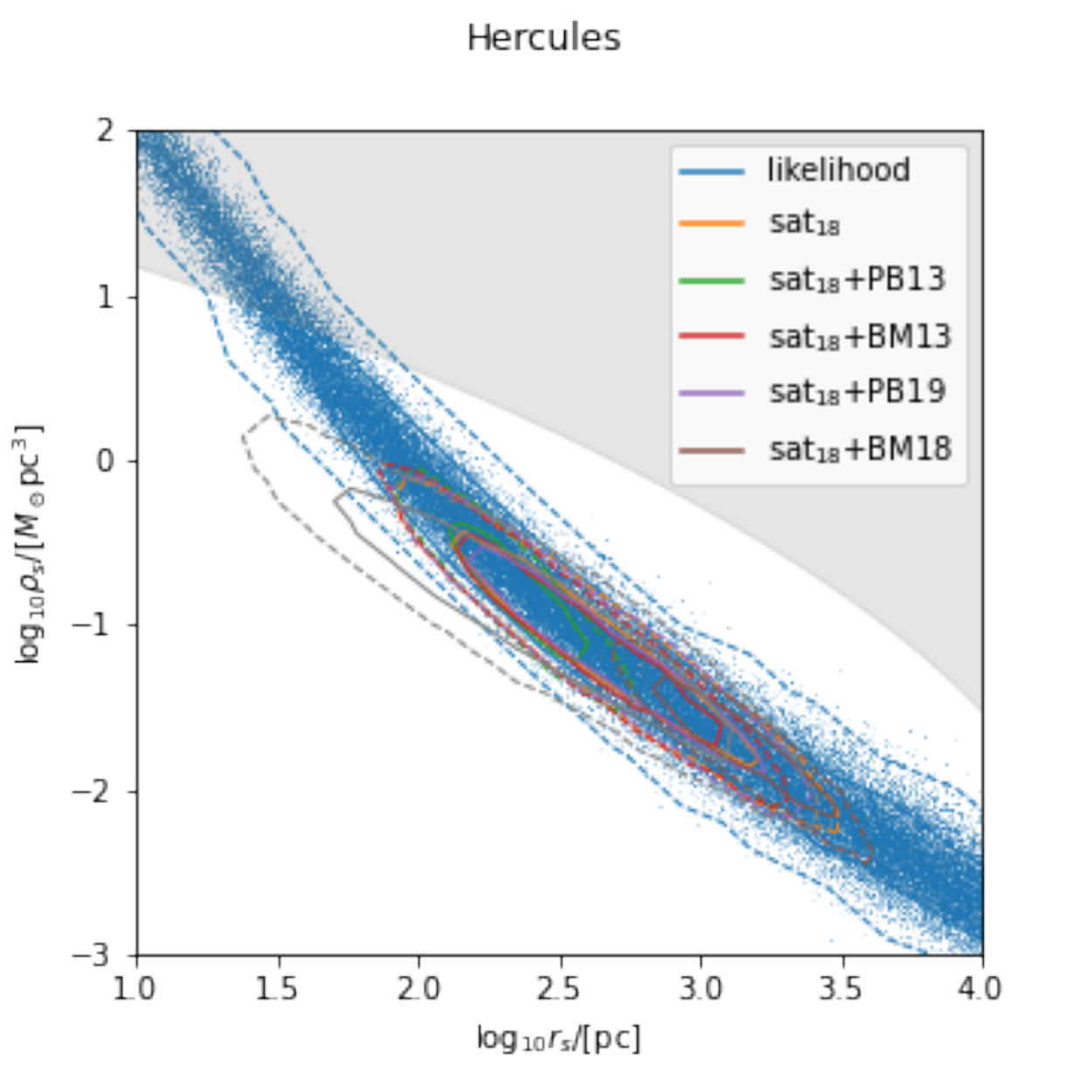}
    \includegraphics[width=\posteriorfigwidth]{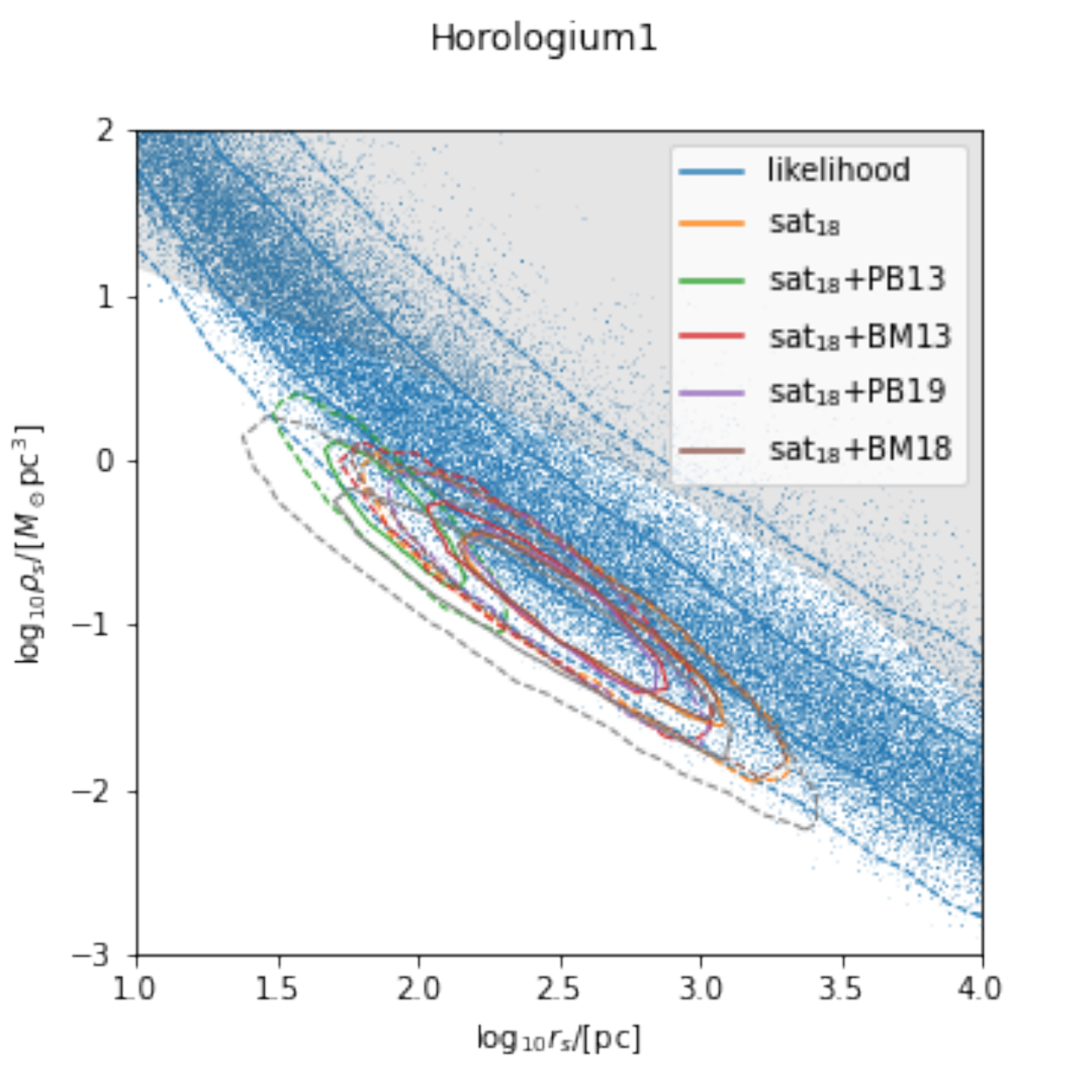}
    \caption{\tableandfigurefontsize
    Same figure as \cref{fig:posteriors_v50_105} but using the satellite prior $\mathrm{sat.}_{18}$.}
    \label{fig:posteriors_v50_180}
\end{figure}

\begin{figure}
    \ContinuedFloat
    \captionsetup{list=off,format=cont}
    \centering
    \tableandfigurefontsize
    \includegraphics[width=\posteriorfigwidth]{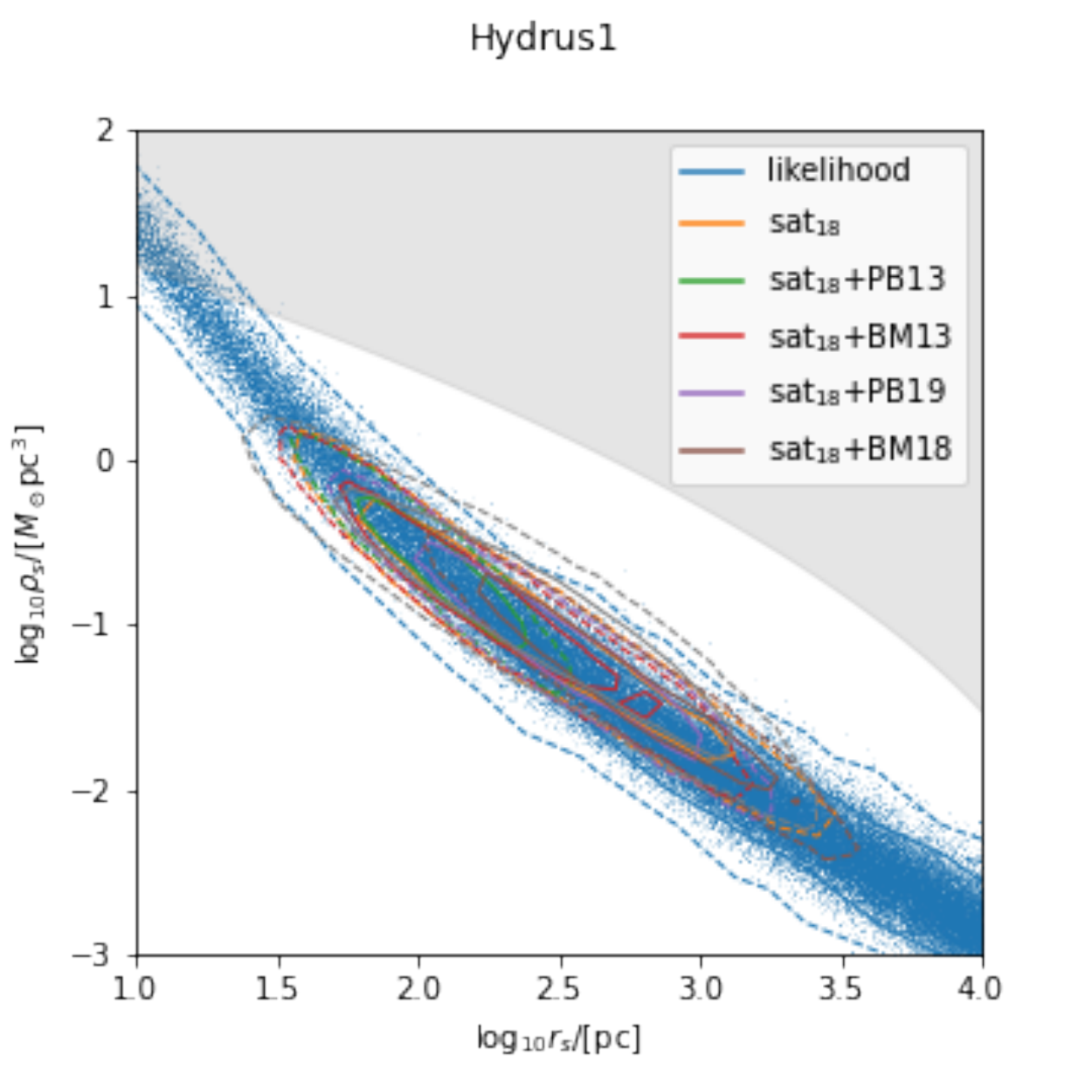}
    \includegraphics[width=\posteriorfigwidth]{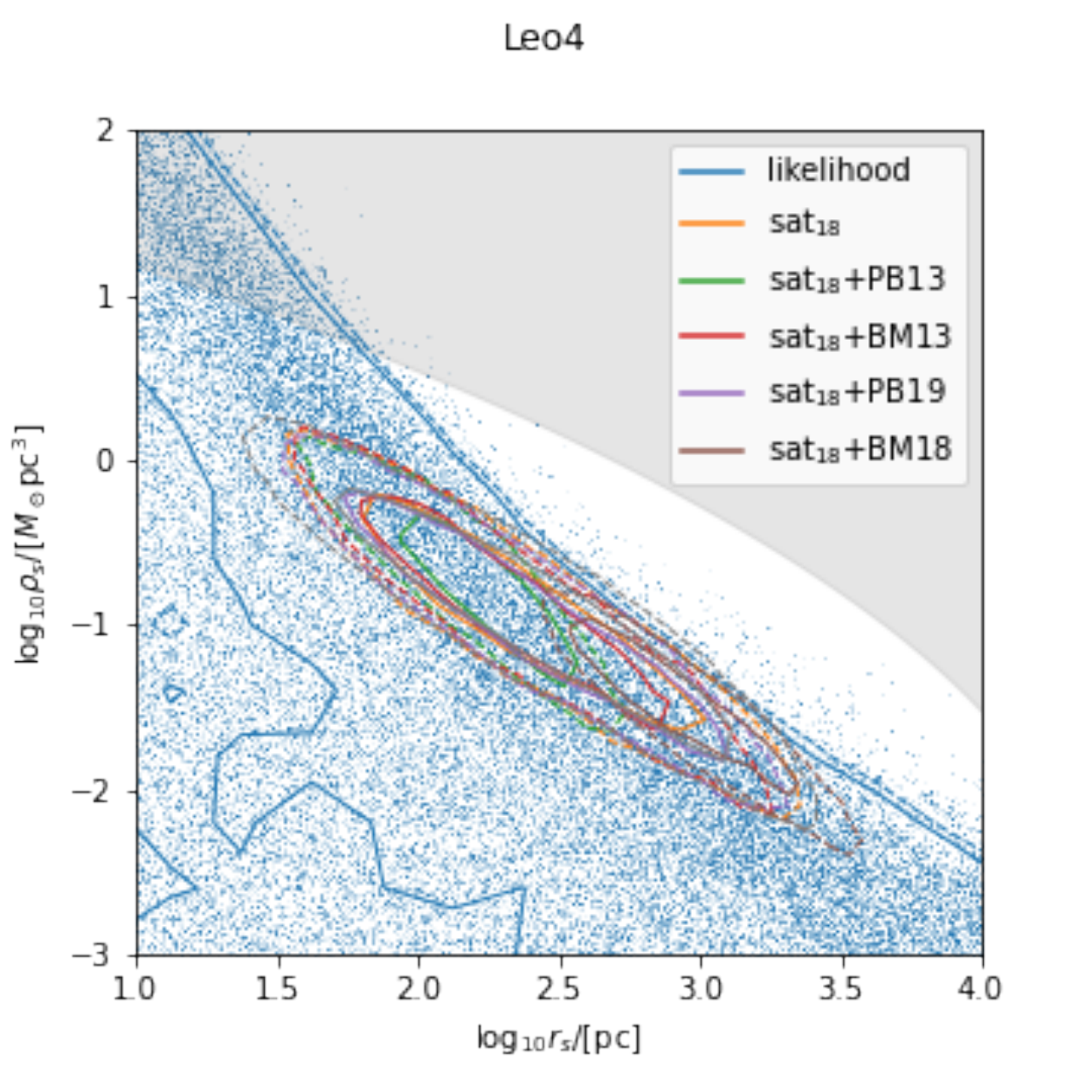}
    \includegraphics[width=\posteriorfigwidth]{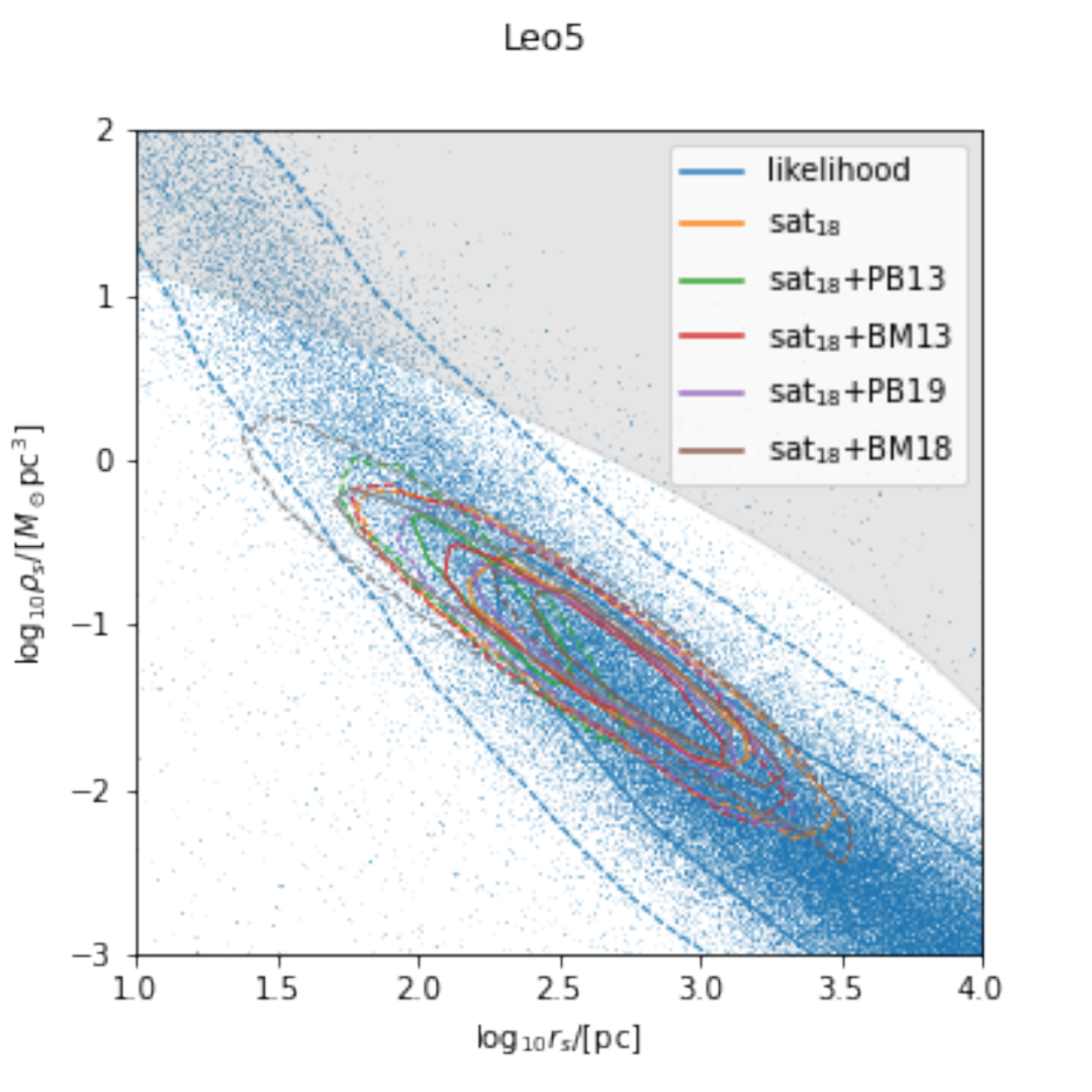}
    \includegraphics[width=\posteriorfigwidth]{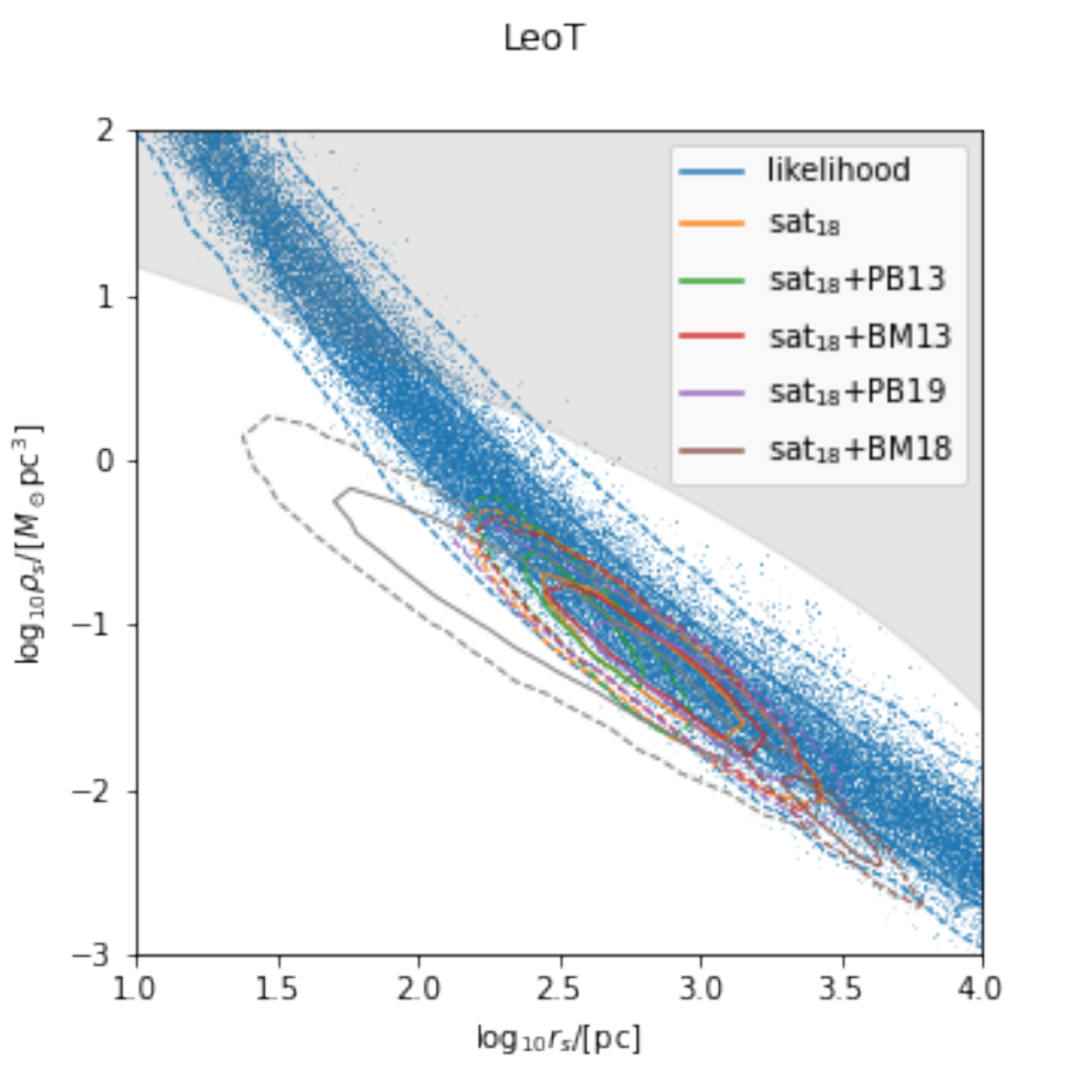}
    \includegraphics[width=\posteriorfigwidth]{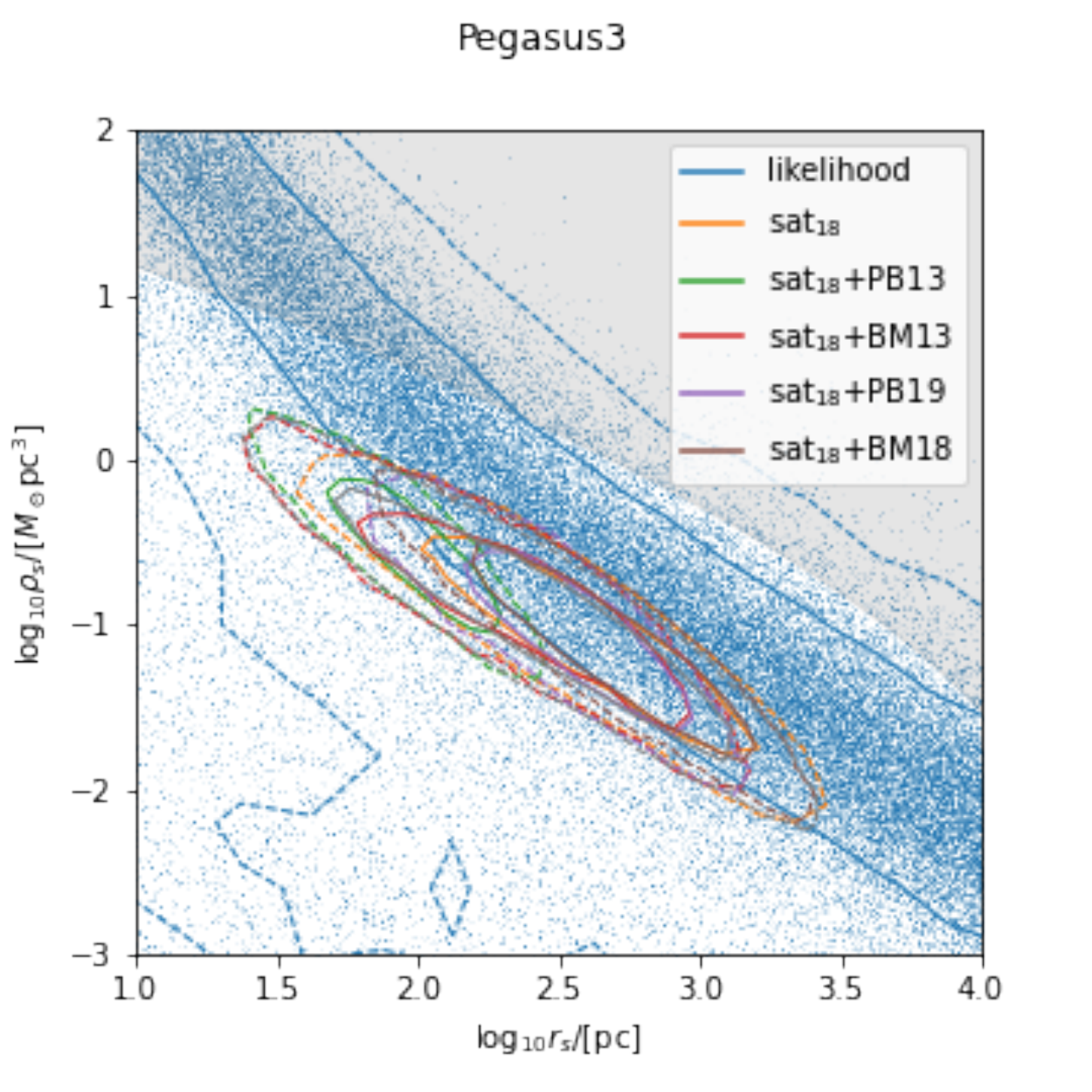}
    \includegraphics[width=\posteriorfigwidth]{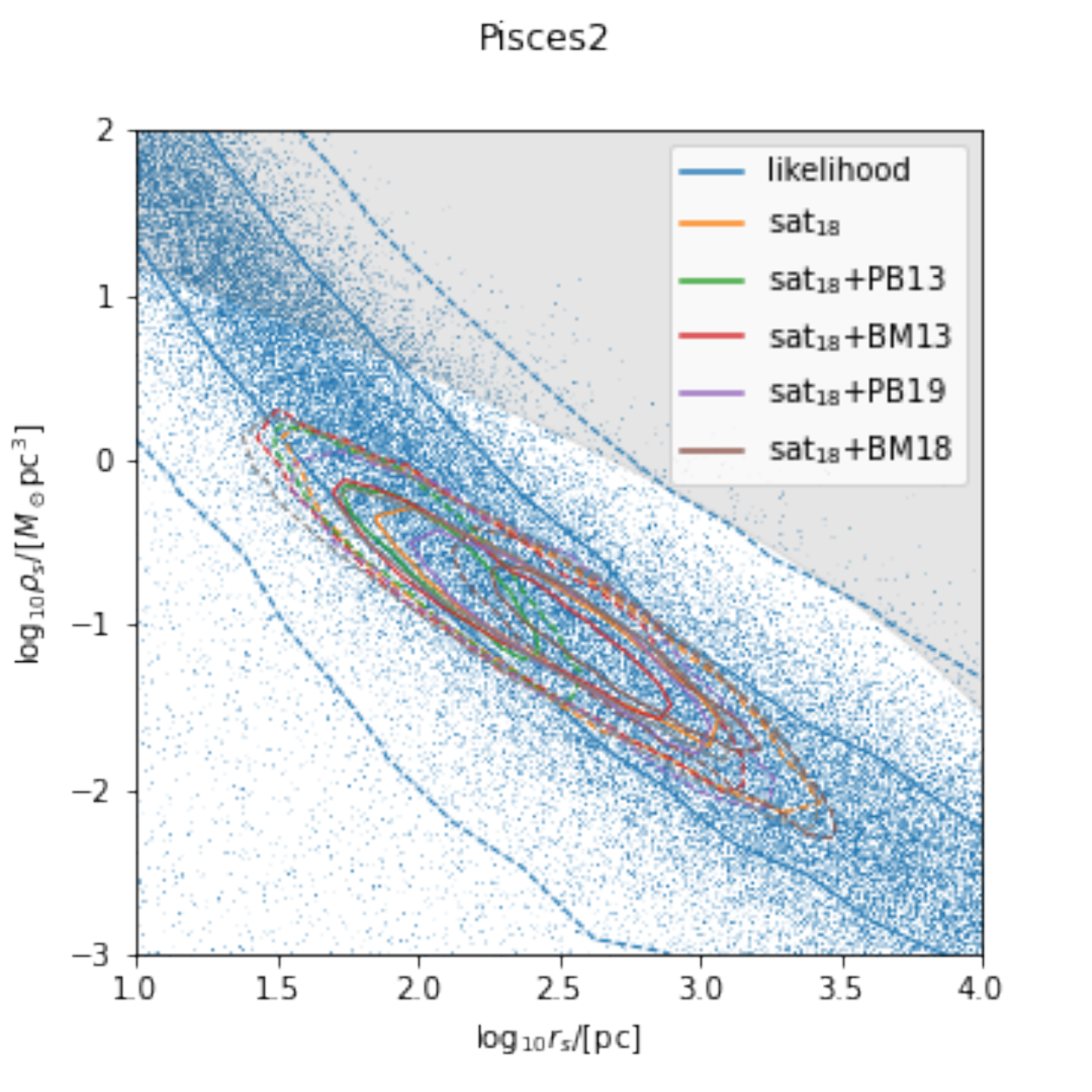}
    \includegraphics[width=\posteriorfigwidth]{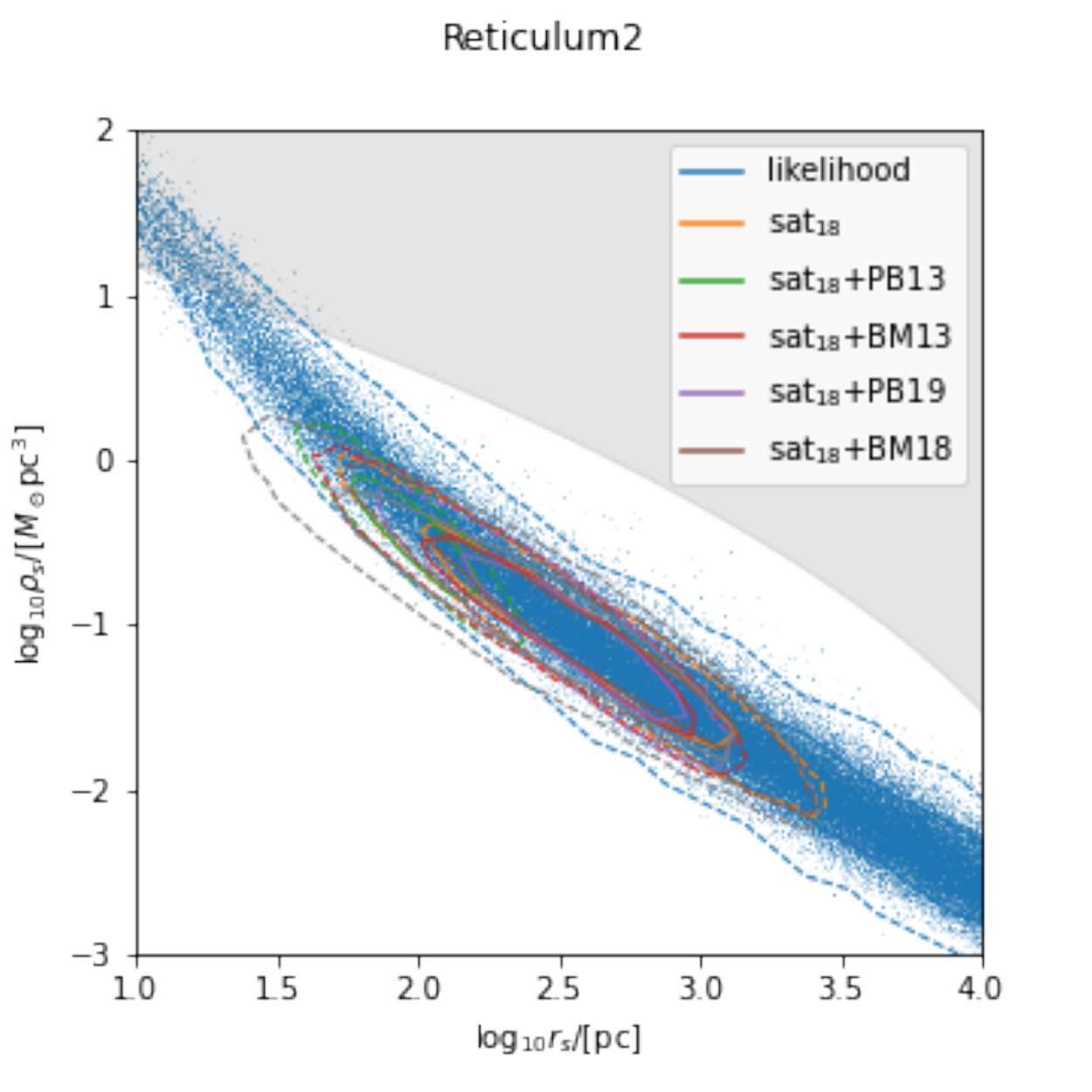}
    \includegraphics[width=\posteriorfigwidth]{fig/Segue1_v50_180.pdf}
    \includegraphics[width=\posteriorfigwidth]{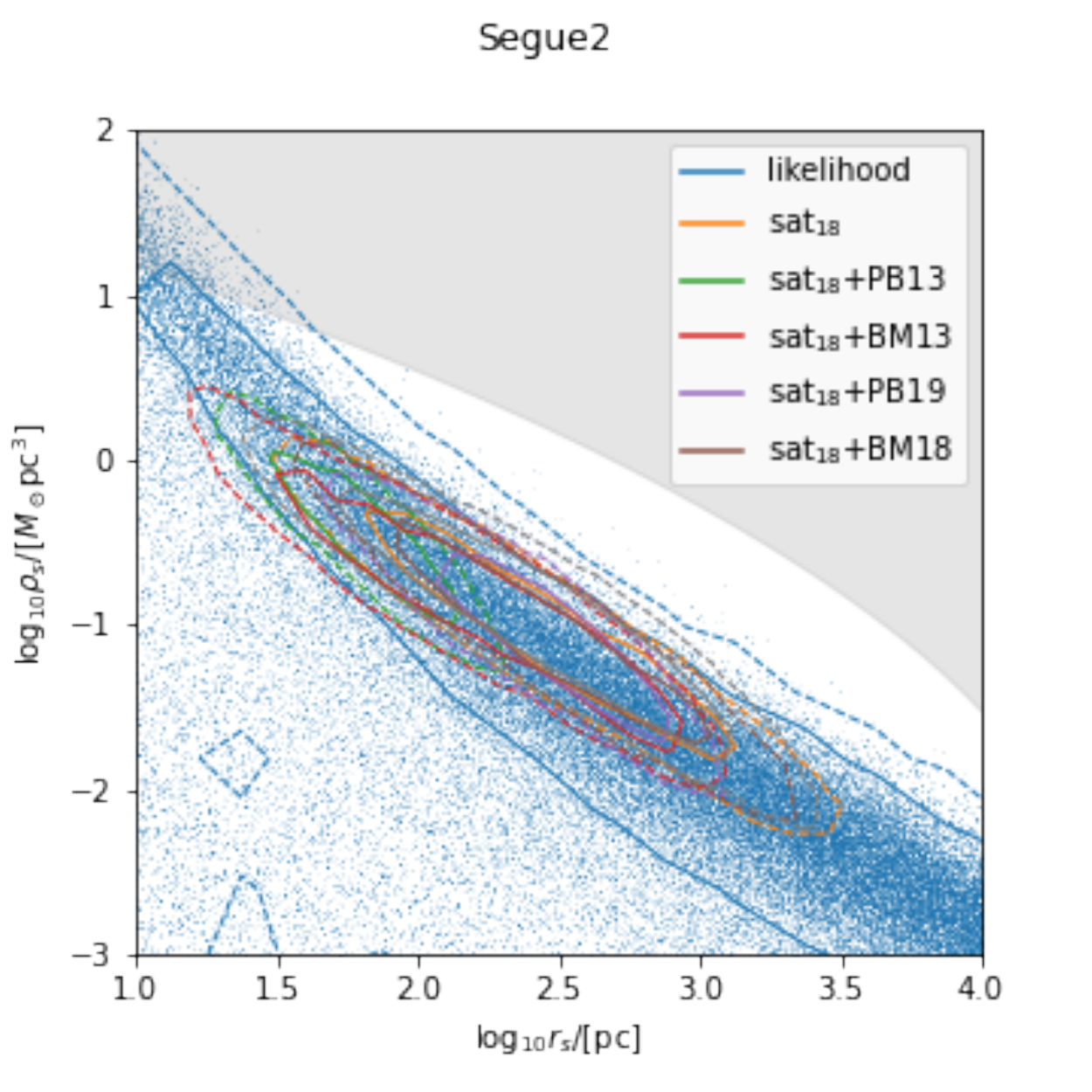}
    \includegraphics[width=\posteriorfigwidth]{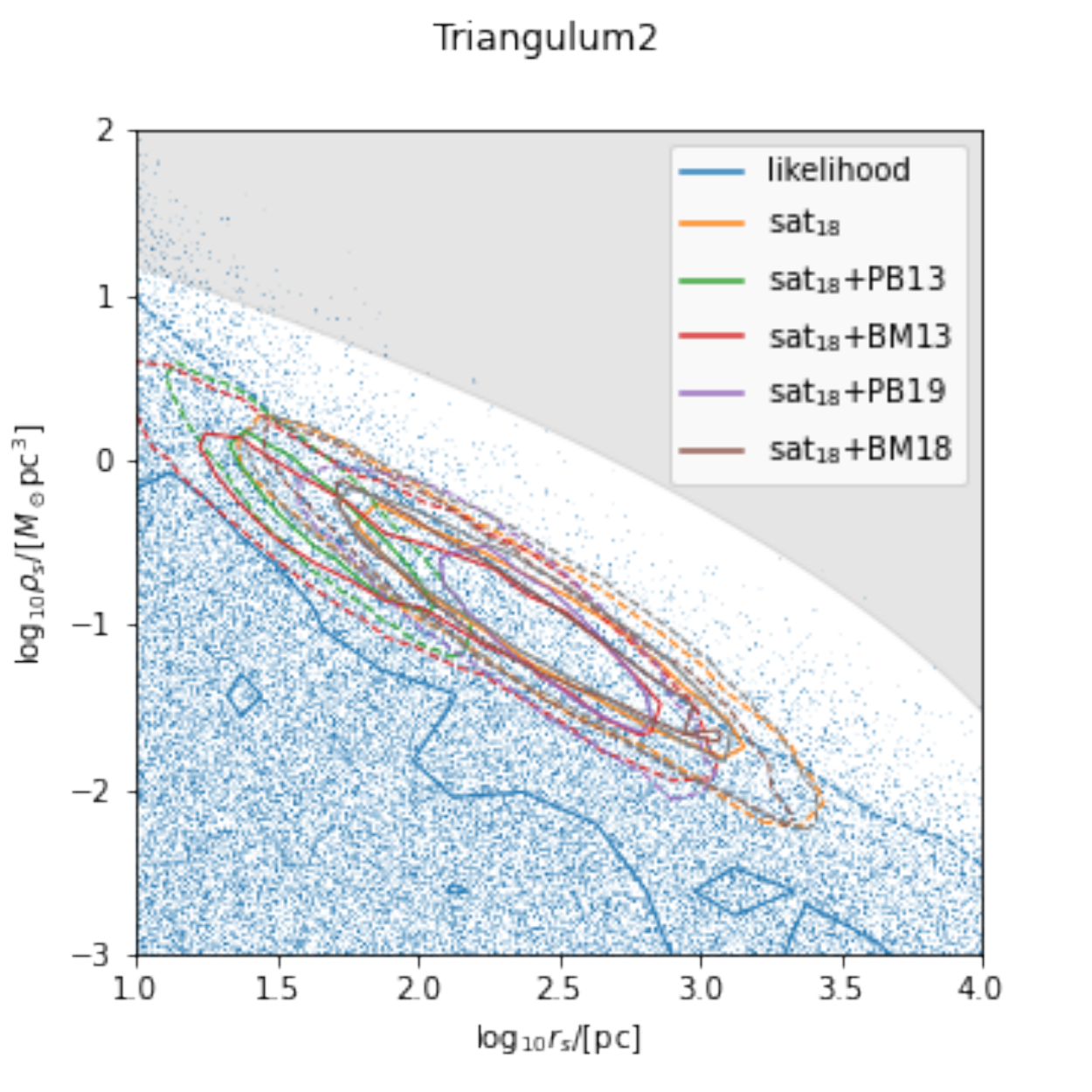}
    \includegraphics[width=\posteriorfigwidth]{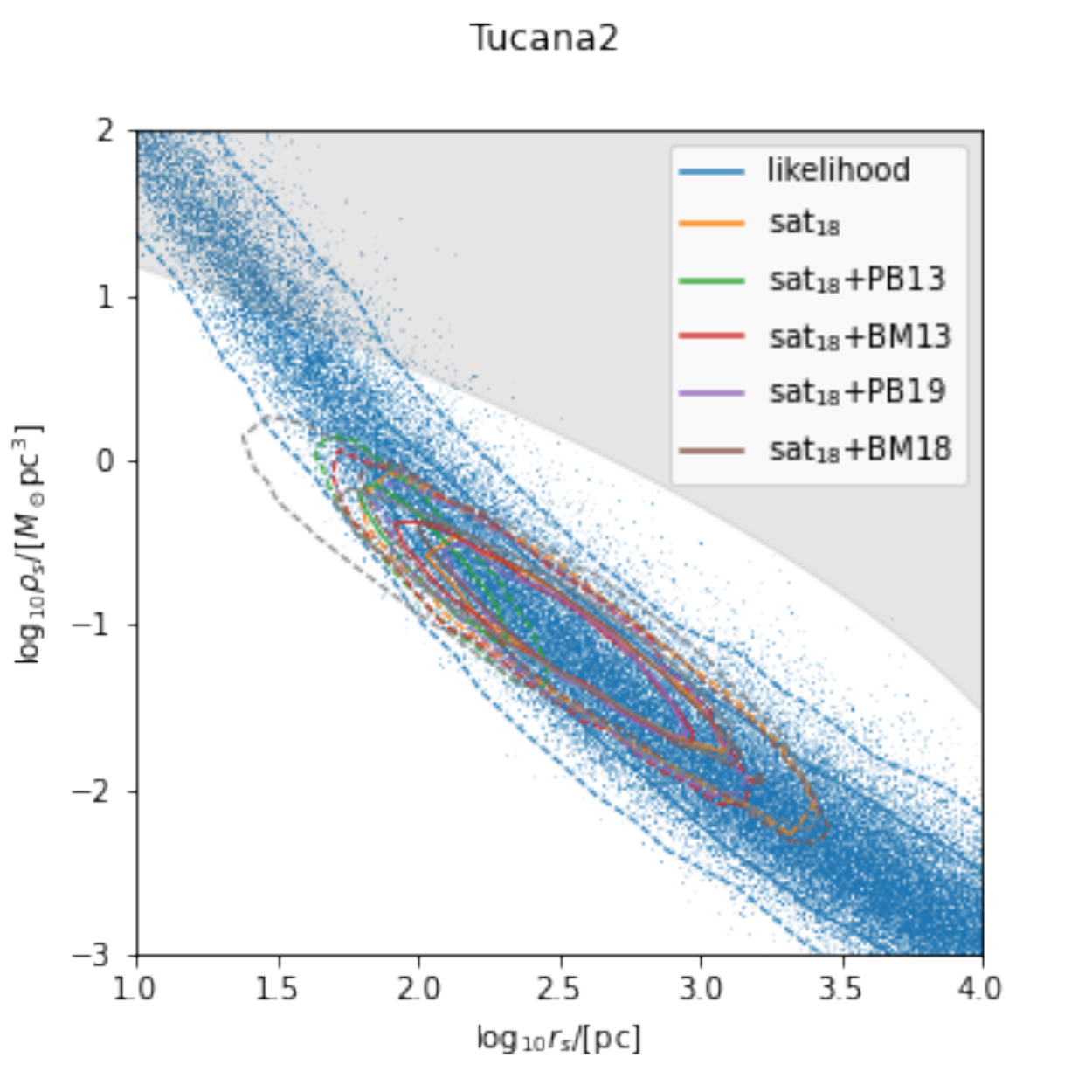}
    \includegraphics[width=\posteriorfigwidth]{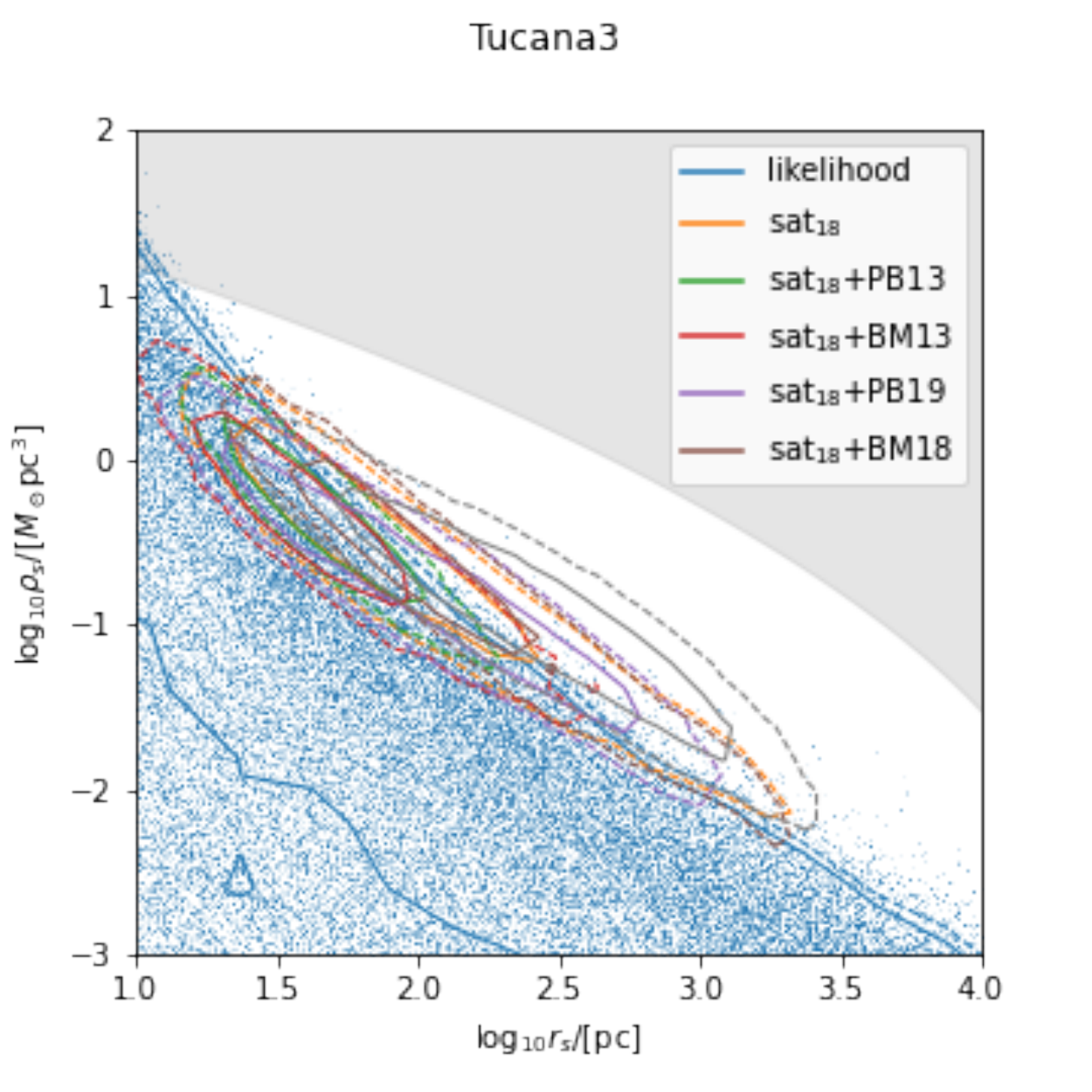}
    \includegraphics[width=\posteriorfigwidth]{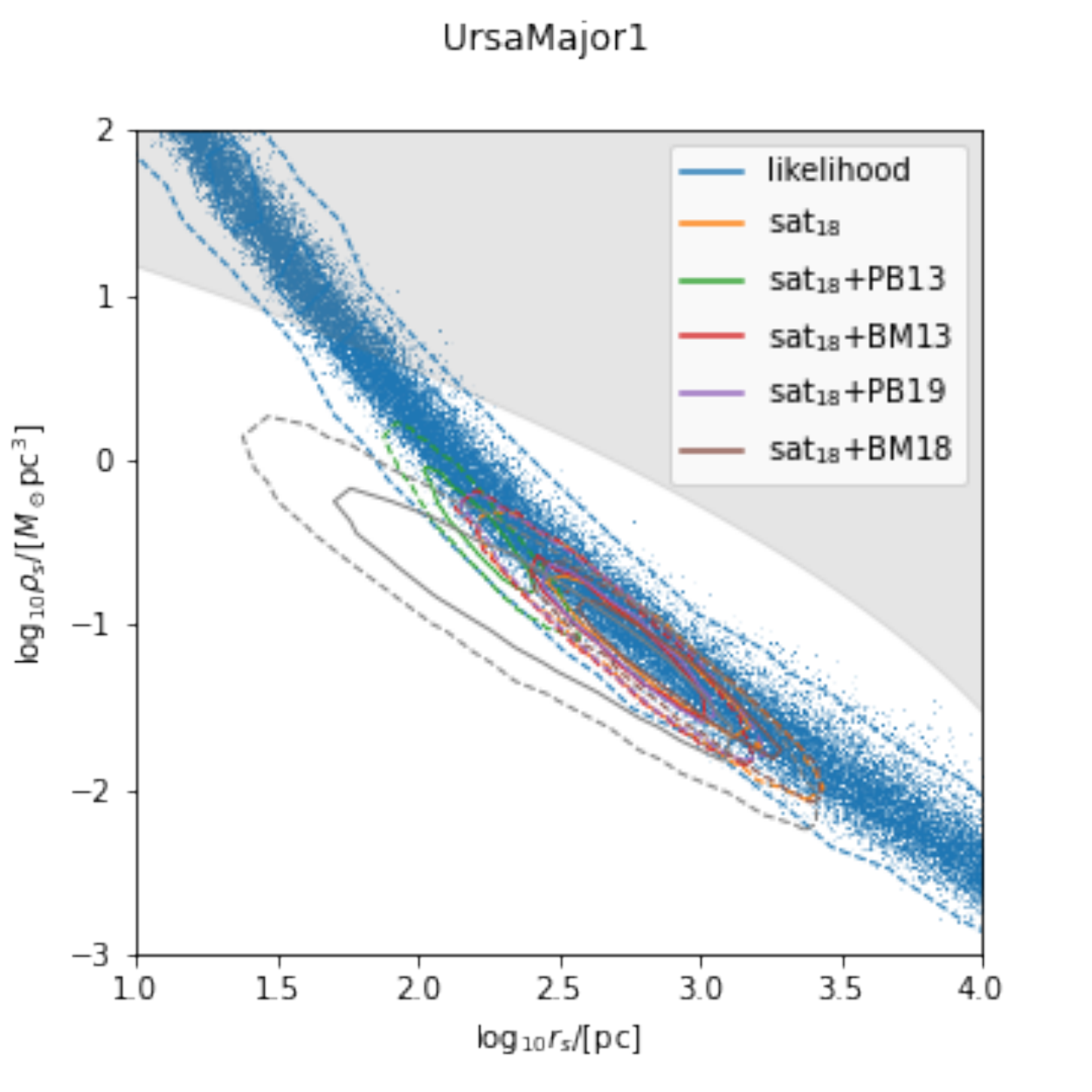}
    \includegraphics[width=\posteriorfigwidth]{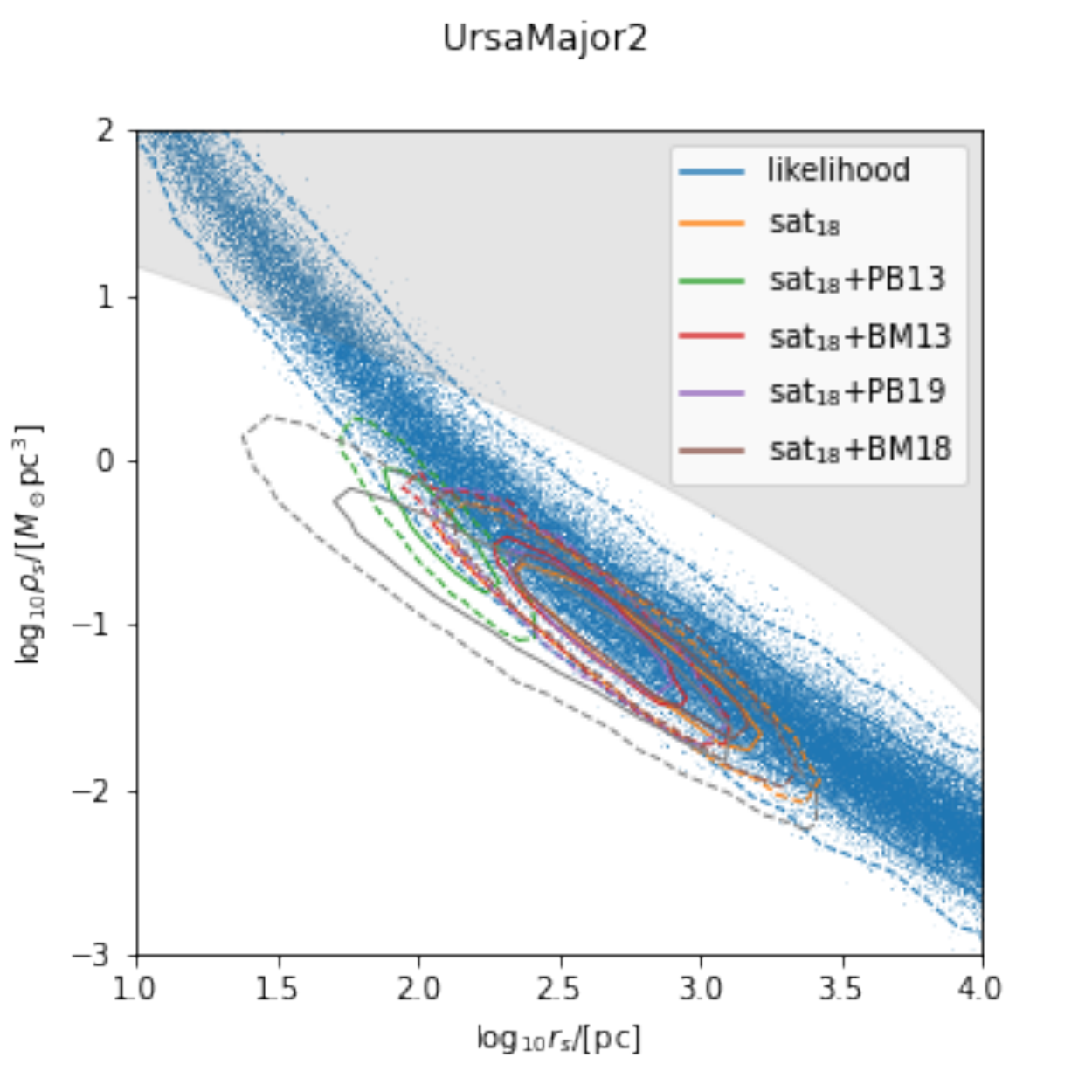}
    \includegraphics[width=\posteriorfigwidth]{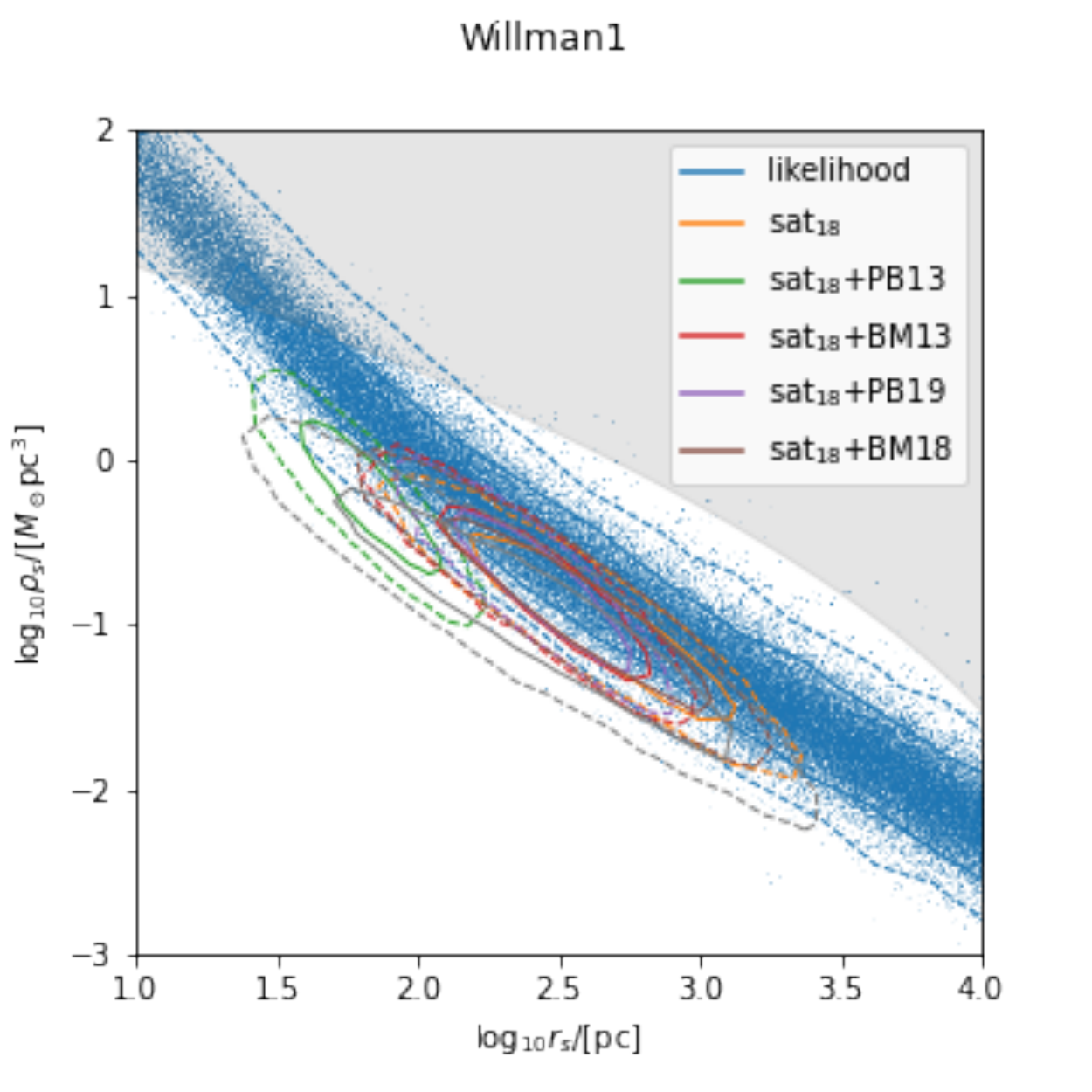}
    \caption{\tableandfigurefontsize}
\end{figure}

\begin{figure}
    \centering
    \tableandfigurefontsize
    \renewcommand{\posteriorfigwidth}{6.0cm}
    \includegraphics[width=\posteriorfigwidth]{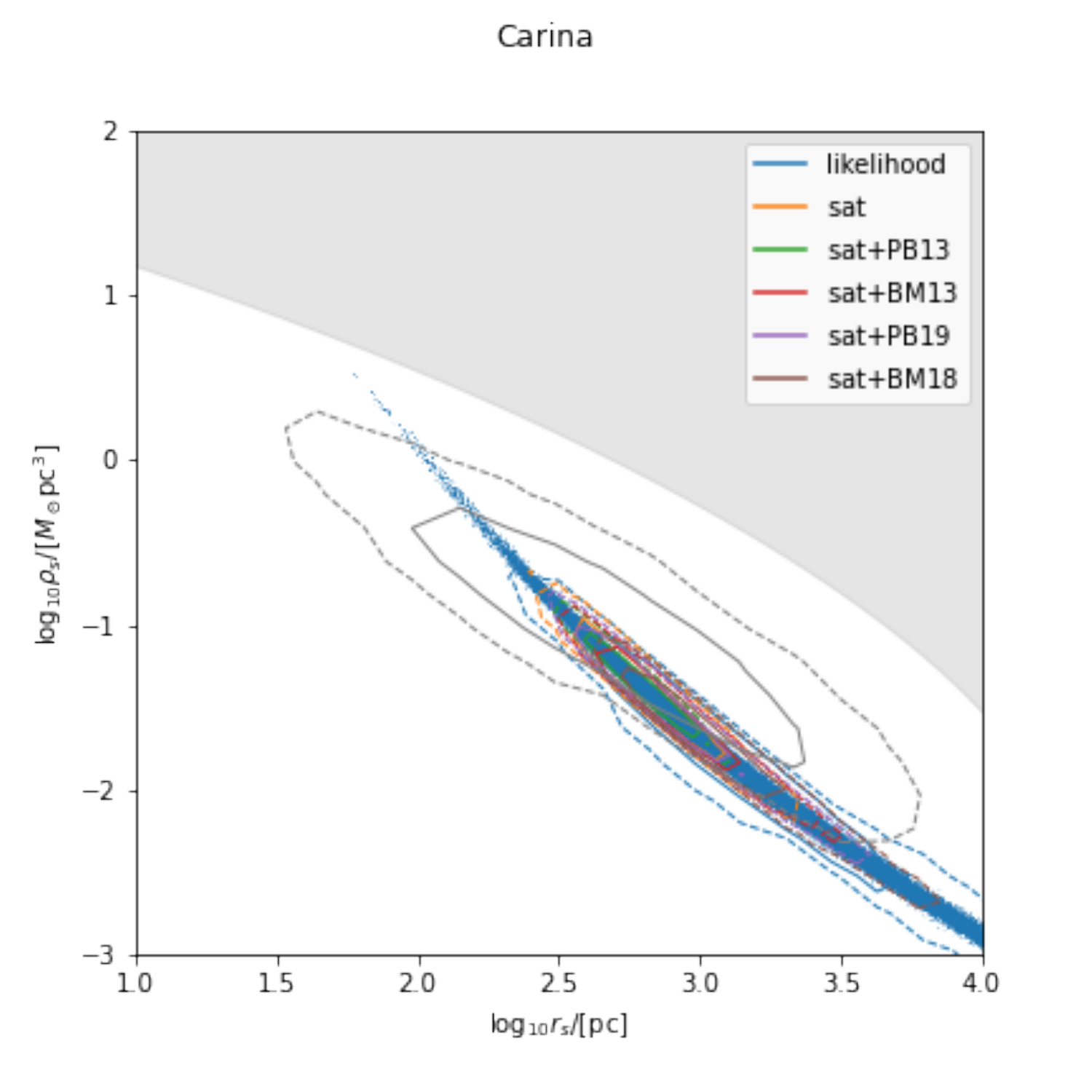}
    \includegraphics[width=\posteriorfigwidth]{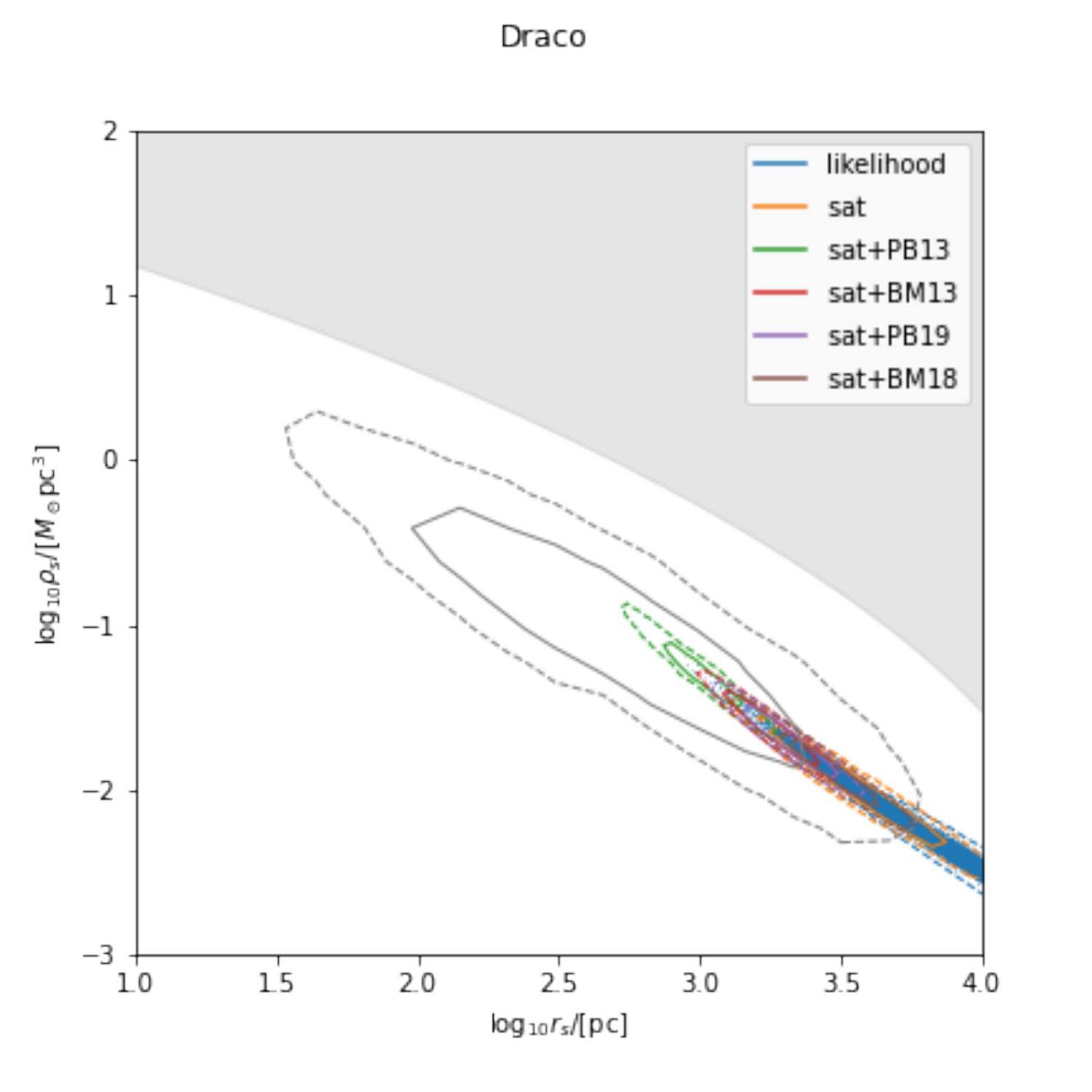}
    \includegraphics[width=\posteriorfigwidth]{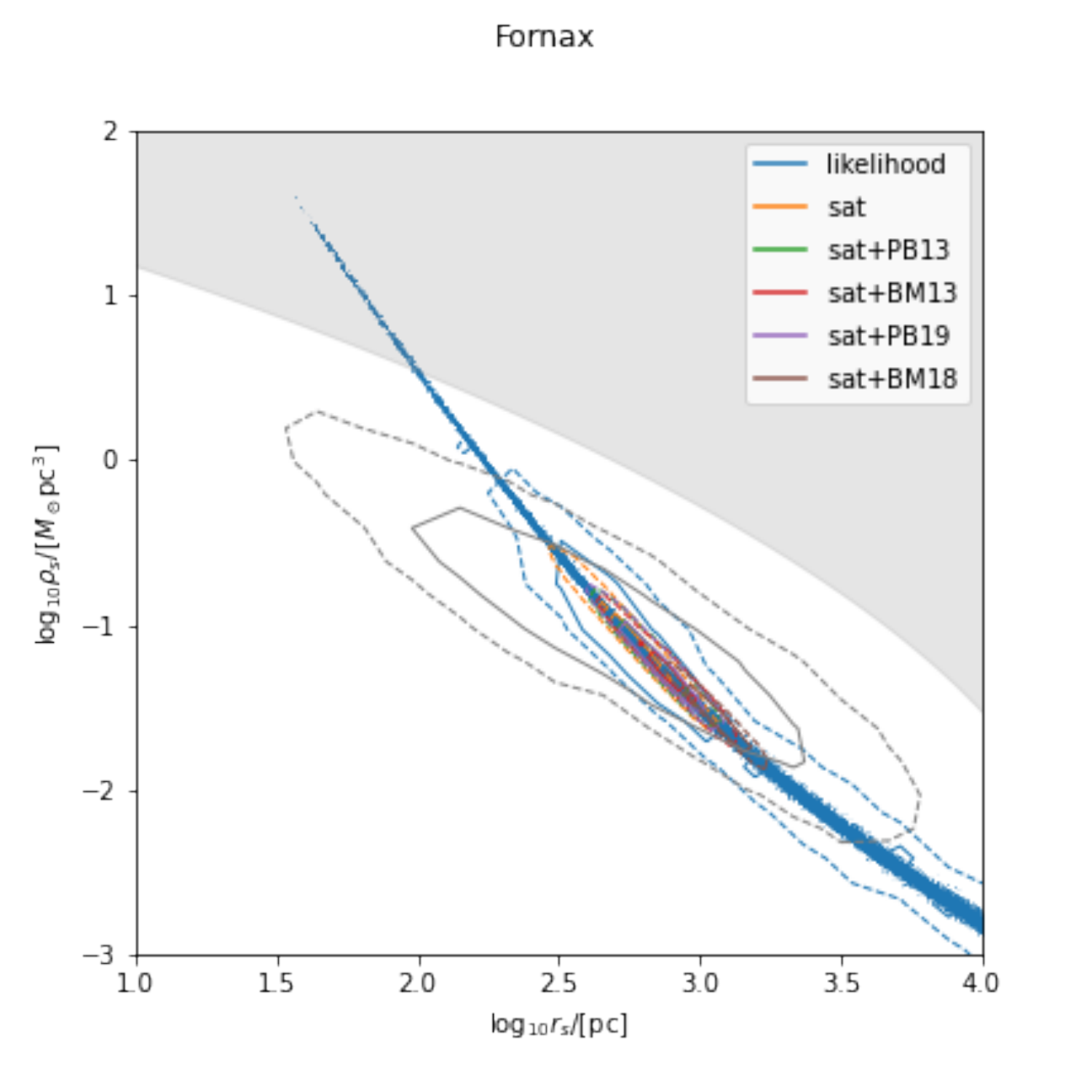}
    \includegraphics[width=\posteriorfigwidth]{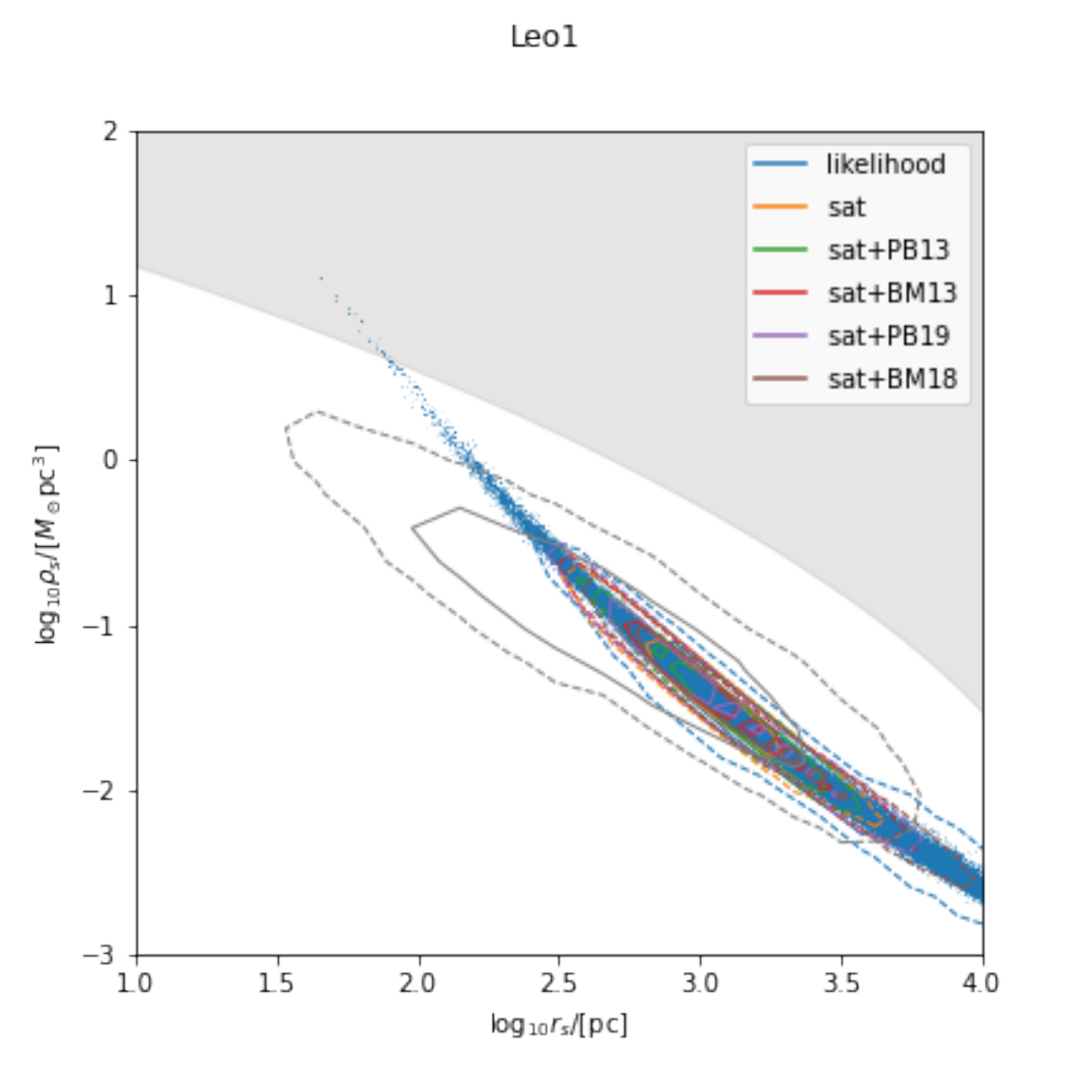}
    \includegraphics[width=\posteriorfigwidth]{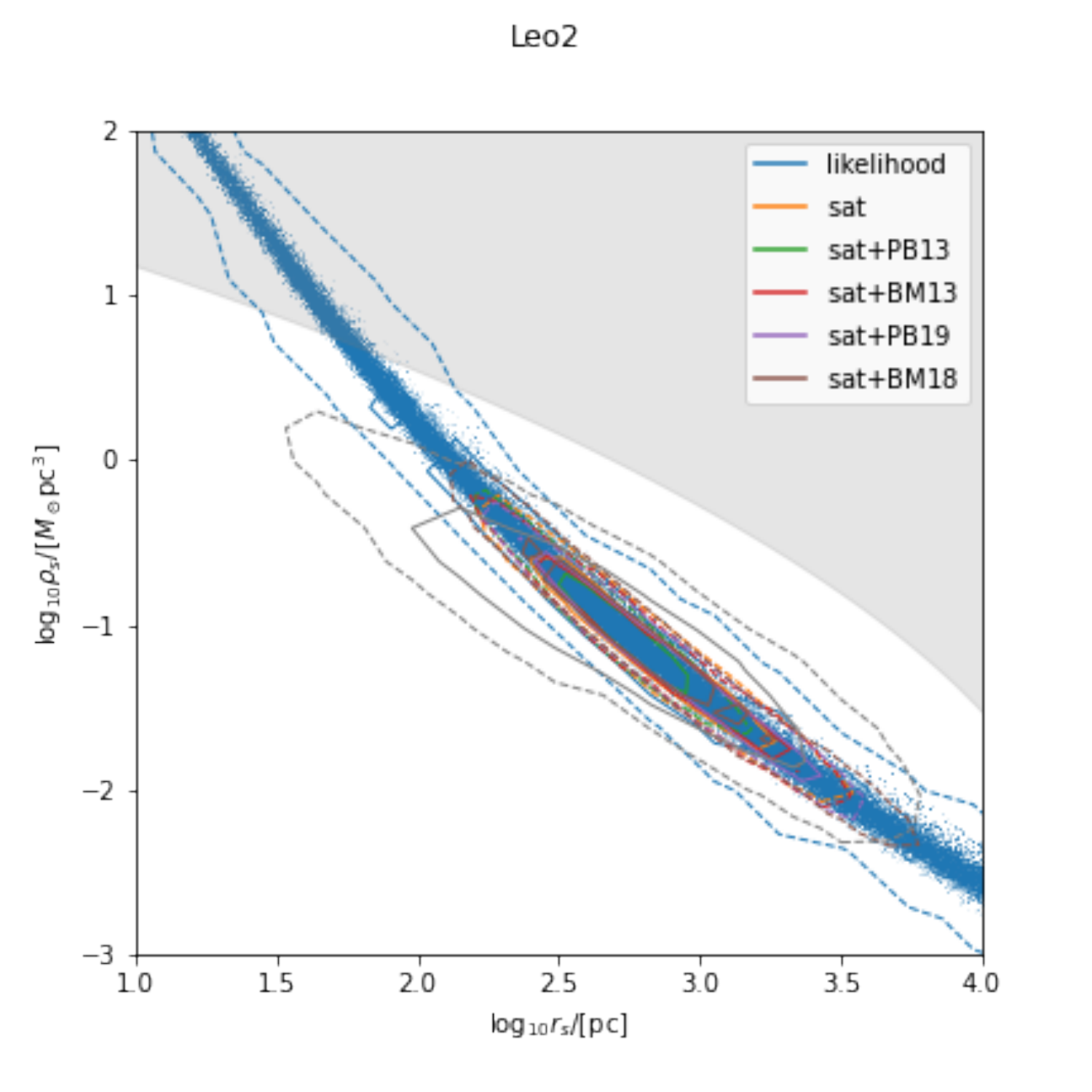}
    \includegraphics[width=\posteriorfigwidth]{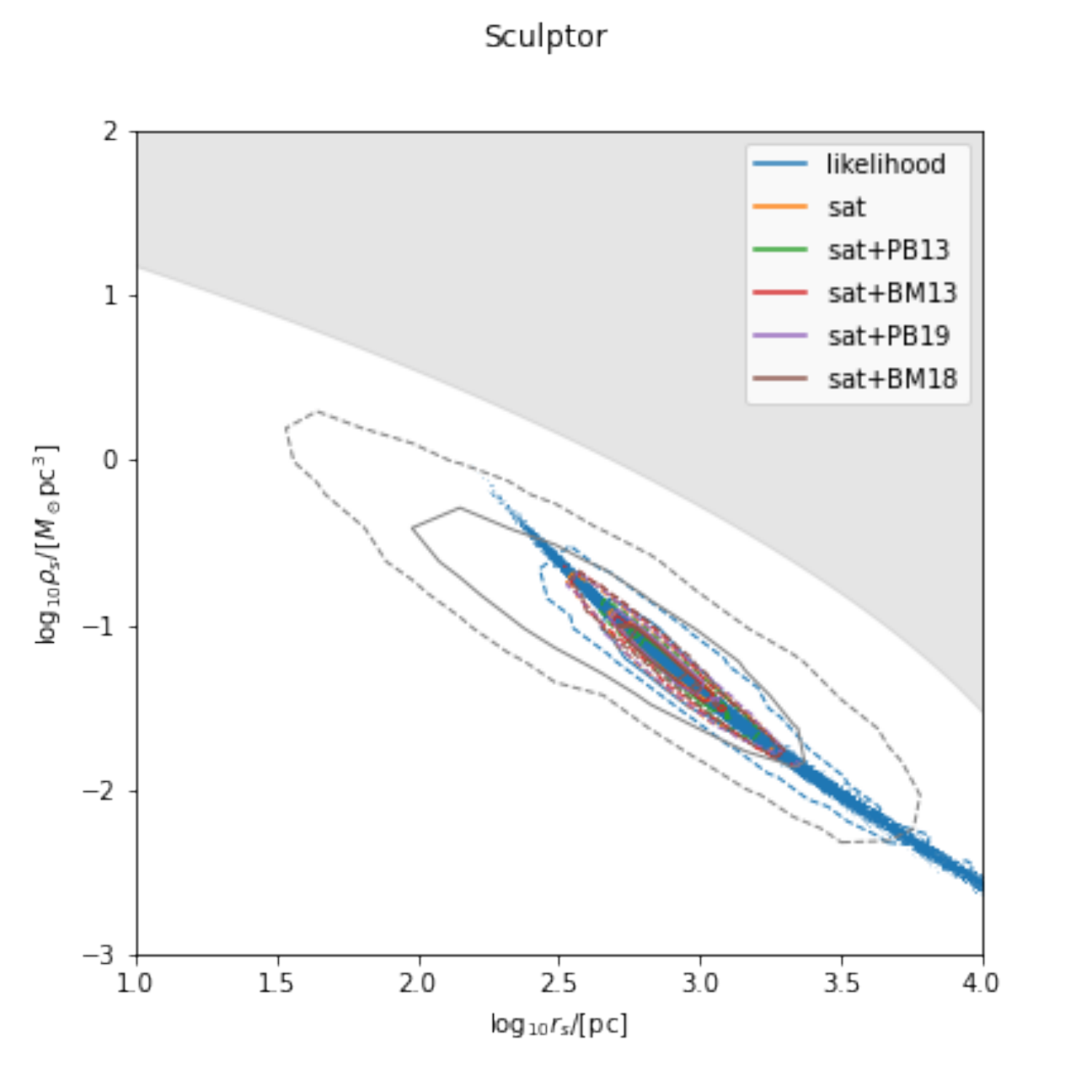}
    \includegraphics[width=\posteriorfigwidth]{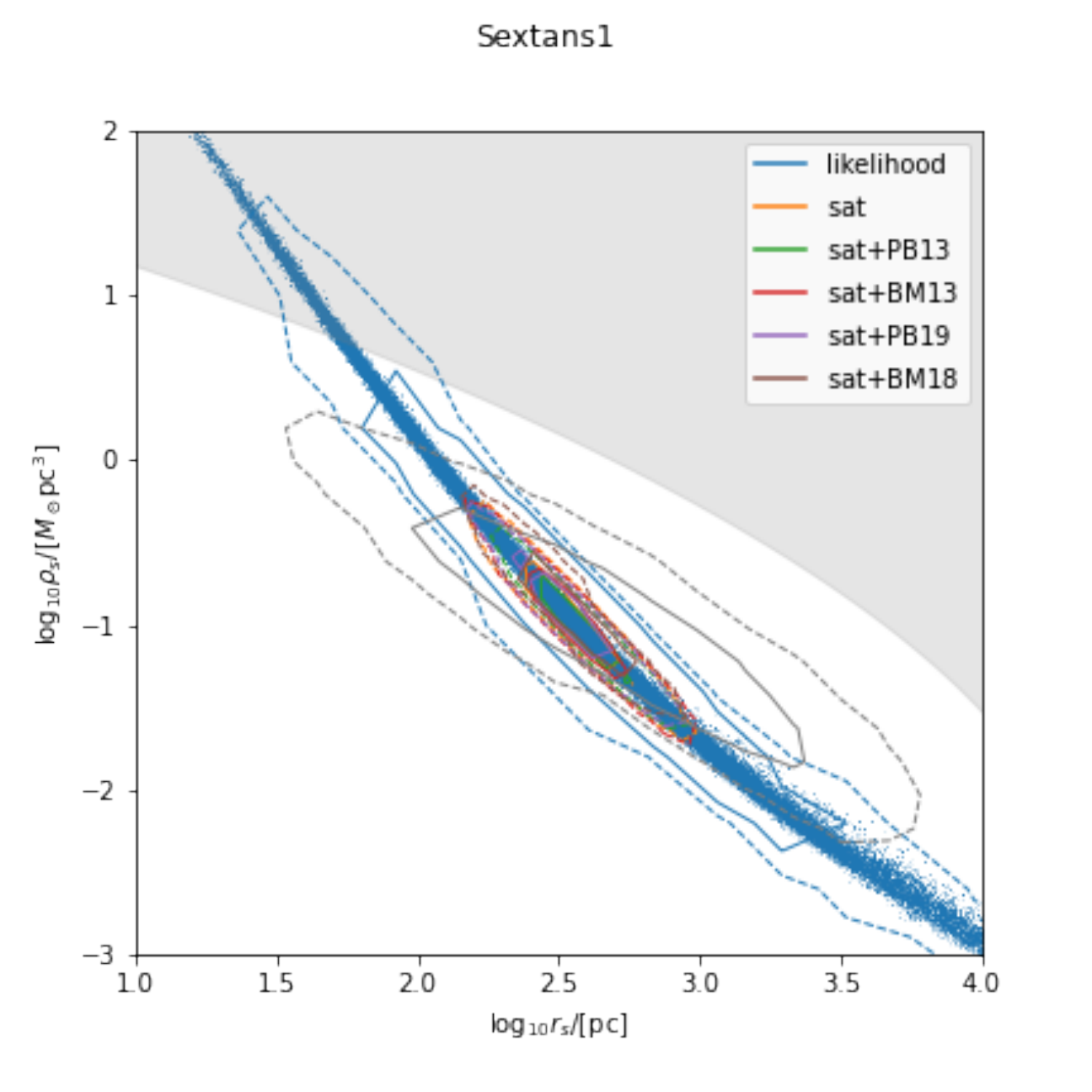}
    \includegraphics[width=\posteriorfigwidth]{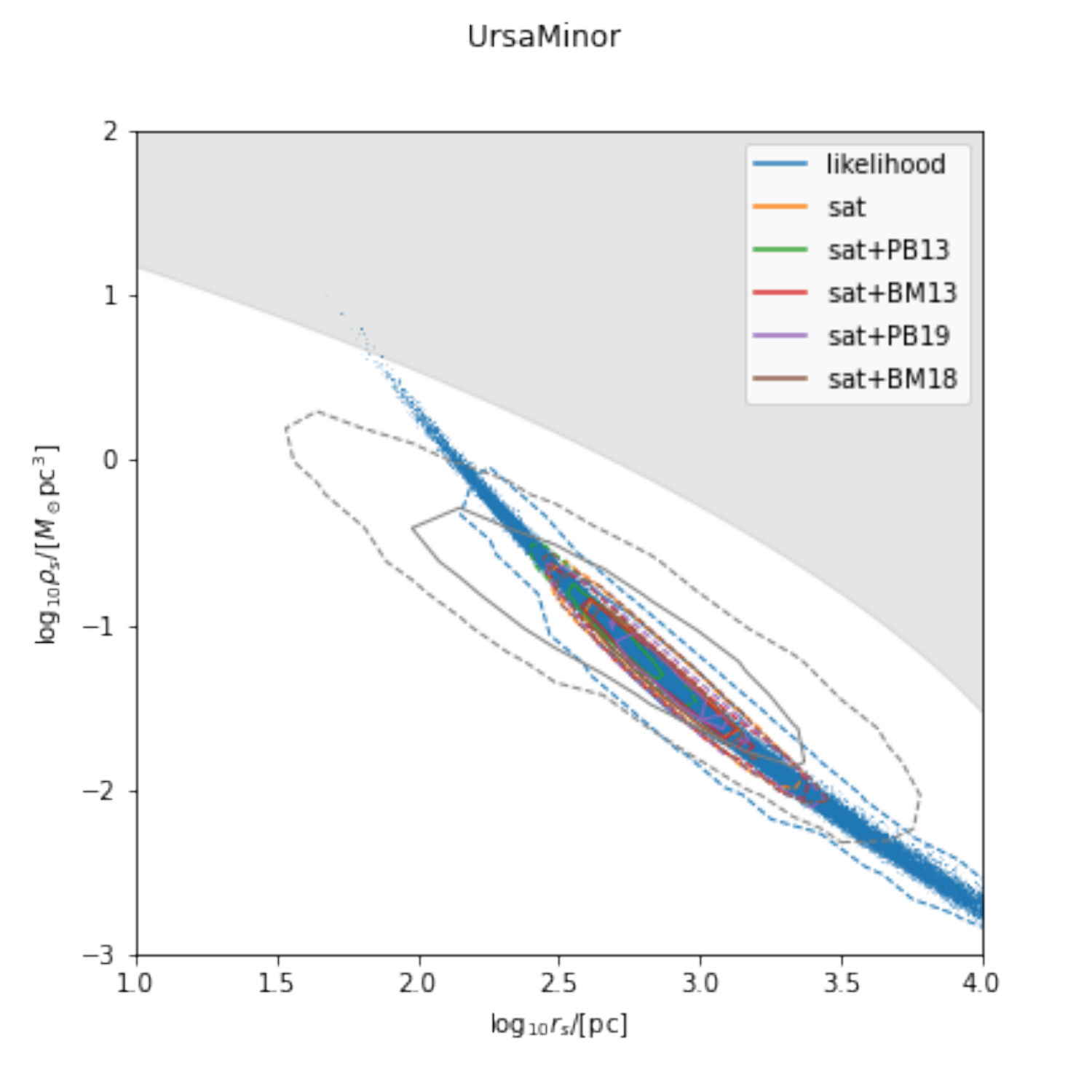}
    \caption{\tableandfigurefontsize
    Same figure as \cref{fig:posteriors_v50_105,fig:posteriors_v50_180} but for classical dSphs.}
    \label{fig:posteriors_classical}
\end{figure}

\newpage
\bibliographystyle{unsrt}
\bibliography{ref}

\end{document}